\documentclass[aps,prx,reprint,showpacs,twocolumn,notitlepage,superscriptaddress]{revtex4-1}
\usepackage[utf8]{inputenc}
\usepackage[american,british]{babel}
\usepackage[T1]{fontenc}
\usepackage[pdftex]{graphicx}  
\usepackage{graphicx, xcolor}
\usepackage{dcolumn}
\usepackage{bm}
\usepackage{amsmath,amsthm,amssymb}
\usepackage{hyperref}
\usepackage{xcolor}
\hypersetup{colorlinks,bookmarksopen,bookmarksnumbered,
citecolor=magenta,
linkcolor=magenta,
pdfstartview=false,
urlcolor=magenta}
\usepackage{verbatim}
\usepackage{algorithm}
\usepackage[noend]{algpseudocode}

\definecolor{va}{rgb}{0.0, 0.5, 0.0}

\begin{document}
\title{Quantum information spreading in random spin chains}

\author{Paola Ruggiero}
\affiliation{King’s College London, Strand, WC2R 2LS London, United Kingdom}
\author{Xhek Turkeshi}
\affiliation{JEIP, USR 3573 CNRS, Coll\`ege de France, PSL Research University,
11 Place Marcelin Berthelot, 75321 Paris Cedex 05, France}

\begin{abstract}
We study the spreading of quantum correlations and information in a one-dimensional quantum spin chain with critical disorder as encoded in an infinite randomness fixed point. 
Specifically, we focus on the dynamics after a quantum quench of the R\'enyi entropies, of the mutual information and of the entanglement negativity in the prototypical XXZ spin chain with random bonds and anisotropy parameters.
We provide analytic predictions in the scaling regime based on real-space renormalization group methods. We support these findings through numerical simulations in the non-interacting limit, where we can access the scaling regime. 
\end{abstract}

\maketitle

\section{Introduction}
Understanding the spreading of quantum information in disordered many-body quantum systems is a formidable but central task in condensed matter and statistical mechanics~\cite{abanin2019colloquium,nandkishore2015manybody,dalessio2016fromquantum,laflorencie2016quantum}. 
Compared to the linear growth of bipartite entanglement under generic \textit{clean} unitary evolution~\cite{calabrese2004entanglement,calabrese2005evolution,eisert2006general,kim2013ballistic,calabrese2016quantum,ho2017entanglementdynamics,nahum2017quantumentanglement,mezei2017onentanglement}, quantum many-body systems with strong disorder are characterized by a slower (sub-linear) rate of quantum information propagation~\cite{dechiara2006entanglement,znidaric2008manybody,bardarson2012unbounded,igloi2012entanglement,levine2012full,bardarson2012unbounded,zhao2016entanglement,iemini2016signatures,huang2017entanglement,znidaric2018entanglement,zhao2019logarithmic,sierant2022stability,maccormack2021operator}.
A prominent example is the many-body localized phase, where the slow logarithmic entanglement growth is a consequence of the effective interactions between the localized orbitals, \textit{i.e.} an infinite number of emergent local integral of motion~\cite{serbyn2013local,huse2014phenomenology,ros2015integrals,abanin2019colloquium,imbrie2016diagonalization,imbrie2016onmanybody,titas2020time,sierant2021challenges}, that arise as effective degrees of freedom from a renormalization group perspective~\cite{vosk2013manybody,vosk2014dynamical}. 

On the other hand, the propagation of quantum information in multipartite many-body quantum systems has remained, at present,  elusive. For standard models of many-body localization transition, these subjects have been partially addressed in Ref.~\cite{herviou2019multiscale} where the generation of entanglement clusters has been addressed, and in Ref.~\cite{sierant2022stability} where the maximum quantum mutual information has been used as a probe for the localization transition.

In this paper, we study the problem of quantum information spreading for spin chains with strong quenched randomness. We use the real-space renormalization group method (RSRG) introduced in Ref.~\cite{vosk2013manybody} to analyze the dynamics of R\'enyi entropy, quantum mutual information and entanglement negativity~\cite{vidal2002computable,shapourian2017partial},
focusing on the archetypal random bond XXZ chain~\cite{vasseur2016particlehole,detomasi2020anomalous}.
At equilibrium, this model is known to host a random singlet phase~\cite{ma1979random,dasgupta1980lowtemperature,doty1992effects}, with the ground state entanglement entropy and negativity exhibiting critical features ascribable to rare singlets connecting arbitrary far regions~\cite{refael2004entanglement,laflorencie2005scaling,refael2009criticality,ruggiero2016entanglement,fagotti2011entanglement,turkeshi2020entanglement,turkeshi2020negativity}.
The dynamics of the entanglement entropy following a quantum quench from an initial product state has been discussed analytically through the real-space renormalization group in Ref.~\cite{vosk2013manybody}, and numerically in Ref.~\cite{zhao2016entanglement,kiefer2020bounds}. Building on these previous works, we use the RSRG~\cite{igloi2005strong,monthus2018strong} to obtain analytic predictions for the propagation of the quantum mutual information and of the entanglement negativity. To test our results, we perform numerical simulations within the limit of no-interaction and find qualitative agreement between our analytic findings and the numerical predictions. 

We find that in the non-interacting limit entanglement is carried only by pair of spins oscillating arbitrary far in the system, which implies the lack of multipartite entanglement beyond two-party. 
One of the main consequences is the logarithmic negativity is proportional to the R\'enyi mutual information (of R\'enyi index $\alpha=1/2$), which generically is only an upper bound for the entanglement. 
Both these quantities exhibit a first $\ln \ln t$ growth in time (up to times that scale exponentially with subsystem size), followed by a decrease in time and finally by a saturation to a non-thermal value in the infinite time limit. 
Additionally, the lack of multipartite entanglement reflects in the growth of these measures being proportional to the time-average contribution of a single oscillating pair. 

Interactions, on the other hand, build up multipartite entanglement clusters, which increase the propagation rate of entanglement.
In this case, we are able to predict the growth in time (in the scaling limit) of quantum mutual information and logarithmic negativity for an initial exponentially extended window of time and show it to be logarithmic-like. 
Moreover, in the late time regime, we expect  saturation to a zero value in this case.

Despite we limit the analysis to the random bond XXZ, we expect similar results to apply in more general strong disorder quantum many-body systems, provided that resonances are irrelevant~\cite{vosk2013manybody}. In particular, the physical phenomenology is fully captured by the flow of the coupling distribution.
Our findings provide analytical and numerical insights into the dynamical multipartite entanglement structure of random spin chains.

The paper is structured as follows. In Sec.~\ref{sec:method} we introduce the model of interest and review the time-dependent strong disorder real space renormalization group method. In Sec.~\ref{sec:propagation} we introduce the entanglement measures we study, and we discuss the consequences of the renormalization group approach for quantum information spreading both for non-interacting and interacting Hamiltonians. 
In Sec.~\ref{sec:numcheck} we present numerical results in the non-interacting limit, supporting the analytic predictions.
We conclude the paper in Sec.~\ref{sec:conclusion}, and postpone two Appendices for the most technical aspects of our work and for further numerical checks.

\section{Real-space renormalization group} 
\label{sec:method}

We are interested in the dynamics of a one-dimensional chain of length $L$ of qubits described by the Hamiltonian 
\begin{align}
    H &= \sum_{i=1}^L H_{i,i+1} \nonumber \\
    H_{i,i+1}&\equiv \frac{J_{i}}{2}\left(\sigma^+_i \sigma^-_{i+1} + \sigma^-_i \sigma^+_{i+1} + 2\Delta_i \sigma^z_i \sigma^z_{i+1}\right)\label{eq:defmodel},
\end{align}
with $\sigma_i^\alpha$ ($\alpha=x,\ y,\ z)$ denoting the Pauli matrices at each site $i$, and  $\sigma^\pm_i=(\sigma^x_i\pm i\sigma^y_i)/2$; $J_i>0$ and $\Delta_i$ are independent random variables drawn randomly from the uncorrelated probability distributions $P_J(J_i)$ and $P_\Delta(\Delta_i)$, respectively. 
For definiteness, in the following we focus on $0\le \Delta_i<1$, but a similar treatment can be generalized to $|\Delta_i|<1$~\cite{fisher1994random}. 
The Hamiltonian in Eq.~\eqref{eq:defmodel} can be mapped to a fermionic interacting model via the Jordan-Wigner transformation~\cite{lieb1961two}. In particular, the limit $\Delta_i=0$ corresponds to the random bond XX model, that is the Dyson model~\cite{dyson1953the}. (In the following, we shall denote the non-interacting limit as Dyson model). 
The interacting model is known to present a quantum spin glass phase~\cite{vosk2013manybody,vasseur2016particlehole} which has been recently numerically investigated in Ref.~\cite{detomasi2020anomalous}. In the following, we use the RSRG method to compute the spreading of quantum information in a quench from an initial product state. 

\subsection{Strong disorder renormalization group}
\label{sec:equilibrium}
Before discussing the time-dependent RSRG, it is convenient to first review the strong disorder renormalization group method for the Hamiltonian in Eq.~\eqref{eq:defmodel} at equilibrium at zero temperature. Here, the key idea is to iteratively integrate out the degrees of freedom interacting through the strongest couplings in the chain $\Omega=\max J_i$. Without loss of generality, we assume the largest bond is $\Omega = J_n$. Denoting $J_\mathrm{L}=J_{n-1}$ and $J_\mathrm{R}=J_{n+1}$ the left and right neighboring bonds, strong disorder implies that typically $\Omega\gg J_\mathrm{L},\ J_\mathrm{R}$~\cite{ma1979random,dasgupta1980lowtemperature}. 
Hence, at leading order in $\Omega$, the spins $S_n$ and $S_{n+1}$ connected by the strongest bond are projected to the ground state of $H_{n,n+1}$ (cf.  Eq.~\eqref{eq:defmodel}), i.e. the singlet state
\begin{equation}
    |s_{n,n+1}\rangle = \frac{|\uparrow_n \downarrow_{n+1}\rangle- |\downarrow_n \uparrow_{n+1}\rangle}{\sqrt{2}}.
\end{equation}
Subleading contributions, that are due to quantum fluctuations and are captured by time-independent perturbation theory~\cite{sakurai}, and for $0\le \Delta_i<1$ lead to an effective coupling~\cite{fisher1994random}
\begin{equation}
    \tilde{J}_n \simeq  \frac{J_L J_R}{\Omega},\qquad \tilde{\Delta}_n \simeq \frac{\Delta_{n-1}\Delta_{n+1}}{4}\label{eq:effsdrg}
\end{equation}
between the spins $S_{n-1}$ and $S_{n+2}$. Importantly, the procedure preserves the XXZ structure of the (effective) Hamiltonian; hence the subsequent renormalization steps are always described by Eq.~\eqref{eq:defmodel} provided the couplings are substituted with the effective couplings in Eq.~\eqref{eq:effsdrg}.

Iterating the above procedure, the disorder strength increases, therefore justifying the validity of the perturbative elimination. In particular, the scaling limit is fully characterized by the probability distribution of the effective couplings, $P_J(\tilde{J}_i)$ and $P_\Delta(\tilde{\Delta}_i)$. 
Those distributions flow towards the so-called infinite randomness fixed point, where the variance of some coupling strength diverges. (For the model Eq.~\eqref{eq:defmodel}, these are the random hoppings $\tilde{J}_i$). Remarkably, it has been proven that for any given distribution for which $J_i>0$, the flow equation has a unique solution, namely that the fixed point is unique~\cite{doty1992effects,fisher1994random}.

A direct consequence of the strong disorder renormalization group is that the ground state of the many-body system gets approximated by the product of singlets connecting spins at arbitrary distances, the so-called random singlet phase (RSP).

Moreover, the interaction $\Delta_i$ plays an \textit{irrelevant} role, provided $0\le \Delta_i< 1$. In particular, these values belong to the same fixed point of $\Delta_i=0$~\cite{fisher1994random}. (This is clear from the decimation rule Eq.~\eqref{eq:effsdrg}, which express the asymptotic freedom for  $\tilde{\Delta}_i$).
As we shall see, the same does not hold for the dynamical system, where interactions are responsible for quantum correlations among \emph{clusters} of spins.

\subsection{Time-dependent real-space renormalization group}
\label{sec:tttt}
The time-dependent real-space renormalization group is the out-of-equilibrium extension of the previously discussed strong disorder renormalization group~\cite{vosk2013manybody,vosk2014dynamical}. The key difference is that the RSRG aims to construct the effective scaling dynamics via the iterative elimination of the degrees of freedom oscillating with the highest frequency~\cite{monthus2018floquet,bukov2015universal}. In particular, the renormalization group decimation does not project the spin pairs into the singlet sector but generates effective degrees of freedom that control the late time dynamics for large systems. In this section, we sketch the ideas behind RSRG, and summarize the key results. We detail in the Appendix~\ref{app:aaa} the derivation of the main equations. 

We consider the quench dynamics generated by preparing the system in a product state $|\Psi_0\rangle$, and letting it evolve through Eq.~\eqref{eq:defmodel}. Most of these results will hold for generic initial product states, provided the total magnetization is the same $S_\mathrm{tot}^z = \sum_i \sigma^z_i = 0$~\cite{vosk2013manybody,igloi2012entanglement}. (The Hamiltonian Eq.~\eqref{eq:defmodel} has a $\mathrm{U}(1)$ symmetry generated by the total magnetization ). 

Assuming strong disorder, we again consider the strongest bond $\Omega$. Without lost of generality, and following the notation in Sec.~\ref{sec:equilibrium}, we assume the strongest bond act between $S_n$ and $S_{n+1}$. We denote $S_\mathrm{L}=S_{n-1}$, $S_\mathrm{R}=S_{n+2}$ respectively the left and right neighboring spins, connected to $S_n$ and $S_{n+1}$ respectively by $J_\mathrm{L}=J_{n-1}$ and $J_\mathrm{R}=J_{n+1}$. In a similar fashion, we denote $\Delta_\Omega$, $\Delta_\mathrm{L}$ and $\Delta_\mathrm{R}$ the interactions associated with, respectively, the bonds $\Omega$, $J_\mathrm{L}$ and $J_\mathrm{R}$. The Hamiltonian~\eqref{eq:defmodel} can be then decomposed as
\begin{equation}
\label{eq:decompos}
    H = H_\Omega + V + H_\mathrm{chain},
\end{equation}
where
\begin{align}
	H_\Omega &= \frac{\Omega}{2} ( \sigma^+_n \sigma^-_{n+1} + \sigma^-_n \sigma^+_{n+1} + 2 \Delta_\Omega \sigma^z_n \sigma^z_{n+1} ),\nonumber\\\label{eq:defdecom}
	H_\textup{chain} &=\sum_{i\neq n\pm 1,n} \frac{J_i}{2}(\sigma^+_i \sigma^-_{i+1} + \Delta_i\sigma^z_i\sigma^z_{i+1} + \text{h.c}),\\
	V &= \frac{J_\mathrm{L}}{2} ( \sigma^+_L \sigma^-_{n} + \sigma^-_L \sigma^+_{n} + 2 \Delta_\mathrm{L} \sigma^z_L \sigma^z_{n} )  \nonumber\\ &\qquad + \frac{J_\mathrm{R}}{2} ( \sigma^+_{n+1} \sigma^-_{R} + \sigma^-_{n+1} \sigma^+_{R} + 2 \Delta_\mathrm{R} \sigma^z_{n+1} \sigma^z_{R} ).\nonumber
\end{align}
The separation of energy scales induced by disorder reflects in the short-time dynamics being dominated by oscillation with frequency $\Omega$ between the two spins on sites $n$ and $n+1$. Their dynamics is hence associated with the strong bond Hamiltonian $H_\Omega$, which entangles them at time $t\approx 1/\Omega$. 
Furthermore, since the frequency of the eigenmodes of  $H_\Omega$ is larger than the scaling of the remaining degrees of freedom, the sites $n$ and $n+1$ are seen by the remaining degrees of freedom as in the stationary (i.e. time-averaged) state. 

\begin{figure}
	\includegraphics[width=\columnwidth]{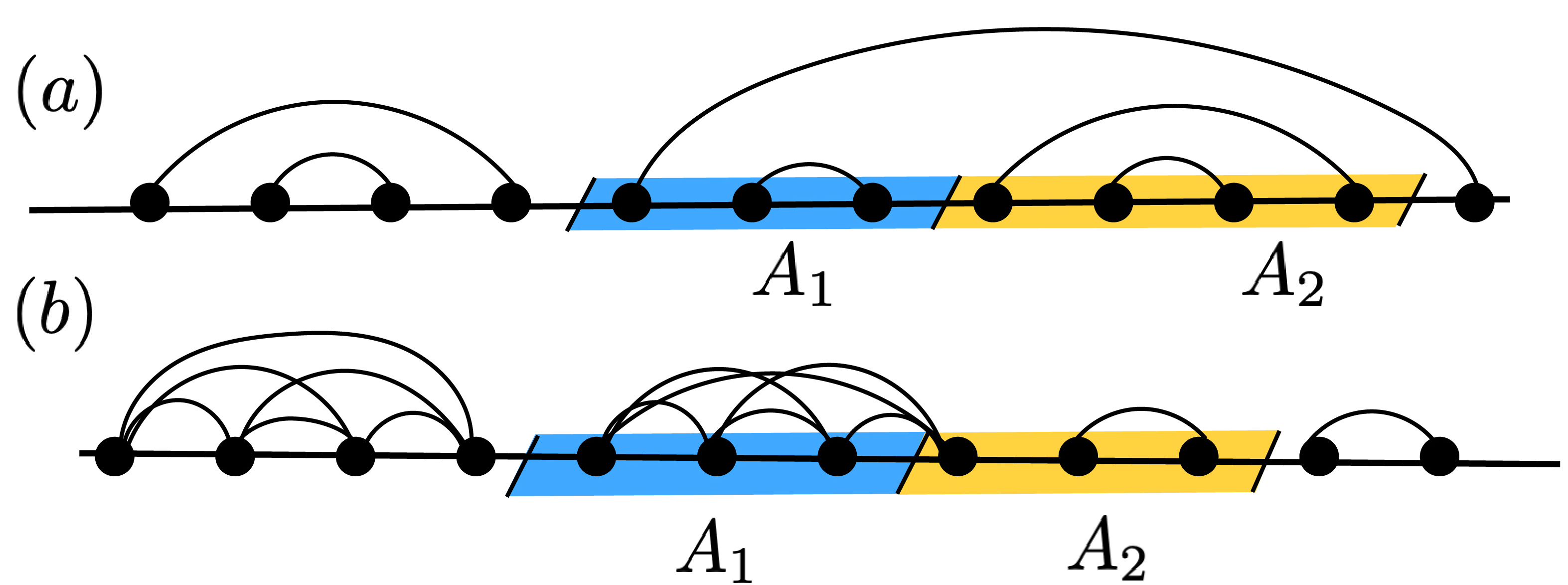}
	\caption{\label{fig:state_t} \textit{Cartoon of the dynamical states---} (a): The state at a given time $t$ within the renormalization group approximation is given as a
	product of oscillating pairs of spins spreading throughout the system, denoted as bonds joining the spins (black dots). 
	(b) Interactions lead to the formation of clusters of spins (dots joined by multiple bonds in the figure), since the system is a coherent superposition of fixed renormalization trajectories. The entanglement shared by two generic subregions $A_{1}$ and $A_2$ is related to the bonds connecting them and is clearly different in the (a) and (b) scenario.} 
\end{figure}

The above idea is formalized within the Floquet high-frequency expansion~\cite{monthus2018floquet,bukov2015universal}, which consists of deriving an effective time-independent Hamiltonian $H_{\rm eff}$ via a second-order expansion of the time-evolving operator in the interaction picture induced by $H_\Omega$. (See also Appendix~\ref{app:aaa}).
For this purpose, it is convenient to reformulate the dynamics within the interaction picture~\cite{sakurai}. The state in the interaction picture is given by
\begin{align}
    |\Psi^I_t\rangle &= e^{-i H_\Omega t} |\Psi_t\rangle;
\end{align}
and the unitary evolution is governed by
\begin{align}
\label{eq:floqqqq}
	U_I(t) = \mathcal{T}\exp\left(-i\int_0^t H_I(t) dt\right) 
\end{align}
with $\mathcal{T}$ denoting time-ordering, and the interaction Hamiltonian is given by
\begin{equation}
H_I(t) = e^{-i H_\Omega t} (H-H_\Omega) e^{+i H_\Omega t}.
\end{equation}
The decimation procedure amounts to finding the effective Hamiltonian $H_{\rm eff}$ of the system defined through 
\begin{equation}
\label{eq:perturbativesh}
U_I (t) \simeq \exp\left(-i H_\textup{eff} t\right) + O\left(\frac{1}{\Omega^2}\right).
\end{equation}
A key consequence is that now we can view the sites connected by the high-frequency bond as a new effective degree of freedom $(n,n+1)\equiv \tilde{n}$, and the high-frequency Hamiltonian as a single qudit operation on this degree of freedom. 
In principle, this state span a 4-dimensional Hilbert space; however, for a given initial state, at leading order in perturbation theory, the effective dynamics will be effectively restricted to subsectors of this Hilbert space.

Suppose, for example, we start from an initial product state, and that the high-frequency bond acts on $|\uparrow \uparrow\rangle$. In this case, the effective dynamics is fully accounted for by a global phase, and the emergent effective degree of freedom (i.e. the aforementioned qudit) is frozen.
If instead the high-frequency bond acts on $|\uparrow \downarrow\rangle$, the effective qudit becomes two-dimensional and oscillates between $| {\pm} \rangle \equiv (|\uparrow \downarrow\rangle \pm \downarrow \uparrow  \rangle )/{\sqrt{2}} $.

Overall, the detailed choice of initial product state would result in a different transient dynamics, but the scaling properties of the system are expected to be the same. (See also the numerical benchmarks in Ref.~\cite{igloi2012entanglement}, and in Appendix~\ref{app:bbb}).

We find convenient to choose as initial condition the N\'eel state
\begin{equation}
    |\Psi_0\rangle = |\uparrow\downarrow\uparrow\downarrow\cdots \uparrow\downarrow\rangle,\label{eq:neel}
\end{equation}
for which ${|\uparrow \uparrow \rangle}$ and ${|\downarrow \downarrow \rangle}$ are never populated due to $\mathrm{U}$(1) conservation constraints. In this case, the effective Hamiltonian is given by
\begin{equation}
\label{eq:Heff}
    \begin{split}
	H_\textup{eff} = H_\textup{chain} + h_{\tilde{n}} \tilde\sigma_{\tilde{n}}^z  
	 +  \frac{ J_\mathrm{L} J_\mathrm{R}}{2\Omega (1-\Delta^2_\Omega)} (\sigma^+_\mathrm{L}\sigma^-_\mathrm{R} + \sigma^-_\mathrm{L}\sigma^+_\mathrm{R}) \tilde{\mathbb{I}}_{\tilde{n}} \\ 
	+ \frac{\Delta_\Omega J_L J_R}{2\Omega} \left[\frac{\sigma^+_\mathrm{L}\sigma^-_\mathrm{R} + \sigma^-_\mathrm{L}\sigma^+_\mathrm{R}}{1-\Delta^2_\Omega} - \frac{\Delta_\mathrm{L} \Delta_\mathrm{R}}{\Delta_\Omega} \sigma^z_\mathrm{L} \sigma^z_\mathrm{R} \right]\tilde\sigma^z_{\tilde{n}},
    \end{split}
\end{equation}
where the following operators act on the effective degree of freedom
\begin{equation}
\label{eq:defgauge}
	\tilde{\mathbb{I}}_{\tilde{n}} = |{+}\rangle\langle {+}| + |{-}\rangle \langle {-}|,\qquad \tilde{\sigma}^z_{\tilde{n}} = |{+}\rangle \langle {+} | - |{-}\rangle \langle{-}|,
\end{equation}
and we further introduced the magnetic field $h_{\tilde{n}} = -(\Delta_\mathrm{L}^2 J_\mathrm{L}^2+ \Delta_\mathrm{R}^2 J_\mathrm{R}^2)/(4\Omega)$. (Note however that these single-spin magnetic contributions can be gauged away because of the conservation of the total magnetization). For notational convenience, when no confusion arises, in the following we sometime omit the tilde in the lattice index. 

The last coupling in Eq.~\eqref{eq:Heff} gives an intuition on why multipartite entanglement is generated in the dynamics. When $\Delta_i\neq 0$, these coupling create a mixed superposition between the states $|{\pm}\rangle$ and the neighbors, at times $t\sim O(2\Omega/(J_\mathrm{L} J_\mathrm{R} \Delta_\Omega))$~\cite{vosk2013manybody}. Eq.~\eqref{eq:Heff} also clarifies that at $\Delta_i=0$ (non-interacting limit), no multipartite entanglement is genuinely generated, as the effective degree of freedom $\tilde{S}_{\tilde{n}}$ is effectively decoupled. Lastly, we note that $\Delta_i=1$ is a peculiar point. Here RSRG fails, since at this point SU(2) invariance is restored, and resonances play a pivotal role in scrambling the degrees of freedom, and in leading to thermalization~\cite{protopopov2017effect,Parameswaran_2018,protopopov2020nonabelian,vasseur3,vasseur2}. 

It is easy to see that $[H_\mathrm{eff},\tilde{\sigma}^z_{\tilde{n}}]=0$, and hence the time evolution can be computed separately for each sector $\tilde{\sigma}^z_{\tilde{n}}=\pm 1$. 
Specifically, the effective Hamiltonian bifurcate into two sectors $H^\pm_\mathrm{eff}= H_\mathrm{eff}(\tilde{\sigma}^z_{\tilde{n}}=\pm1)$, which have the same form of the Hamiltonian in Eq.~\eqref{eq:defmodel}, with renormalized couplings
\begin{align}
    \tilde{J}^\pm_{\tilde{n}} &= \frac{J_\mathrm{L} J_\mathrm{R}}{\Omega (1\mp \Delta_\Omega)}\\
    \tilde{\Delta}^\pm_{\tilde{n}} &= \mp\frac{\Delta_\mathrm{L}\Delta_\mathrm{R}}{4}(1\pm \Delta_\Omega).\label{eq:exactdecimation}
\end{align}
(We stress the overall sign in the coupling $\Delta$ in Eq.~\eqref{eq:exactdecimation}  has only the effect of reversing the effective degree of freedom $|{+}\rangle\leftrightarrow |{-}\rangle$.)
Since this bifurcation arise at any renormalization group time, a fixed choice of $\{\tilde{\sigma}^z_n\}$ fixes a unique flow of the parameters $\{\tilde{J}_i\}$ and $\{\tilde{\Delta}_i\}$. We denote each such realization as a \textit{renormalization group trajectory}. 

Importantly, the asymptotic form of any trajectory is the same. 
Indeed, Eq.~\eqref{eq:exactdecimation} shows that the couplings $\Delta$ are reducing in strength at subsequent iterations (asymptotic freedom).  This has the relevant consequence that the renormalized coupling distribution are asymptotically independent. 
In particular, the late-time effective decimation within each renormalization group trajectory is captured by the same decimation rules occurring for the equilibrium strong disorder renormalization group Eq.~\eqref{eq:effsdrg}. 
(We refer to  Ref.~\cite{fisher1994random} for an indepth discussion of those flow equations and their stability, while here we pinpoint the key ideas.)

Starting from the joint distribution  $P_{J,\Delta}(\tilde{J},\tilde{\Delta}; \Omega)$, where $\Omega$ is the largest energy scale at a specific renormalization group step, the decimation rule Eq.~\eqref{eq:effsdrg} induces a differential equation for $P_{J,\Delta}$ as $\Omega$ is decreased
\begin{equation} \label{eq:flowJDelta}
	\frac{\partial{P_{J, \Delta}}}{\partial\Omega}= \mathcal{F}[P_{J, \Delta}].
\end{equation}
Here $\mathcal{F}$ is a functional of $P_{J,\Delta}$ (see Ref.~\cite{vosk2013manybody} for the full expression). 
Switching to the logarithmic variables
\begin{equation} \label{eq:log_var_dyn}
\zeta_i= \log \frac{\Omega}{J_i} ,\quad \beta_i= -\log |\Delta_i|,\quad \Gamma=\log \frac{\Omega_0}{\Omega} =\log \Omega_0 t,
\end{equation}
with $\Omega_0,\Omega\sim 1/t$ the initial and the reduced energy scale, respectively,
and integrating Eq.\eqref{eq:flowJDelta} over the interaction variable $\beta$, we obtain
\begin{equation} \label{eq:flow_zeta}
\frac{\partial \rho_{\zeta}}{\partial \Gamma} = \frac{\partial \rho_{\zeta}}{\partial \Gamma} + \rho_{\zeta} (0)\int_0^{\infty} d\zeta_L d\zeta_R \delta (\zeta - \zeta_L - \zeta_R) \rho_{\zeta} (\zeta_L)  \rho_{\zeta} (\zeta_R)
\end{equation}
where $\rho_{\zeta} (\zeta; \Gamma) = \int d \beta P_{\zeta,\beta} (\zeta,\beta; \Gamma)$ is the distribution of $\zeta$ at renormalization group time $\Gamma$.
At equilibrium, the flow variable $\Gamma$ is related to the decimation length scale $\Gamma\sim \sqrt{\ell}$, while in this time-dependent RSRG it is related to the real time scale (via the high-frequency integration) $\Gamma\sim \log t$ (see also Sec.~\ref{sec:propagation}). 
The solution of the flow equation is given by 
\begin{equation} \label{eq:solution_flowEq}
\rho_{\zeta} (\zeta; \Gamma) = a (\Gamma) e^{-a (\Gamma) \zeta},\quad a(\Gamma) = \frac{1}{\Gamma + 1/a_0} \ .
\end{equation}
which holds for any real $a_0$. 


\subsection{Predictions and limitations of the renormalization group approach} 
\label{sec:predilim}
We now briefly discuss the predictions and limitations of the dynamical strong disorder renormalization group. 

In the non-interacting case, it follows from the effective Hamiltonian in Eq.~\eqref{eq:Heff} that $\tilde{\sigma}^z_n$ are effectively decoupled degrees of freedom. Their role is completely accounted for by the virtual interactions between their neighboring sites, which oscillates with a renormalized frequency. 
Starting from the N\'eel state, it is hence clear that in the scaling time limit, the state of the system gets approximated by a product of oscillating singlets extending at arbitrary range~\footnote{The most general case of random product state is a product of singlets (cf. Eq.~\eqref{eq:singlets}) and frozen degrees of freedom.}
\begin{equation}
\label{eq:singlets}
    |\Psi_t\rangle = \left(\prod_{(i,j)\in I} |s_{i,j}(t)\rangle\right) \left(\prod_{k\in I_c} |s_k\rangle\right) ,
\end{equation}
with $|s_{i,j}(t)\rangle = (|\uparrow_i\downarrow_j\rangle \pm e^{-i \omega_{i,j} t} |\downarrow_i\uparrow_j\rangle)/\sqrt{2}$, and $\omega_{i,j}$ the oscillating frequency of the pair; moreover,
the set $I$ contains all the decimated pairs, while $I_c$ the non-decimated spins; finally  $|s_k\rangle=|\uparrow\rangle,|\downarrow\rangle$. (See Fig.~\ref{fig:state_t} (a) for a pictorial representation.)

In the interacting case, instead, the system state is in a superposition of different states of the form~\eqref{eq:singlets}, each one fixed by a specific value of the conserved number $\tilde{\sigma}^z_n$ in Eq.~\eqref{eq:Heff}. Therefore one has
\begin{equation}
\label{eq:clustr}
    |\Psi_t\rangle = \sum_{\mu} a_\mu(t) \left[\left(\prod_{(i,j) \in I^\mu} |s_{i,j}^\mu(t)\rangle\right)\left(\prod_{k \in I_c^\mu}|s_k\rangle \right)\right].
\end{equation}
Here $\mu$ is a collective index fixing a renormalization group trajectory, i.e. a set of choices of $\{\tilde{\sigma}^z_n\}$ for subsequent renormalization times. 
For intermediate times, the state in Eq.~\eqref{eq:clustr} is described by a product of independent clusters (see Fig.~\ref{fig:state_t} (b)). 
Indeed, multiple renormalization group trajectories share the same product structure in Eq.~\eqref{eq:singlets}, with exception of localized space regions where interactions couple different singlets. 
This scenario holds for a diluted number of clusters and breaks down when large clusters interact between themselves.
We shall discuss the consequences of these results for the quantum information spreading in Sec.~\ref{sec:interactingcase}.

We conclude by discussing the limitation of strong disorder renormalization group, mainly coming from resonances between high-frequency modes integrated out in different regions of the chain. When resonances are relevant, the perturbative expansion in Eq.~\eqref{eq:perturbativesh} fails, and the system rapidly thermalizes.
Within the strong disorder regime, which is the one of interest in this paper, we discuss the conditions for which the resonances are irrelevant in the renormalization group sense for the random bond XXZ (see also Ref.~\cite{vosk2013manybody}). Suppose two pairs oscillating around $\Omega_0$ have their typical frequency separated by $\delta\Omega$. 
We use the probability distribution of bond strength Eq.~\eqref{eq:solution_flowEq} to estimate the density of couplings mismatched by $\delta \Omega$, that is given by $\varrho \simeq a(\delta\Omega/\Omega_0)$. Hence, their typical distance scales as $L_\mathrm{res} \propto a^{-1}(\Omega/\delta\Omega)$, leading to $\delta\Omega = \Omega_0/(L_\mathrm{res} a)$. 
The condition for two pairs to be resonant is for $\delta\Omega$ to be of the same order (or smaller) than their effective interaction $J_\mathrm{eff}$.
In this scenario, the resonant bonds will escape the reduced dynamics, and explore arbitrary regions of the Hilbert space. 

To estimate $J_\mathrm{eff}$, we use the renormalization group energy scale associated to the average separation (cf. Sec.~\ref{sec:tttt}) length $L_\mathrm{dist} \simeq \ln^2(\Omega_0/J_\mathrm{eff})$ ~\cite{vosk2013manybody,fisher1994random}.


Imposing that $L_\mathrm{res}\simeq L_\mathrm{dist}$, we obtain the relation
\begin{equation}
    J_\mathrm{eff}\sim \Omega_0 e^{-\sqrt{L_\mathrm{res}/a}}.
\end{equation}
Then, the condition $J_\mathrm{eff}\gtrsim \delta\Omega$ translates to a condition on the typical separation between resonant bonds
\begin{equation}
    a L_\mathrm{res} e^{-\sqrt{L_\mathrm{res}}/a}\gtrsim 1.
\end{equation}
Hence, resonances will play no relevant role provided that $a$ is sufficiently small. We recall that $a$ depends on the disorder strength, and on the choice of initial conditions.

\section{Propagation of quantum information} 
\label{sec:propagation}
We are now in position to discuss the renormalization group asymptotic predictions for the quantum information dynamics generated by Eq.~\eqref{eq:defmodel}, focusing on the N\'eel state in Eq.~\eqref{eq:neel} as our initial state (while generalization to different product states is straightforward, as already mentioned). After introducing the information measures of interest, we first discuss the case of non-interacting spin chains, and then interacting ones.

\subsection{Observables of interest}
\label{sec:qinfo}
In the remaining of this paper, we are going to consider the R\'enyi entanglement entropies, the R\'enyi mutual information, the logarithmic negativity, and fermionic negativity. 
In this subsection, we briefly review their definition, and refer to the literature for additional details.

\subsubsection{R\'enyi entropy and mutual information} 
Given a quantum system and a bipartition $A\cup B$, the entanglement entropy is defined by~\cite{Nielsen2009}
\begin{equation} \label{eq:entent}
S_{A}= - {\rm tr} \, \rho_A \ln \rho_A
\end{equation}
where $\rho_A = {\rm tr}_{B} \rho$ is the reduced density matrix associated with the subsystem $A$, and $\rho = |\Psi\rangle\langle \Psi|$ the density matrix of the whole system.
For $\alpha>0$, one can also define the R\'enyi entropies as~\cite{laflorencie2016quantum}
\begin{equation} \label{eq:renyis}
S^{(\alpha)}_{A}= \frac{1}{1-\alpha} \log {\rm tr} {\rho_A^\alpha}
\end{equation}
such that $\lim_{\alpha \to 1} S^{(\alpha)}_{A}= S_{A}$.
All R\'enyi entropies are good entanglement measures for any bipartition, provided the global state of the system is pure~\cite{berta2015renyi}. If the system is mixed, a scenario of interest when considering a tripartite (or more generally, a multipartite) system, the R\'enyi entropy is not an entanglement measure as it embodies also classical correlations. 

In the case of a tripartition $A_1\cup A_2\cup B$, a useful measure of correlations between the parties $A_1$ and $A_2$ is given by the quantum (R\'enyi) mutual information, defined from Eq.~\eqref{eq:renyis} as
\begin{equation} \label{eq:minfo}
I_{A_1:A_2}^{(\alpha)}= S_{A_1}^{(\alpha)} + S_{A_2}^{(\alpha)} - S_{A_1 \cup A_2}^{(\alpha)}
\end{equation}
As for the R\'enyi entropies in Eq.~\eqref{eq:renyis}, the mutual information is not an entanglement measure being sensitive to the total correlations, both quantum and classical. However, it provides relevant information on how the information propagates in the system. For instance, we recall that the $\alpha=1/2$ mutual information was found to be proportional to the negativity (see below for a definition) in many settings, ranging from integrable models~\cite{alba2017entanglement,alba2018entanglement,alba2019quantum_information,alba2019quantum} to conformal field thoeries~\cite{maccormack2021operator,kudlerflam2020the,kudlerflam2020thequasi}. We note that this relationship has been recently investigated also within quantum circuits~\cite{bertini2022entanglement}, where it was found to hold at early times, and within monitored and open quantum systems~\cite{shi2020entanglement, sang2021entanglement,sharma2022,weinstein2022,turkeshi2022enhanced,alba2022logarithmic}, where it was shown to hold and breakdown in specific setups and regimes. 

\subsubsection{Logarithmic and fermionic negativity} 
A computable and genuine entanglement measure for generic mixed states is the so-called entanglement negativity~\cite{vidal2002computable,plenio2005logarithmic}, which is defined in terms of the partial transposition. 

Given the tripartition $A_1\cup A_2\cup B$ and the reduced density matrix $\rho_{A}$ with $A=A_1\cup A_2$, the partial transpose $\rho_A^{T_2}$ is defined as
\begin{align}
\label{eq:patialtrans}
	\langle \varphi_1, \varphi_2 |\rho^{T_2}_A|  \varphi'_1, \varphi'_2 \rangle
	\equiv \langle \varphi_1, \varphi_2' |\rho_A| \varphi_1', \varphi_2 \rangle.
\end{align}
where $\{ \varphi_i \}$ are bases in $A_i \, (i=1,2)$.
The logarithmic negativity is defined from $\rho_A^{T_2}$ as
\begin{equation}
\label{eq:defnegbos}
\mathcal{E}_{A_1: A_2} = \log {\rm tr} \| \rho_A^{T_2} \|.
\end{equation}
where $ \| O \| = {\rm tr} \sqrt{ O O^\dagger } $ denotes the trace norm of the operator $O$.

In the following we shall consider also the \emph{fermionic} negativity, recently proposed as an entanglement monotone of particular relevance for fermionic systems~\cite{shapourian2017partial}. This is of interest to us for two reasons:
First, as mentioned above, Eq.~\eqref{eq:defmodel} maps to an interacting fermionic model via the Jordan Wigner transformation.
Furthermore, in the non-interacting limit, the fermionic negativity can be computed using polynomial resources (see Sec.~\ref{sec:numcheck}). 

For convenience, we recall here the expression of the partial time reversal in the basis $|\{s_i\}\rangle$ of the operators $\sigma^z_i$~\footnote{Note, however, the fermionic negativity, as any entanglement measure, is independent of the choice of basis.}. (Here $s_j=1$ for ${|\uparrow_j\rangle}$ and $s_j=-1$ for ${|\downarrow_j\rangle}$.)
Compared to the partial transpose in  Eq.~\eqref{eq:patialtrans}, the partial time reversal of the density matrix $\rho_A$ contains an additional phase factor~\cite{shapourian2017partial,shapourian2019entanglement,shapourian2019twisted}
\begin{align}
\label{eq:partialtrev}
	\langle \{s_i\}_1,\{s_i\}_2 &|\rho^{R_2}_A|  \{s_i'\}_1, \{s_i'\}_2 \rangle 	\equiv (-1)^{\phi(\{s_i\},\{s'_i\})} \times \nonumber \\
 &\langle  \{s\}_1, \{s'\}_2 |\rho_A| \{s'\}_1, \{s\}_2 \rangle ,
\end{align}
which is defined in terms of $\tau_k = \sum_{j\in A_k} (1+s_j)/2$ and $\tau'_k = \sum_{j\in A_k} (1+s_j')/2$ (with $k=1,2$) as 
\begin{align}
   \phi(\{s_i\},\{s'_i\}) &= \frac{\tau_1(\tau_1+2)}{2} + \frac{\tau_2'(\tau_2'+2)}{2} + \tau_2'\tau_2 \nonumber \\
    &\quad + \tau_1\tau_2 + \tau_1'\tau_2' + (\tau_1'+\tau_2')(\tau_1+\tau_2).
\end{align}
The fermionic negativity is then given by
\begin{equation}
    \label{eq:fermneg}
    \mathcal{E}^f = \log\mathrm{tr}||\rho_A^{R_2}||.
\end{equation}
In general, the fermionic and the logarithmic negativity are different. However, as we show in the next subsection, they coincide within the scaling limit for states of the form ~\eqref{eq:singlets}~\cite{shapourian2017partial}.
Using this equality, we shall use the fermionic negativity for the numerical checks in Sec.~\ref{sec:numcheck}. 

\subsection{Quantum information dynamics in the non-interacting limit} 
\label{sec:nonnnn}
In this subsection we discuss the dynamics when $\Delta_i=0$. As already observed, in this limit the model in Eq.~\eqref{eq:defmodel} reduces to a non-interacting problem of free fermions with random hoppings (the Dyson model). 
As discussed in Sec.~\ref{sec:predilim}, the state in the scaling limit is described by a product of oscillating pairs extending arbitrarily far in space (cf. Eq.~\eqref{eq:singlets}). Hence, it is convenient to first recall the dynamics of pairs formation, which is the building block for computing all the entanglement measures introduced above. 

\subsubsection{Production rate of oscillating pairs} \label{sec:pair_production}
Given an arbitrary bipartition of an infinite system 
in $A$ (of length $\ell \gg 1$) and its complement $A_c$, 
we are interested in the number of mutually shared oscillating pairs forming up to a given renormalization group time $\Gamma$. We are going to exploit the equilibrium results derived in Ref.~\cite{refael2004entanglement} for the singlets production rate, and extend them in the out-of-equilibrium scenario, where singlets are replaced by oscillating pairs.

At equilibrium, the average rate $ \gamma_n =\log( \Gamma_{n+1}/\Gamma_n )$ of singlets formation is given by $\overline{\gamma_n} =3$. In the scaling limit, this object does not depend on the renormalization step  $\overline{\gamma} \equiv \overline{\gamma_n} $. Such rate, together with the number of singlets forming over a given bond at renormalization group time $\Gamma$, i.e. $n_b \simeq \int^{\Gamma} d \Gamma' a(\Gamma')$ (cf. Eq.~\eqref{eq:solution_flowEq}), gives direct access to the total number of singlets
\begin{equation} \label{eq:nAB}
 \overline{n}_{A: A_c}  = b_{A: A_c} \cdot \frac{n_b}{\overline{ \gamma}} \simeq \frac{ b_{A: A_c}}{3} \ln (\Gamma+ 1/a_0) + O(1),
\end{equation}
where $O(1)$ takes into account subleading corrections, and $b_{X:Y}$ denotes the number of bonds shared by two subsystems $X$ and $Y$. For instance, when the subsystem $A$ consists of a single interval in the middle of the system we have $b_{A: A_c} = 2$, while when it is attached to a boundary $b_{A:A_c}=1$.

The crucial result of the RSRG procedure in Sec.~\ref{sec:method} is Eq.~\eqref{eq:flow_zeta}, namely the fact that, in the proper variables,  the renormalization group flow describing the system after a quantum quench is the same one would get at equilibrium.

Consequently, when moving to the quench setup Eq.~\eqref{eq:nAB} is still valid. What changes is the interpretation of the renormalization group time, now related to the real time through $\Gamma (t)= \ln( {\Omega_0} t )$ (cf. Eq.~\eqref{eq:log_var_dyn}). We note, however, that the number of shared oscillating pairs forming between a \textit{finite} interval $A$ and its complement $A_c$ will saturate. This is clear from the renormalization group perspective, since the saturation occurs when all the spins of the smallest interval are integrated out. On average, this happens when $\Gamma \sim \sqrt{\ell}$, with $\ell$ being the length of the interval~\cite{refael2004entanglement,refael2009criticality,vosk2013manybody}. 
Overall, the above observations can be summarized in the following expression
\begin{align}
    \overline{n}_{A:A_c}(t) &= \frac{ b_{A: A_c}}{3}\ln \left(\Gamma_{\ell} (t)+1/a_0\right)+O(1),\label{eq:tdepnab}
\end{align}
with
\begin{align}
    \label{eq:Gamma_L_t}
\Gamma_{\ell} (t) &=
\begin{cases}
 \ln( {\Omega_0} t )  & \Gamma(t)  \lesssim \sqrt{\ell}\\
 \sqrt{\ell} & \textrm{otherwise} \ .
\end{cases}
\end{align}
The above results are sufficient to compute the R\'enyi entropies in Eq.~\eqref{eq:renyis}.

For the mutual information and the negativity we need to generalize Eq.~\eqref{eq:tdepnab} to arbitrary intervals $X$ and $Y$. In full generality, we consider a $2k$-multipartite system ${\bigcup_{X\in \mathcal{G}_0}}X$ with ${\mathcal{G}_0 =\{A_1,B_1,\dots,A_k,B_k\}}$. We define ${\mathcal{G}}$ as the set of all possible compact subintervals of the chain. 
For each ${X\in \mathcal{G}}$, of length $\ell_X$, we may first use the additivity of the number of oscillating pairs (as it was the case for the number of singlets at equilibrium) shared between $X$ and its complementary $X_c$ to decompose $n_{X: X_c}$ as follows
\begin{equation}
\label{eq:bababa}
	n_{X:X_c} = \sum_{Y,Z\in \mathcal{G}} n_{X\cap Y: X_c \cap Z}.
\end{equation}
After taking the disorder average, we have a set of linear equations, 
\begin{equation} \label{eq:average_nXbX}
	 \overline{n}_{X:X_c} = \sum_{Y,Z\in \mathcal{G}} \overline{n}_{X\cap Y: X_c\cap Z} \ ,
\end{equation}
whose solution gives $\overline{n}_{X:Y} $ for any ${X,Y\in \mathcal{G}_0}$.
Now, the left-hand side of \eqref{eq:average_nXbX} is the average of pairs between complementary regions, given by Eq.~\eqref{eq:tdepnab}.
Neglecting subleading $O(1)$ contributions, we have
\begin{equation}
\label{eq:system}
	\frac{b_{X:X_c}}{3}\ln \Gamma_{\ell_X}(t) \simeq \sum_{Y,Z\in \mathcal{G}} \overline{n}_{X\cap Y: X_c\cap Z}.
\end{equation}
This set of equations can be straightforwardly solved for the variables $\overline{n}_{X\cap Y: X_c\cap Z}$, and a unique solution can be extracted for any chosen partition of the system. 

\subsubsection{Dynamics of the R\'enyi entropies}
Using the pair structure of the dynamical state, it is clear that the average bipartite entanglement entropy of a subsystem $A$ is proportional to the number of shared singlets between $A$ and $A_c$~\cite{refael2004entanglement,vosk2013manybody}, i.e., 
\begin{equation}
    \overline{S}_A(t) = s_p \overline{n}_{A:A_c}(t).
\end{equation}
Now, for the same reason, also the average R\'enyi entanglement entropy will be proportional to $\overline{n}_{A:A_c}(t)$, but with the R\'enyi-index prefactor $s^{(\alpha)}_p$. Its contribution can be estimated as follows.

Consider a pair of spins oscillating at high-frequency $\omega$ between the two configuration $|\uparrow \downarrow \rangle$ and $|\downarrow \uparrow \rangle$ described by a state of the form $ |s\rangle=(|-\rangle + e^{i\omega t} |+\rangle)/\sqrt{2}$, with $|\pm \rangle$ defined in Sec.~\ref{sec:method}. Within the renormalization group approach, this effective emergent degree of freedom is seen as frozen by the remaining of the chain. Thus, $s^{(\alpha)}_p$ is estimated as the time-average, over one period, of its R\'enyi entropy. 

Given the associated density matrix reads in the basis $\{|\uparrow\uparrow\rangle,|\uparrow\downarrow\rangle, |\downarrow\uparrow\rangle,|\downarrow\downarrow\rangle \}$
\begin{align} \label{eq:rho2spins}
	\rho_{s} = \frac{1}{2}\begin{pmatrix}
		0 & 0 & 0 & 0 \\
		0 & 1+\cos\omega t & -i\sin\omega t & 0\\
		0 & +i\sin\omega t & 1-\cos\omega t & 0\\
		0 & 0 & 0& 0
	\end{pmatrix}.
\end{align}
the spectrum of the reduced density matrix $\rho_1(t)= \text{Tr}_2 \rho_s$ is given by 
\begin{equation}
	\lambda_{\pm} = \frac{1}{2}(1\pm \cos(\omega t)).
\end{equation}
It follows, from the definition in Eq.~\eqref{eq:renyis}, that 
\begin{equation} \label{eq:sp_t}
	s^{(\alpha)}_p = \frac{1}{1-\alpha}\frac{\omega}{2\pi }\int_0^{2\pi/\omega}dt \ln\left(\sum_{\mu=\pm} \lambda_\mu^\alpha(t)\right).
\end{equation}
In particular, Eq.~\eqref{eq:sp_t} gives the following contribution for the  entanglement entropy~\footnote{For comparison, we note that in Ref.~\cite{vosk2013manybody} the density of entanglement per pair $S_p$ was instead considered, with $S_p =s_p/\log2 \simeq 0.557$).}
\begin{equation} \label{eq:sp_pair}
	s_p = \lim_{\alpha\to 1} 	s^{(\alpha)}_p  = (2\log 2-1) \simeq 0.386.
\end{equation}
Combining  Eq.~\eqref{eq:sp_t} with the scaling of the singlet formation Eq.~\eqref{eq:tdepnab}, we obtain
\begin{align} \label{eq:entRSPdynL}
\overline{ S_A^{(\alpha)} }(t)  = s^{(\alpha)}_p  \frac{b_{A: A_c}}{3} \log \left( \Gamma_{\ell} (t) + 1/a_0 \right).
\end{align}

We conclude the discussion on the entropies with some remarks. First, the entanglement propagation in Eq.~\eqref{eq:entRSPdynL} obeys the logarithmic bound of information propagation proven in Ref.~\cite{burrell2007bounds}. Furthermore, the stationary value of the entanglement entropy saturate to a non-thermal value, namely
\begin{equation}
    \overline{S_A^{(\alpha)}}(\infty) \simeq s^{(\alpha)}_p\frac{b_{A: A_c}}{6}\ln \ell.
\end{equation}
Lastly, the predictions presented here are compatible with the numerical results presented in Ref.~\cite{igloi2012entanglement} for the random transverse field Ising chain, and for the random XX chain presented in Ref.~\cite{zhao2016entanglement}. We shall review some related aspects in Sec.~\ref{sec:numcheck}.

\subsubsection{Logarithmic negativity, fermionic negativity and mutual information}
\label{sec:logneg}
In this section, we discuss the observables of main interest for this manuscript, i.e. the entanglement negativity and the mutual information.

As for the R\'enyi entropies, the starting point of our analysis is the pair product structure of the time-dependent state. From this, it follows that, given the tripartition $A_1\cup A_2\cup B$, the average logarithmic negativity is
\begin{equation} \label{eq:neg_eq}
\overline{\mathcal{E}}_{A_1 : A_2} (t)= \epsilon_p \overline{n}_{A_1 : A_2}(t)
\end{equation}
where $\epsilon_p$ is the negativity associated to the oscillating pair, and $\overline{n}_{A_1 : A_2}(t)$ are the shared pairs between ${A_1}$ and ${A_2}$. The latter is obtained solving Eq.~\eqref{eq:system} for a specified partition (examples follow below), whereas $\epsilon_p$ is estimated from the oscillating pair density matrix $\rho_s$  (cf. Eq.~\eqref{eq:rho2spins}) and using the definition in Eq.~\eqref{eq:patialtrans}. The partial transpose of Eq.~\eqref{eq:rho2spins} with respect to the second spin is given by 
\begin{align}\label{eq:partialtsing}
	\rho_s^{T_2} = \frac{1}{2}\begin{pmatrix}
		0 & 0 & 0 &  i\sin\omega t\\
		0 & 1+\cos\omega t & 0 & 0\\
		0 & 0  & 1-\cos\omega t & 0\\
		-i\sin\omega t & 0 & 0& 0
	\end{pmatrix}.
\end{align}
The trace norm is easily computed by diagonalizing the above matrix, and the negativity contribution reads
\begin{align}
	\mathcal{E}_s(t)\equiv \ln ||\rho_s^{T_2}|| = \ln (1+|\sin\omega t|).
\end{align}
Time-averaging over a period we get
\begin{align} \label{eq:epsilonp}
	\epsilon_p = \frac{\omega}{2\pi}\int_0^{2\pi/\omega} dt \mathcal{E}_s(t) = \frac{4 \mathrm{Cat}}{\pi} - \log 2.
\end{align}
Here $\mathrm{Cat}\simeq 0.915\dots$ is the Catalan constant. 

Two remarks are in order here. First, by explicit computation, the prefactor $\epsilon_p$ is the same for the fermionic negativity (cf.~Eq.~\eqref{eq:fermneg}), and for the R\'enyi entropy at $\alpha=1/2$ (cf.~Eq.~\eqref{eq:renyis}). 
Furthermore, since the state is a product state of singlets, it is easy to compute fermionic negativity and the mutual information explicitly, and verify that
\begin{equation} \label{relations_free}
    \overline{\mathcal{E}}_{A_1:A_2} = \overline{\mathcal{E}^f}_{A_1:A_2} = \frac{\overline{I^{(1/2)}}_{A_1:A_2}}{2}.
\end{equation}

We now provide two explicit examples by choosing the specific geometry of the tripartition. The solution of Eq.~\eqref{eq:system} for these scenarios follows from the equilibrium calculations presented in Ref.~\cite{ruggiero2016entanglement,turkeshi2020negativity}.

\paragraph{Adjacent intervals.} We consider a tripartition $A_1\cup A_2\cup B$ of an infinite systems, with $A_1$ and $A_2$ two adjacent intervals of finite length, $\ell_1$ and $\ell_2$, respectively.
In this case, according to Eq.~\eqref{eq:system}, we get the following linear equations
\begin{equation}
\begin{cases}
 \overline{n}_{A_1: ({A}_1)_c}= \overline{n}_{A_1: A_2} +  \overline{n}_{A_1: B} = \frac{2}{3} \ln \Gamma_{\ell_1} (t) \\
 \overline{n}_{A_2: ({A}_2)_c}=  \overline{n}_{A_1: A_2} +  \overline{n}_{A_2: B}= \frac{2}{3} \ln \Gamma_{\ell_2} (t) \\
 \overline{n}_{A_1 \cup A_2: B}= \overline{n}_{A_1: B} +  \overline{n}_{A_2: B}= \frac{2}{3} \log \Gamma_{\ell_1 + \ell_2} (t)
\end{cases}
\end{equation}
The solution is given by
\begin{equation} \label{eq:nA1A2_t}
\overline{n}_{A_1: A_2}(t) = \frac{1}{3} \log\left( \frac{\Gamma_{\ell_1} (t) \Gamma_{\ell_2} (t) }{\Gamma_{\ell_1+\ell_2} (t) } \right).
\end{equation}
Eq.~\eqref{eq:nA1A2_t} together with Eqs.~\eqref{eq:neg_eq} and \eqref{eq:epsilonp} is the complete result for the evolution of the entanglement negativity in the Dyson model. 

\begin{figure}
	\includegraphics[width=\columnwidth]{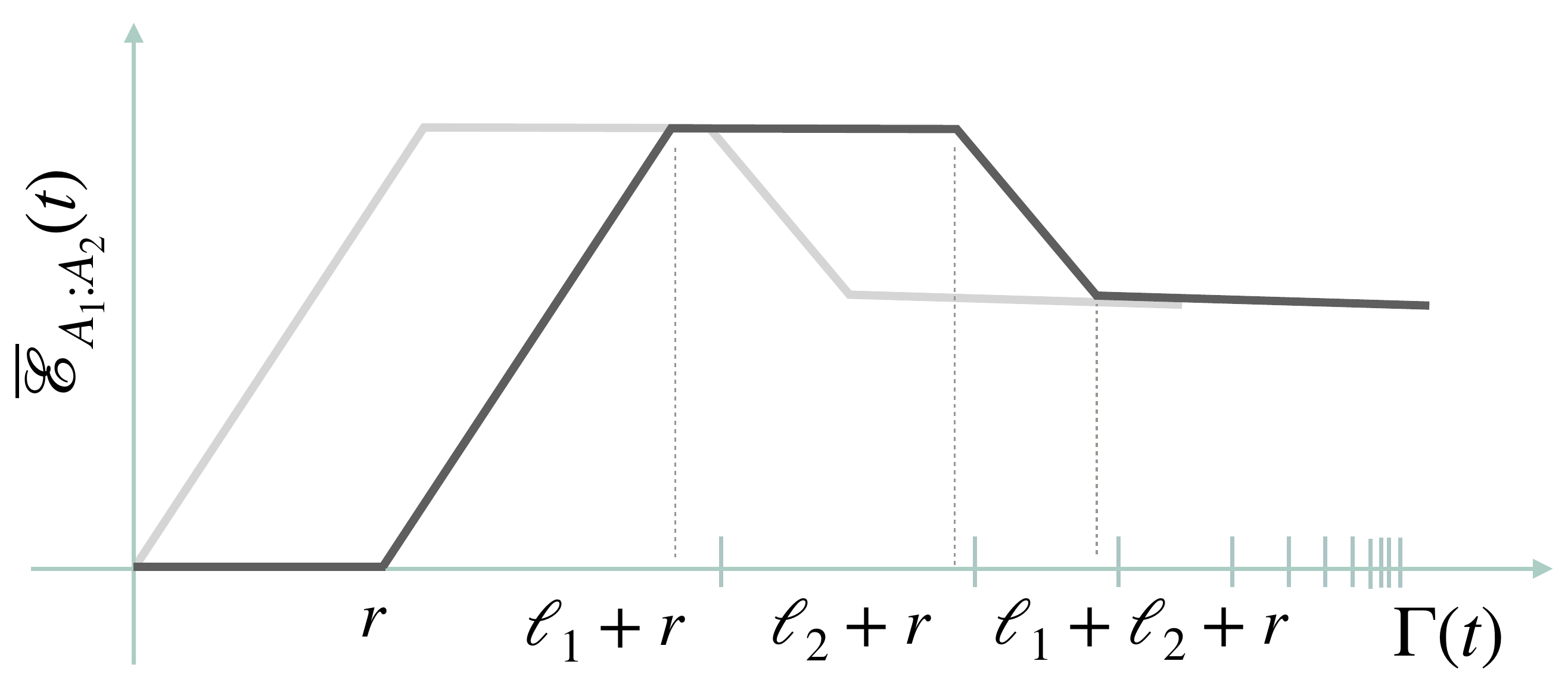}
	\caption{\label{fig:neg_free} \textit{Typical behaviour  of the logarithmic negativity dynamics in the Dyson model ---} We plot in logarithmic scale (with $\Gamma (t) = \log(\Omega_0 t)$ being the coordinate on the horizontal axis) the RSRG predictions for the Dyson model Eqs.~\eqref{nA1A2_t2}-\eqref{eq:nA1A2_dis}. The generic case of two intervals of length $\ell_1$ and $\ell_2$ separated by a distance $r$ is displayed in dark gray. The equivalent picture for adjacent interval is simply obtained by setting $r=0$, as shown in light gray. Moreover, when $\ell_1=\ell_2$ the intermediate plateau dissolves into a cusp, giving rise to a triangular-like shape. Numerically, only the initial growth can be accessed (see Sec.~\ref{sec:numcheck}). The same curve actually describes the fermionic negativity $\overline{\mathcal{E}}^f_{A_1:A_2} (t)$ and the mutual information $\overline{I^{(1/2)}}_{A_1 : A_2}$, according to Eq.~\eqref{relations_free}.}
\end{figure}

Let us now discuss the consequences on the entanglement negativity. Assuming without loss of generality that $\ell_1< \ell_2$, we get (cf. Eq.~\eqref{eq:Gamma_L_t})
\begin{align} \label{nA1A2_t2}
\overline{\mathcal{E}}_{A_1:A_2} = \frac{\epsilon_p}{3}
\begin{cases}
 \log \log (\Omega_0 t)  & t \lesssim e^{\sqrt{\ell_1}} \\
\log \sqrt{\ell_1} &  e^{\sqrt{\ell_1}}\lesssim t \lesssim e^{\sqrt{\ell_2}} \\
 \log (\frac{\ell_1 \ell_2 }{\log (\Omega_0 t)})  & e^{\sqrt{\ell_2}} \lesssim t \lesssim e^{\sqrt{\ell_1 + \ell_2}} \\
 \log \left( \frac{\ell_1 \ell_2}{\ell_1 + \ell_2} \right)  &   t \gtrsim e^{\sqrt{\ell_1 + \ell_2}}.
\end{cases}
\end{align}
Eq.~\eqref{nA1A2_t2} shows four interesting and unusual regimes: (i) an initial double-log growth (as for the entanglement entropy); (ii) a first plateau (dissolving into a cusp for $\ell_1 = \ell_2$); (iii) a new time dependence, where entanglement decreases; (vi) a final saturation to a value which is non extensive, and different from the ground state one (even if with the same logarithmic scaling). The typical behaviour is reported in Fig.~\ref{fig:neg_free}.

While the initial growth, as well as the final saturation, are reminiscent of the entanglement entropy behavior, the intermediate regimes are clearly different: in particular an intermediate plateau is followed by a decrease of entanglement.
This is similar to what happens for the entanglement negativity after a quench in clean systems: the same behavior, indeed, was found both in the context of conformal field theory~\cite{coser2014entanglement} and for integrable models~\cite{alba2019quantum_information}. The difference is the scale of time: the linear rate at which the entanglement changes in clean systems is now replaced by a logarithmic rate. 
Once again, this is consistent with the logarithmic bound of entanglement growth in Ref.~\cite{burrell2007bounds}.

\paragraph{Disjoint intervals.}

A similar approach can be used for a (infinite) system $B_2\cup A_1\cup B_1 \cup A_2\cup B_2$ where we focus on two disjoint intervals $A_1$ and $A_2$ of finite length, respectively $\ell_{1}$ and $\ell_2$, separated by an interval $B_1$ of length $r$. Following the calculations in Ref.~\cite{ruggiero2016negativity} we get
\begin{equation} \label{eq:nA1A2_dis}
\overline{\mathcal{E}}_{A_1: A_2} (t) = \frac{e_p}{3} \log\left( \frac{\Gamma_{\ell_1+r} (t) \Gamma_{\ell_2+r} (t) }{\Gamma_{r} (t)\Gamma_{\ell_1+\ell_2+r} (t) } \right),
\end{equation}
whose profile is again given in Fig.~\ref{fig:neg_free}.

\subsection{Quantum information dynamics in the interacting system}
\label{sec:interactingcase}
Switching on interactions ($\Delta_i\neq 0$), the qualitative picture described breaks down at a characteristic time $t_\mathrm{ent}^0 \sim 2 \Omega_0/(J_0^2 \Delta_0)$. (Here the subscript $0$ denotes the characteristic couplings at initial time).  
As described in Sec.~\ref{sec:method}, interactions lead to separate renormalization group trajectories, fixed by the values of the effective degrees of freedom $\tilde{\sigma}^z_n$, which are conserved by the effective Hamiltonian $H_\mathrm{eff}$ at each time-step. After the time $t_\mathrm{ent}^0$, two different trajectories related by different values of a specific $\tilde{\sigma}^z_n$ start generating mutual entanglement and the system state is in a superposition of multiple oscillating pairs. In particular, after a time $t^0_\mathrm{ent}$, the state start to form multispin entanglement clusters.

As we discuss below, we are not able to work out the negativity dynamics in all the regimes discussed in Sec.~\ref{sec:nonnnn} due to the complex coherent pattern arising in the interacting case. Nonetheless, some relevant phenomenological results can still be deduced.

\begin{figure}
	\includegraphics[width=\columnwidth]{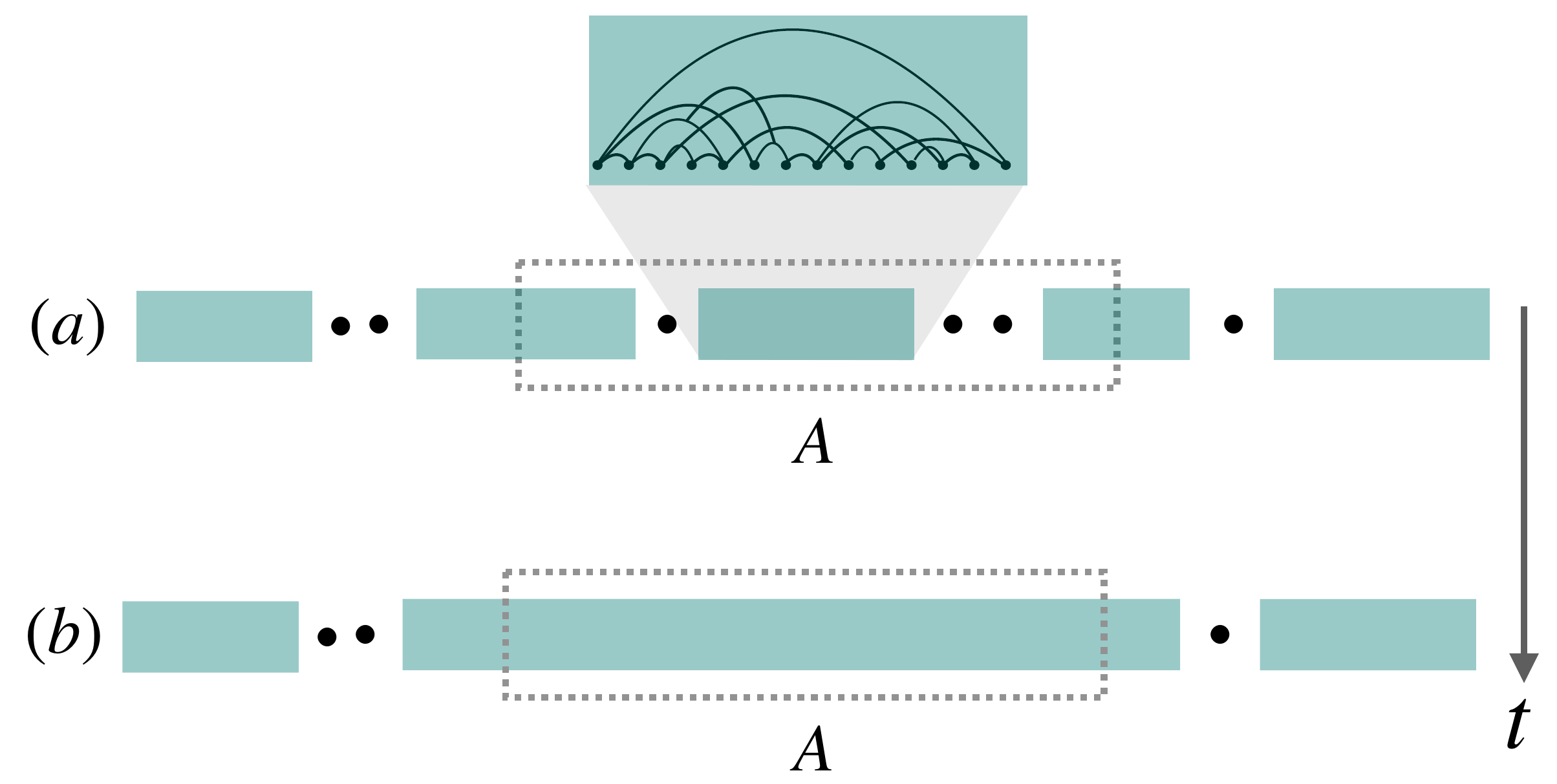}
	\caption{\label{fig:ent_renyi_interacting} \textit{Cartoon of the bipartite entanglement dynamics in the interacting case. ---}
	The state at time $t$ in the RG approximation in the interacting model, is represented via green bricks, denoting clusters of spins that extend to average dimension $d_0 (t)$ (cf. Eq.~\eqref{Ldistance2}); they are separated by black dots, representing spins which are unpaired at time $t$. When computing entanglement and R\'enyi entropies of a subsystem $A$ (in the figure being a single interval) of length $\ell$ we have two limiting regimes. (a): In the regime $d_0 (t) \ll \ell$, only the clusters around the boundaries of $A$ ($b_{A:A_c}=2$ in this specific case) will contribute to the entanglement. (b): At later times, instead, when $d_0 (t) \gg \ell$, we eventually expect the entropy to saturate. The entanglement behaviour describing these two regimes is given in Eq.~\eqref{renyis_int}.
	}
\end{figure}

\subsubsection{Entanglement and R\'enyi entropies}
First, it is instructive to review the entanglement entropy spreading estimated in Ref.~\cite{vosk2013manybody}. 

An important observation is that an effective spin will get entangled with the neighboring spins at the characteristic time $t^1_{\rm ent}\sim 2 \Omega_1/(J_1^2\Delta_1)$, where the subscript $1$ denotes the typical couplings at the time of formation of the effective spin. This reasoning is straightforwardly generalized to a cluster of decimated spins as follows. 
Consider a given time $t_\mathrm{in}$ such that the average distance between non-decimated spins is given by (cf. Sec.~\ref{sec:predilim})
\begin{equation} \label{Ldistance}
d ( t_{\rm in} ) = \left[ a_0 \log \left(\Omega_0 t_{\rm in} \right)+1\right]^2.
\end{equation}
Then, by the time ${t \sim t_\mathrm{in} + t_\mathrm{ent}}$ (${t_{\rm ent} \sim 2 \Omega_\mathrm{ent}/(J_\mathrm{ent}^2 \Delta_\mathrm{ent})}$ is the characteristic time fixed by the typical renormalized interactions between the neighboring sites of the cluster), we can assume all the spins within $d (t_\mathrm{in})$ to be entangled, namely entanglement clusters of length $d (t_{\rm in})$ have formed. 
Thus, the entanglement entropy at time $t$ is estimated~\cite{vosk2013manybody} to be $\overline{S_A}(t) \simeq d(t_\mathrm{in})/2 $. This is obtained as the mean entropy of a random state~\cite{page1993average}, conditioned to the available degrees of freedom, giving the factor $1/2$ which takes into account the fact that the states with aligned spins in each eliminated pair are not populated (cf. Sec.~\ref{sec:method}). Next, one need to express $d ( t_\mathrm{in})$ as a function of $t$. This leads to
\begin{multline} \label{Ldistance2}
d_0 (t) \equiv d ({t}_\mathrm{in}(t)) \approx 
 \left(\frac{\log (t/t^0_{\rm ent})}{\log (\Omega_0/J_0)} +1 \right)^2 \theta (t-t^0_{\rm ent}) \theta (t^* -t)  \\
+ \left(\frac{\log (t/t^0_{\rm ent})}{\log (\Omega_0/J_0)} +1 \right)^{2 \phi} \theta(t-t^*) - 1
\end{multline}
where $\theta (x)$ denotes the Heaviside theta function, $\phi=(1+\sqrt{5})/2$ is the golden ration, and $t^*$ is a crossover time which depends on the initial conditions (see Ref.~\cite{vosk2013manybody} for details).
Remarkably, this shows both the unbounded logarithmic growth of the entanglement entropy seen in the numerical simulations~\cite{dechiara2006entanglement,znidaric2008manybody}, as well as the delay of this interaction-induced growth by a time that scales as the inverse interaction strength~\cite{bardarson2012unbounded}. 

Now, one can actually extend the exact same reasoning to R\'enyi entropies $S_A^{(\alpha)}$. In fact, for large subsystem one has $\overline{S_A^{(\alpha)}} = \overline{S_A}$, namely the average do not depend on the R\'enyi index $\alpha$~\cite{zyczkowski2001induced}.
We conclude that for all $\alpha>0$ it holds
\begin{equation}
    \overline{S_A^{(\alpha)}}(t)\simeq  d_0 (t)/2
\end{equation}
with $d_0(t)$ in \eqref{Ldistance2}. 

The above predictions are valid for an infinite system cut in two via a single cut. 
In the case of a finite subsytem, we expect the entanglement to saturate: a conjecture for the saturation to an extensive non-thermal value is given, again, in Ref.~\cite{vosk2013manybody}.

Following the guidelines in Refs.~\cite{mbeng2017negativity,blondeau2016universal}, the above discussion can be further generalized to the subsystems made of several intervals.
We consider the generic case of a subsystem $A=\cup_i A_i$ made of $k$ disjoint intervals $A_i$ of length $\ell_i$, separated by distances $r_i$, within an infinite system, so that the entangling surface is made by $b_{A:A_c}=2 k$ points (boundaries). We denote by $r_\mathrm{min}= \min_i r_i$ the minimum distance between intervals, and $\ell_\mathrm{min}=\min_i \ell_i$ and $\ell_\mathrm{max} = \max_i \ell_i$ the length of the smallest and the largest intervals, respectively.
In the limit $d_0(t)\ll r_\mathrm{min}, \ell_\mathrm{min}$ -- a regime that typically holds on exponential times $t$ --, the contributions to the entanglement from each boundary add up. The clusters, indeed, become the ``independent units'' in the interacting case, and the state can be approximated (at leading order) as a tensor product of such clusters (cf.~Sec.~\ref{sec:predilim}). Conversely, in the limit $d_0 (t) \gg \ell_\mathrm{max}$, the subsystem $A$ reaches its non-thermal entanglement stationary value. 
The above considerations can be summarized in
\begin{equation} \label{renyis_int}
    \overline{S_A^{(\alpha)}}(t)\simeq
    \begin{cases}
        \frac{b_{A: A_c}}{2} d_0 (t) & d_0 (t) \ll r_\mathrm{min}, \ell_\mathrm{min} \\
        c \sum_i \ell_i  & d_0 (t) \gg \ell_\mathrm{max},
    \end{cases}
\end{equation}
for some positive (unknown) constant $c$. A pictorial representation is given in Fig.~\ref{fig:ent_renyi_interacting}.

\begin{figure}
	\includegraphics[width=\columnwidth]{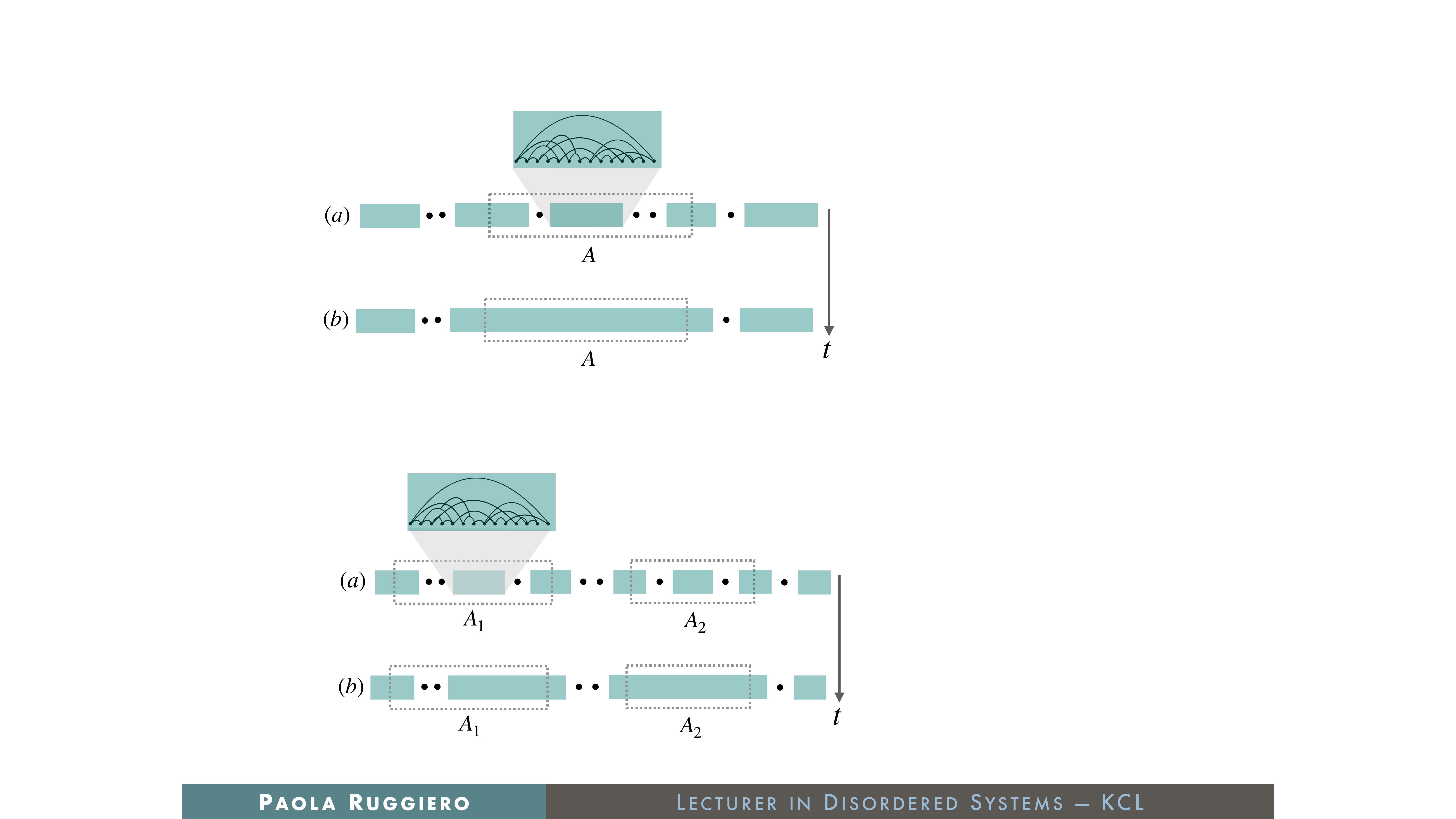}
	\caption{\label{fig:neg_info_interacting}
	\textit{Cartoon of the multipartite entanglement dynamics in the interacting case. ---}
	The state at time $t$ is represented with the same legend as in Fig.~\ref{fig:ent_renyi_interacting}, and we consider a multipartite systems, focusing on quantum correlation between two non-complementary subsystems $A_{1,2}$ of length $\ell_{1,2}$. (a) 
	In the regime $\ell_{1,2} \gg d_0 (t)$ only the clusters at the shared boundaries will contribute to entanglement (in the figure $b_{A_1:A_2}=0$): this gives the predictions in Eq.\eqref{prediction_int}. (b) At later times, instead, when clusters becomes comparable to subsystems' size, the multipartite structure of entanglement does not allows anymore to give an easy RG prediction in the intermediate regimes for both mutual information and negativity, until the very late time saturation regime.
	}
\end{figure}

\subsubsection{Logarithmic negativity and mutual information}
Similarly to the non-interacting case, but now restricting to the time regime where clusters are much smaller than all characteristic lengths in the systems,
the behavior of the logarithmic and fermionic negativity, and of the mutual information can be simply deduced from those of R\'enyi and entanglement entropies above. 

Consider in particular the case of two intervals $A_{1,2}$ of length $\ell_{1,2}$, embedded in an infinite system (see pictorial representation in Fig.~\ref{fig:neg_info_interacting}). In the small clusters regime we find

\begin{equation} \label{prediction_int}
    \overline{\mathcal{E}}_{A_1:A_2} =  \overline{\mathcal{E}}^f_{A_1:A_2} = \frac{\overline{I^{(\alpha)}}_{A_1 :A_2} }{2}= \frac{b_{A_1:A_2}}{2} d_0 (t)
\end{equation}
with $b_{A_1: A_2}$ the number of shared boudaries between $A_1$ and $A_2$, namely
\begin{equation} \label{bA1A2}
b_{A_1 :A_2} =
        \begin{cases}
         2 & A_{1},A_2 \; {\rm adjacent} \\
         0 & A_{1},A_2 \; {\rm disjoint} .
     \end{cases}
\end{equation}
We note, again, that the times where such prediction holds extend to exponentially long times (cf. Eq.~\eqref{Ldistance2}).

Let us note that the relation between negativity and mutual information in Eqs.~\eqref{prediction_int}-\eqref{bA1A2}, valid for any index $\alpha$, may be an artifact of the rough estimate in Eq.~\eqref{renyis_int}. These stem from the average entanglement contribution of a random state, as considered in Ref.~\cite{vosk2013manybody}.
We leave for future work to investigate the fine-structure properties of the R\'enyi entropies of the interacting model.
Furthermore, when the cluster size is comparable to the subsystems, the simplified discussion proposed here fails, since the additivity property breaks down. In particular, we cannot estimate through simple arguments the intermediate regimes, due to the complex multipartite structure. 

We can however estimate the expression for both the negativity and the mutual information at $t\gg \ell_1,\ell_2$. In this case, the spins in the  system $A$ belong to a same entanglement cluster. Assuming finite $\ell_1,\ell_2$, it follows from Ref.~\cite{shapourian2021entanglement} that the negativity at large time is $\overline{\mathcal{E}}_{A_1:A_2}=0$. Similarly, the contributions of the mutual information are estimated from Eq.~\eqref{renyis_int}, which again give in the late time regime $\overline{I^{(\alpha)}}_{A_1:A_2} = 0$.

\begin{figure}
    \centering
    \includegraphics[width=\columnwidth]{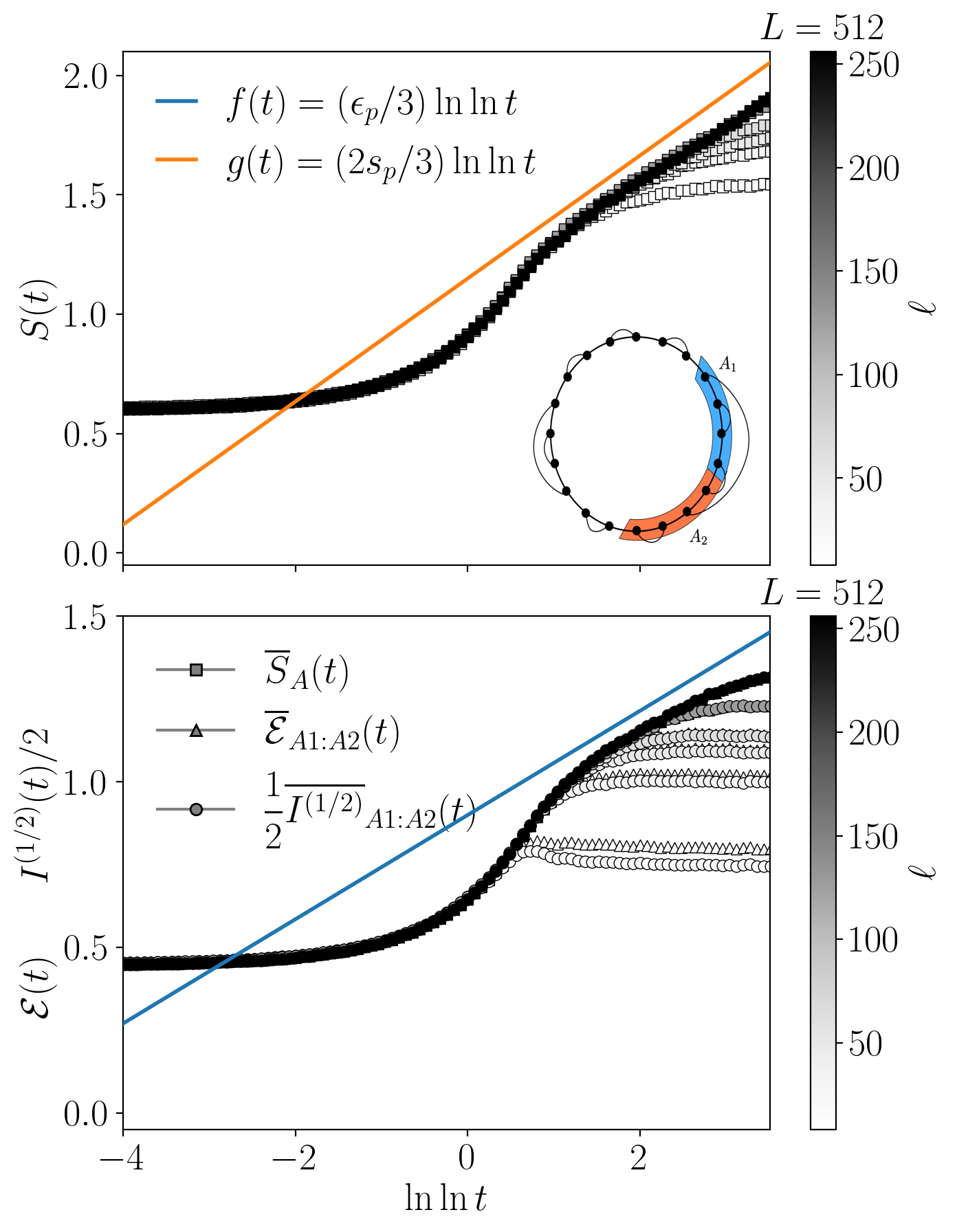}
    \caption{\textit{Quantum information spreading in the Dyson model.---} The time dependence of different entanglement measures is plotted for $L=512$ and varying $2\ell=|A_1|=|A_2|$ and $\delta=4$ in a system with periodic boundary conditions. Specifically, in the top panel, we plot the average entanglement entropy, while in the bottom one the mutual information and the entanglement negativity. The functions $f(t)$ and $g(t)$, which are the expected scaling behavior (cf. Sec.~\ref{sec:nonnnn}), are plotted to guide the eye. We note that the entanglement negativity and the mutual information collapse on each other in the thermodynamic limit.  }
    \label{fig:propagation}
\end{figure}

\section{Numerical benchmarks}
\label{sec:numcheck}
To support the renormalization group results presented in  this manuscript, in this section we provide numerical tests for the Dyson model.

We map Eq.~\eqref{eq:defmodel} for $\Delta_i=0$ through a Jordan Wigner transformation~\cite{lieb1961two} to
\begin{equation}
\label{eq:fermionin}
    H = \sum_{i}^\mathrm{BC} J_i (c_{i} c_{i+1}^\dagger + h.c.)\equiv  (\vec{c})^\dagger \cdot \mathcal{H} \cdot \vec{c},
\end{equation}
where $c_i$ and $c_i^\dagger$ are respectively the annihilation and creation fermionic operators, and $\mathcal{H}$ is the Hamiltonian density. We note that for periodic boundary conditions in Eq.~\eqref{eq:defmodel}, antiperiodic boundary condition (ABC) apply in Eq.~\eqref{eq:fermionin} (BC=ABC, i.e. $J_{L+1} = -J_1$), whereas for open boundary conditions (BC=OBC) the sum is restricted to $i<L$.
To study the effect of disorder, we shall employ the convenient choice of distribution 
\begin{equation}
\label{eq:delta}
    P_J(J) = \frac{1}{\delta} J^{-1+1/\delta},\qquad J\in[0,1].
\end{equation}
In Eq.~\eqref{eq:delta}, $\delta=0$  corresponds to the clean case, while $\delta\to\infty$ is the fixed point of the flow (cf.~Eq.~\eqref{eq:solution_flowEq}). We shall consider $\delta=2\div 16$ in our numerics, for which we expect the strong disorder renormalization group results to qualitative capture the physics. 

The quadratic nature of the Hamiltonian in Eq.~\eqref{eq:fermionin} allows efficient numerical computations, provided the initial state is Gaussian. (This is the case for any product state in the $\sigma^z$ basis, which is of interest for this paper). 
In this scenario, the evolution of the full state is faithfully encoded in that of the correlation matrix~\cite{Peschel_2009}
\begin{equation}
    \label{eq:corrmat}
    C_{i,j} (t) = \langle \Psi_t| c_i^\dagger c_j|\Psi_t\rangle. 
\end{equation}
Using fermionic algebra, it is easy to show that
\begin{equation}
\label{eq:timedepcorrr}
    C(t) = e^{-i \mathcal{H}t} C(t=0) e^{+i \mathcal{H}t}.
\end{equation}
The entanglement entropy and the fermionic negativity are then computable in polynomial resources~\footnote{The same does not hold true for the logarithmic negativity, since the partial transpose of a fermionic Gaussian state is \textit{not} Gaussian.}. For completeness, we briefly summarize how these are computed from the correlation matrix (cf. Eq.~\eqref{eq:corrmat}) and refer to the literature for additional details. 

\paragraph{Entanglement entropy} Given a bipartition $A\cup B$, the entanglement and R\'enyi entropy are encoded in the reduced correlation matrix $C^A_{i,j} = C_{i,j}$ for $i,j\in A$
\begin{equation}
\label{eq:ferminetpry}
    S^{(\alpha)} = \frac{1}{1-\alpha} \mathrm{tr}\ln \left[C^\alpha + (\openone - C)^\alpha\right]
\end{equation}
where $\openone$ is the matrix identity of dimension $|A|\times |A|$~\cite{Peschel_2009}. In a similar fashion, one can compute the R\'enyi mutual information. 

\paragraph{Fermionic negativity} Given a tripartition $A_1\cup A_2\cup B = A\cup B$, we define the correlation matrix $ G = 2 C^A-\openone$. It can be proven~\cite{shapourian2017partial,eisert2018entanglement,alba2022logarithmic} that the partial time reversal density matrix is still a Gaussian state. The latter, and its hermitian conjugate, both needed in the evaluation of the operator norm in Eq.~\eqref{eq:fermneg}, are given by
\begin{equation}
   G_\pm = \begin{pmatrix} G_{A_1,A_1} & \pm i G_{A_1,A_2}\\
   \pm iG_{A_2,A_1} & -G_{A_2,A_2}\end{pmatrix}.
\end{equation}
Introducing the matrix 
\begin{equation}
    G_\star = \frac{1}{2}\left[\openone - (\openone + G_+ G_-)^{-1}(G_++G_-)\right],
\end{equation}
the fermionic negativity is given by
\begin{equation}
\label{eq:fermioncanoisaj}
    \mathcal{E}^f = \sum_j \left(\ln\left(\sqrt{\mu_j} + \sqrt{1-\mu_j}\right) + \frac{1}{2}\ln[1-2\lambda_j + 2\lambda_j^2]\right),
\end{equation}
where $\mu_j$ are the eigenvalues of $G_\star$ and $\lambda_j$ are the eigenvalues of $C^A$. Since the logarithmic negativity $\mathcal{E}$ and the fermionic negativity $\mathcal{E}^f$ are expected to coincide for the non-interacting case and in the scaling limit (cf.~Eq.~\eqref{relations_free}), we shall consider in this section only the fermionic negativity, and denote it simply (entanglement) negativity.

We conclude this subsection with a remark. In the interacting case ($\Delta_i\neq 0$) the system is not Gaussian, and the computational complexity for the time-evolution requires exponential resources. Since the expectation of the strong disorder renormalization group requires the scaling limit, we do not attempt a comparison between numerics and the renormalization group results presented in Sec.~\ref{sec:interactingcase}.

\subsection{Results}
We use Eq.~\eqref{eq:timedepcorrr} to evaluate the time evolution to times $t_\mathrm{max}\sim 10^{15}$ starting from a N\'eel state, and use Eq.~\eqref{eq:ferminetpry} to compute the entropic measures and Eq.~\eqref{eq:fermioncanoisaj} to compute the negativity. The maximum time $t_\mathrm{max}$ is fixed by the constraint of using double precision floating point numerics.
We shall consider $\mathcal{N}=10^5$ disorder realization for each choice of $L$, subsystem size $\ell$, and disorder strength $\delta$.

In Fig.~\ref{fig:propagation} we plot the propagation of entanglement entropy, mutual information and negativity for a chain of length $L=512$ and considering adjacent intervals of size $\ell_1=\ell_2$. (In Appendix~\ref{app:bbb}, we discuss the robustness of such results by varying the total system size). We compare these results with the analytic estimates obtained in Sec.~\ref{sec:nonnnn}, finding good agreement. 

We note that, due to the double logarithmic behavior of the entanglement measures, we can only explore the first regime in Eq.~\eqref{eq:Gamma_L_t}.
Nevertheless, these results are informative, and show that the product ansatz explored in Eq.~\eqref{eq:singlets} is valid. Indeed, the mutual information and the negativity collapse onto each other for sufficiently large system sizes (whereas they present finite size corrections for small systems). 

\begin{figure}
    \centering
    \includegraphics[width=\columnwidth]{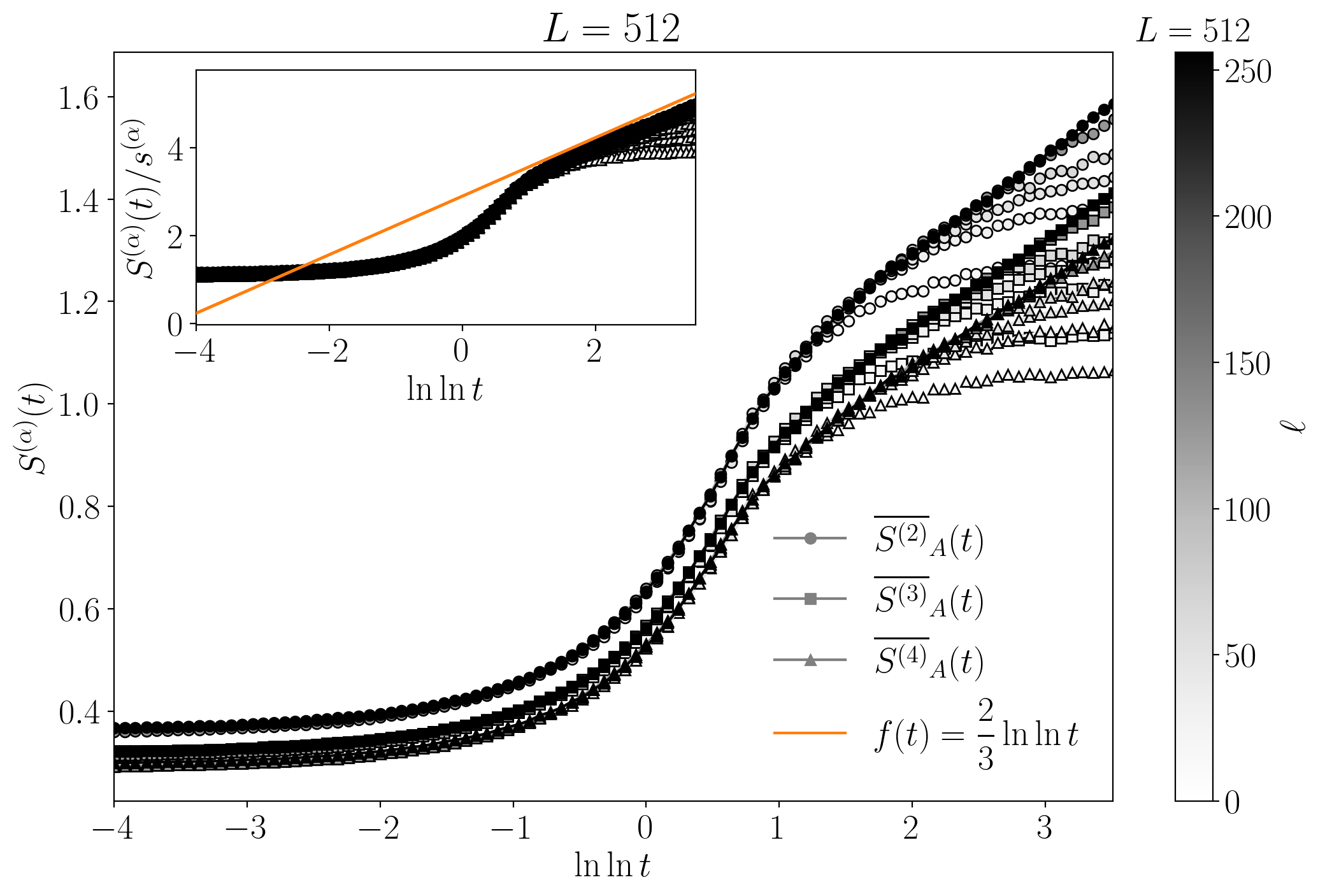}
    \caption{\textit{R\'enyi entropies in the Dyson model.---} The time dependence of the R\'enyi entropy is plotted for $L=512$, varying $\ell=|A|$, and varying $\alpha=2\div 4$  in a system with periodic boundary conditions with $\delta=6$. We use the geometry depicted in Fig.~\ref{fig:propagation}, and plot in the main figure the bare data. (Inset) The R\'enyi entropy rescaled by the pair contribution $s^{(\alpha)}$ (cf. Eq.~\eqref{eq:sp_t}). Here, $f(t)$ is the expected scaling behavior (cf.~Sec.~\ref{sec:nonnnn}), plotted to guide the eye.  }
    \label{fig:renyimm}
\end{figure}

To cross-validate the pair picture, in Fig.~\ref{fig:renyimm} we plot different values of the R\'enyi entropy, and show the curves at different $\alpha$ all collapse onto a single curve when rescaled by Eq.~\eqref{eq:sp_t}.

Overall, these numerical results demonstrate the strong disorder renormalization group analysis captures the relevant physics of the quantum information dynamics.

\section{Conclusions}
\label{sec:conclusion}
In this paper we have discussed the propagation of quantum information in systems with strong disorder. We have considered an archetypal model (the random bond XXZ spin chain) to derive analytical predictions through the time-dependent real space renormalization group method.

In the absence of interactions, we showed the dynamical state is described by a dynamical generalization of the RSP, namely it is well approximated by a product state of oscillating pairs in the scaling limit. From this, we were able to estimate the spreading of the R\'enyi entropy, the mutual information and the entanglement negativity. For the regime accessible by double-precision numeric, we benchmark these results with numerical simulations, finding good agreement.

In the interacting case, we were able to generalize the results in Ref.~\cite{vosk2013manybody} for the entanglement entropy to provide qualitative predictions for R\'enyi entropies and (R\'enyi) mutual information. 
For negativity and mutual information, the complex multipartite entanglement structure allows us to estimate the case of adjacent intervals for the (exponentially extended) early time regime, and in the late time limit. Within the former, we find the mutual information and the entanglement negativity are proportional, and scale logarithmically in time. Instead, at late times, both these quantities vanish, similarly to what occurs in integrable quantum systems~\cite{alba2019quantum_information}. 

Our results provide a qualitative understanding of the quantum information spreading in random spin chains. 
It would be interesting to investigate the spreading on symmetry resolved entanglement quantifiers~\cite{goldstein2018symmetryresolved,cornfeld2018imbalance,sara,riccarda,Parez_2021,kiefer2020bounds,kiefer2020evidence,kiefer2021slow,kiefer2021unlimited,kiefer2022particle,luitz2020absence,znidaric2022resonance}, which provide suitable tools for the experimental entanglement detection, as demonstrated in Ref.~\cite{Lukin_2019,neven2021symmetry}. 
Similar techniques can be also applied in systems with long range interactions~\cite{vasseur1,Pappalardi_2019,Monthus_2015}, where the dynamics of entanglement is known to be atypical also in pure systems~\cite{prxent,pappalardi2,pappalardi3,pappalardi4}.
Additionally, it would be interesting to study the quantum information dynamics of open quantum system disordered media, whose clean dynamics has recently been considered~\cite{carollo2021emergent,alba2022logarithmic,wellnitz2022rise}. Strong disorder renormalization group has been already applied to dissipative systems in Ref.~\cite{monthus2017dissipative}, and more recently in the context of measurement-induced entanglement transitions~\cite{zabalo2022disordered}.

\begin{acknowledgments}
    We thank V. Alba, P. Calabrese and V. Vitale for discussions. We acknowledge computational resources from King's College HPC Cluster "Rosalind", and Coll\`ege de France IPH cluster. XT acknowledges support from the ANR grant ``NonEQuMat'' (ANR-19-CE47-0001).
\end{acknowledgments}

\appendix

\section{High-frequency Floquet integration and derivation of the effective Hamiltonian}
\label{app:aaa}
In this section we review the high-frequency Floquet expansion, and how to derive the effective Hamiltonian Eq.~\eqref{eq:Heff}. We refer to Ref.~\cite{bukov2015universal,monthus2018floquet} for additional details. 

\subsection{High-frequency Floquet expansion}
We consider the generalization of the decomposition in Eq.~\eqref{eq:decompos}, that is $H=H_\Omega + V + H_\textup{rest}$. In full generality, we assume
\begin{equation}
	\label{eq:a1.2}
	H_\Omega = \sum_{\phi\in \sigma(H_\Omega)} P_\phi
\end{equation}
where $\sigma(H_\Omega)$ is the spectrum of $H_\Omega$, $P_\phi$ are the projectors onto the eigenvalues spaces. We simplify the description to the case of non-degenerate spectrum, so that $P_\phi$ map onto a one dimensional manifold. (Similar considerations can be extended to the degenerate case, combining the following analysis with ordinary degenerate perturbation theory).
We assume that $H_\textup{rest}$ is unaffected, at leading order, by $H_\Omega$, hence entering trivially the effective Hamiltonian $H_\textup{eff}$ in Eq.~\eqref{eq:Heff}. We can therefore only focus on the action of $H_\Omega$ on $V$. 

In the interaction picture
\begin{equation}
	\label{eq:a1.3}
	V_I(t) = e^{i H_\Omega t} V e^{-i H_\Omega t} = \sum_{\phi,\chi} e^{-i (E_\chi - E_\phi) t} P_\phi V P_\chi.
\end{equation}
We introduce the energy differences $\Delta E_j = E_\chi - E_\phi$, labelled by an integer $j$. We have:
\begin{equation}
	\label{eq:a1.4}
	V_I(t) =\sum_j e^{i \Delta E_j t} V_j,\qquad V_j = \sum_\phi \sum_\chi P_\phi V P_\chi \delta_{\Delta E_j, E_\chi - E_\phi}.
\end{equation}
\begin{widetext}
The unitary evolution in the interaction picture in Eq.~\eqref{eq:floqqqq} is given by
\begin{equation}
	U_I(t) = \mathcal{T} e^{-i \int_0^t d\tau V_I(\tau)} = 1 - i \int_0^t d\tau V_I(\tau) - \int_0^t d\tau_1 \int_0^{\tau_1} d\tau_2 V_I(\tau_1) V_I(\tau_2)+ \mathcal{O}(t^3)
\end{equation}
We recall the renormalization step is given by 
\begin{equation}
	U_I(t) \mapsto U_\textup{eff}(t) = e^{-i H_\textup{eff} t}  = 1 - i H_\textup{eff} t -\frac12 H_\textup{eff} t^2 + \mathcal{O}(t^3).
\end{equation}
This is done by expanding both unitary operators $U_I(t)$ and $U_\textup{eff}(t)$ and matching the same order terms.
One has
\begin{equation}
	H_\textup{eff} = H^{(0)}_\textup{eff} + H^{(1)}_\textup{eff} + \mathcal{O}\left(\left(\frac{1}{\max\{\Delta E_j\}}\right)^2\right).
\end{equation}
where
\begin{align}
	H^{(0)}_\textup{eff} & = \frac{1}{t}\int_0^t d\tau V_I(\tau) = V_0\\
	H^{(1)}_\textup{eff} & = -\frac{i}{2 t}\int_0^t d\tau_1 \int_0^{\tau_1} d\tau_2 V_I(\tau_1) V_I(\tau_2)  =   \sum_{j, \Delta E_j >0 } \frac{1}{\Delta E_j}\left[V_j, V^\dagger_{j}\right]
\end{align}
After simple algebra, we get
\begin{equation}
	\label{eq:effectiveness}
	H_\textup{eff} = \sum_\phi P_\phi V P_\phi + \sum_\phi \sum_{\chi\neq \phi} P_\phi V \frac{P_\chi}{E_\phi - E_\chi} V P_\phi.
\end{equation}
Let us notice that, although the projectors act on the full Hilbert space of $H_\Omega$, in practice depending on the interactions, only a reduced manifold of degrees of freedom is populated. See also the discussion in Sec.~\ref{sec:method}.

\subsection{Analysis for the Hamiltonian in Eq.~\eqref{eq:defmodel} }
We now specialize to the Hamiltonian in  Eq.~\eqref{eq:defmodel}, and apply the rules obtained in the previous section (cf.~Eq.~\eqref{eq:effectiveness}).
Let us write the explicit form of $H_\Omega$
\begin{equation}
	H_\Omega = \frac{\Omega}{2}\begin{pmatrix}
		\frac{\Delta_\Omega}{2} & 0 & 0 & 0\\
		0 & -\frac{\Delta_\Omega}{2} & 1 & 0\\
		0 & 1 & -\frac{\Delta_\Omega}{2} & 0\\
		0 & 0 & 0 & \frac{\Delta_\Omega}{2}
	\end{pmatrix},
\end{equation}
whose eigendecomposition is given by
\begin{align}
	E_s = -\frac{\Omega}{2}\left(1+\frac{\Delta_\Omega}{2}\right) & \qquad |s\rangle = \frac{1}{\sqrt{2}}\left(|\uparrow\downarrow\rangle - |\downarrow\uparrow\rangle\right)\\
	E_{t_0} = +\frac{\Omega}{2}\left(1-\frac{\Delta_\Omega}{2}\right) & \qquad |t_0\rangle = \frac{1}{\sqrt{2}}\left(|\uparrow\downarrow\rangle + |\downarrow\uparrow\rangle\right) \\
	E_{t_\pm} = +\frac{\Omega\Delta_\Omega}{4} & \qquad |t_+\rangle = |\uparrow\uparrow\rangle,\quad |t_-\rangle = |\downarrow\downarrow\rangle
\end{align}
Using Eq.~\eqref{eq:effectiveness}, we obtain the effective Hamiltonian. For convenience, we report here the non-zero leading contributions
\begin{align}
	\phi = s, \chi = t_0: & \qquad  \frac{-1}{\Omega}|\langle s| V |t_0\rangle|^2 |s\rangle\langle s| = -\frac{\Delta_\mathrm{L} \Delta_\mathrm{R} J_\mathrm{L} J_\mathrm{R}}{2\Omega} \sigma_\mathrm{L}^z \sigma^z_\mathrm{R} |s\rangle \langle s| - \frac{\Delta_\mathrm{L}^2 J_\mathrm{L}^2+ \Delta_\mathrm{R}^2 J_\mathrm{R}^2}{4\Omega} |s\rangle \langle s| \\
	\phi = s, \chi = t_\pm: & \qquad  \frac{-1}{\Omega(1 + \Delta_\Omega)/2 }|\langle s| V |t_\pm\rangle|^2 |s\rangle\langle s| = \frac{ J_\mathrm{L} J_\mathrm{R}}{2\Omega (1+\Delta_\Omega)} (\sigma^+_\mathrm{L}\sigma^-_\mathrm{R} + \sigma^-_\mathrm{L}\sigma^+_\mathrm{R}) |s\rangle \langle s|\\
	\phi = t_0, \chi = s: & \qquad  \frac{1}{\Omega}|\langle t_0| V |s\rangle|^2 |t_0\rangle\langle t_0| = \frac{\Delta_\mathrm{L} \Delta_\mathrm{L} J_\mathrm{R} J_\mathrm{R}}{2\Omega} \sigma_\mathrm{L}^z \sigma^z_\mathrm{R} |t_0\rangle \langle t_0| + \frac{\Delta_\mathrm{L}^2 J_\mathrm{L}^2+ \Delta_\mathrm{R}^2 J_\mathrm{R}^2}{4\Omega} |t_0\rangle \langle t_0|\\
	\phi = t_0, \chi = t_\pm: & \qquad  \frac{+1}{\Omega(1 - \Delta_\Omega)/2 }|\langle s| V |t_\pm\rangle|^2 |t_0\rangle\langle t_0|= \frac{ J_\mathrm{L} J_\mathrm{R}}{2\Omega (1-\Delta_\Omega)} (\sigma^+_\mathrm{L}\sigma^-_\mathrm{R} + \sigma^-_\mathrm{L}\sigma^+_\mathrm{R}) |t_0\rangle \langle t_0|.
\end{align}
Summing them up, and using the definition in Eq.~\eqref{eq:defgauge}, we obtain the final expression Eq.~\eqref{eq:Heff}.

\begin{figure*}[t!]
    \centering
    \includegraphics[width=\textwidth]{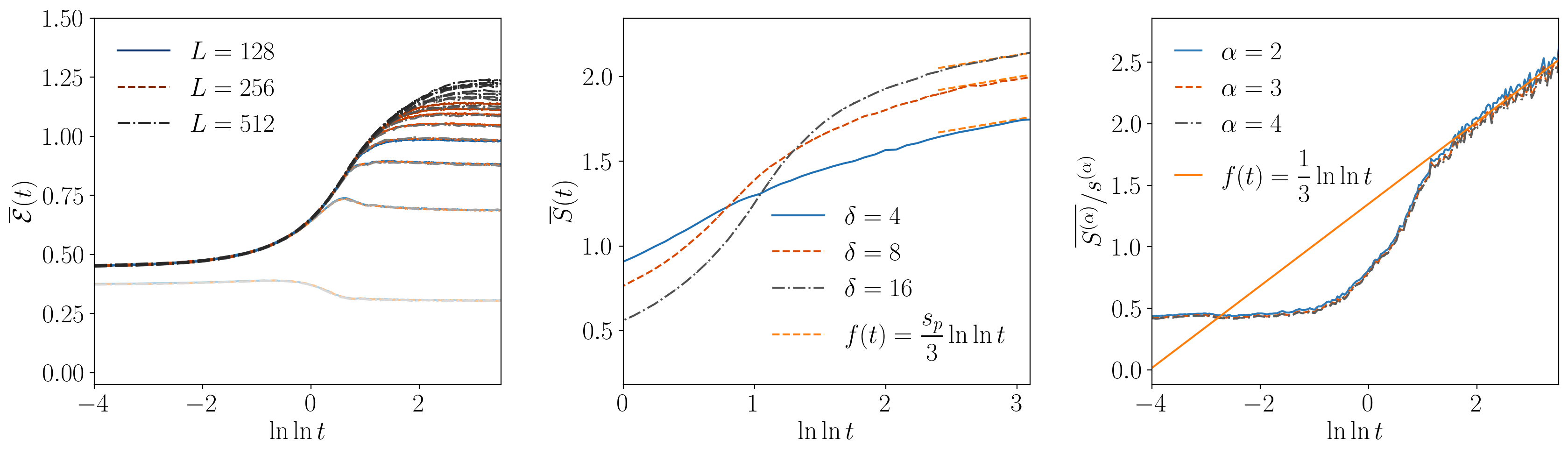}
    \caption{\textit{Robustness of numerical results---} We test the numerical results presented in the main text varying the parameters of the model. (Left) We consider $\ell=16\div L/4$ for $L=128,256,512$ and $\delta=4$. Our data show the values of the entanglement measure are the same when $\ell\ll L$, hence they are independent of the specific value of the system size $L$. (Center) Scaling of the entanglement entropy for open boundary conditions, varying the disorder strength $\delta$. After a non-universal transient time, the scaling of the entanglement entropy reach the expected form $f(t)=(s_p/3)\ln\ln t$ (dashed orange lines). 
    (Right) Scaling of the R\'enyi entropy for different values of $\alpha$ for open boundary condition and averaging over random initial states.}
    \label{fig:syssize}
\end{figure*}

\end{widetext}

\section{Additional numerical results}
\label{app:bbb}
In this section we consider additional numerical checks. Specifically we show how the results presented in Sec.~\ref{sec:numcheck} change when (i) considering different system sizes $L$; (ii) different disorder strength $\delta$, (iii) different initial conditions. Our results are summarized in Fig.~\ref{fig:syssize}.
(For readability, we avoid including all the tests, and report only a selection of them.) 

First, we note that for a fixed $\ell\ll L$, the value of $L$ is irrelevant, and all the curves fall on top of each other (See Fig.~\ref{fig:syssize}(Left)). 

In Fig.~\ref{fig:syssize}(Center,Right), we compare the results using open boundary conditions, for which $b_{A:A_c} = 1$ (cf.~Eq.~\eqref{eq:tdepnab}). We can appreciate the agreement between numerical results and the theoretical prediction. 

The values of the disorder give instead non-universal contributions in the dynamics, but the late time behavior is the same for subsystems that are sufficiently large. These are shown in Fig.~\ref{fig:syssize} (Center). 

Lastly, we note that the initial condition is irrelevant to the overall physical picture. As shown in Fig~\ref{fig:syssize} (Right), averaging over the initial random product state conditions, and on the disorder configurations. Contrary to the remaining of the paper (where we use $\mathcal{N}=10^6$ realizations), in total $\mathcal{N}=10^5$ for this plot. It is clear from the plot that the qualitative predictions of the RSRG are unaffected by the choice of the initial state.

\bibliography{neg_biblio}

\begin{thebibliography}{118}%
\makeatletter
\providecommand \@ifxundefined [1]{%
 \@ifx{#1\undefined}
}%
\providecommand \@ifnum [1]{%
 \ifnum #1\expandafter \@firstoftwo
 \else \expandafter \@secondoftwo
 \fi
}%
\providecommand \@ifx [1]{%
 \ifx #1\expandafter \@firstoftwo
 \else \expandafter \@secondoftwo
 \fi
}%
\providecommand \natexlab [1]{#1}%
\providecommand \enquote  [1]{``#1''}%
\providecommand \bibnamefont  [1]{#1}%
\providecommand \bibfnamefont [1]{#1}%
\providecommand \citenamefont [1]{#1}%
\providecommand \href@noop [0]{\@secondoftwo}%
\providecommand \href [0]{\begingroup \@sanitize@url \@href}%
\providecommand \@href[1]{\@@startlink{#1}\@@href}%
\providecommand \@@href[1]{\endgroup#1\@@endlink}%
\providecommand \@sanitize@url [0]{\catcode `\\12\catcode `\$12\catcode
  `\&12\catcode `\#12\catcode `\^12\catcode `\_12\catcode `\%12\relax}%
\providecommand \@@startlink[1]{}%
\providecommand \@@endlink[0]{}%
\providecommand \url  [0]{\begingroup\@sanitize@url \@url }%
\providecommand \@url [1]{\endgroup\@href {#1}{\urlprefix }}%
\providecommand \urlprefix  [0]{URL }%
\providecommand \Eprint [0]{\href }%
\providecommand \doibase [0]{http://dx.doi.org/}%
\providecommand \selectlanguage [0]{\@gobble}%
\providecommand \bibinfo  [0]{\@secondoftwo}%
\providecommand \bibfield  [0]{\@secondoftwo}%
\providecommand \translation [1]{[#1]}%
\providecommand \BibitemOpen [0]{}%
\providecommand \bibitemStop [0]{}%
\providecommand \bibitemNoStop [0]{.\EOS\space}%
\providecommand \EOS [0]{\spacefactor3000\relax}%
\providecommand \BibitemShut  [1]{\csname bibitem#1\endcsname}%
\let\auto@bib@innerbib\@empty
\bibitem [{\citenamefont {Abanin}\ \emph {et~al.}(2019)\citenamefont {Abanin},
  \citenamefont {Altman}, \citenamefont {Bloch},\ and\ \citenamefont
  {Serbyn}}]{abanin2019colloquium}%
  \BibitemOpen
  \bibfield  {author} {\bibinfo {author} {\bibfnamefont {D.~A.}\ \bibnamefont
  {Abanin}}, \bibinfo {author} {\bibfnamefont {E.}~\bibnamefont {Altman}},
  \bibinfo {author} {\bibfnamefont {I.}~\bibnamefont {Bloch}}, \ and\ \bibinfo
  {author} {\bibfnamefont {M.}~\bibnamefont {Serbyn}},\ }\href
  {https://link.aps.org/doi/10.1103/RevModPhys.91.021001} {\bibfield  {journal}
  {\bibinfo  {journal} {Rev. Mod. Phys.}\ }\textbf {\bibinfo {volume} {91}},\
  \bibinfo {pages} {021001} (\bibinfo {year} {2019})}\BibitemShut {NoStop}%
\bibitem [{\citenamefont {Nandkishore}\ and\ \citenamefont
  {Huse}(2015)}]{nandkishore2015manybody}%
  \BibitemOpen
  \bibfield  {author} {\bibinfo {author} {\bibfnamefont {R.}~\bibnamefont
  {Nandkishore}}\ and\ \bibinfo {author} {\bibfnamefont {D.~A.}\ \bibnamefont
  {Huse}},\ }\href {https://doi.org/10.1146/annurev-conmatphys-031214-014726}
  {\bibfield  {journal} {\bibinfo  {journal} {Annu. Rev. Condens. Matter
  Phys.}\ }\textbf {\bibinfo {volume} {6}},\ \bibinfo {pages} {15} (\bibinfo
  {year} {2015})}\BibitemShut {NoStop}%
\bibitem [{\citenamefont {D'Alessio}\ \emph {et~al.}(2016)\citenamefont
  {D'Alessio}, \citenamefont {Kafri}, \citenamefont {Polkovnikov},\ and\
  \citenamefont {Rigol}}]{dalessio2016fromquantum}%
  \BibitemOpen
  \bibfield  {author} {\bibinfo {author} {\bibfnamefont {L.}~\bibnamefont
  {D'Alessio}}, \bibinfo {author} {\bibfnamefont {Y.}~\bibnamefont {Kafri}},
  \bibinfo {author} {\bibfnamefont {A.}~\bibnamefont {Polkovnikov}}, \ and\
  \bibinfo {author} {\bibfnamefont {M.}~\bibnamefont {Rigol}},\ }\href
  {https://doi.org/10.1080/00018732.2016.1198134} {\bibfield  {journal}
  {\bibinfo  {journal} {Adv. Phys.}\ }\textbf {\bibinfo {volume} {65}},\
  \bibinfo {pages} {239} (\bibinfo {year} {2016})}\BibitemShut {NoStop}%
\bibitem [{\citenamefont {Laflorencie}(2016)}]{laflorencie2016quantum}%
  \BibitemOpen
  \bibfield  {author} {\bibinfo {author} {\bibfnamefont {N.}~\bibnamefont
  {Laflorencie}},\ }\href {\doibase 10.1016/j.physrep.2016.06.008} {\bibfield
  {journal} {\bibinfo  {journal} {Phys. Rep.}\ }\textbf {\bibinfo {volume}
  {646}},\ \bibinfo {pages} {1} (\bibinfo {year} {2016})}\BibitemShut {NoStop}%
\bibitem [{\citenamefont {Calabrese}\ and\ \citenamefont
  {Cardy}(2004)}]{calabrese2004entanglement}%
  \BibitemOpen
  \bibfield  {author} {\bibinfo {author} {\bibfnamefont {P.}~\bibnamefont
  {Calabrese}}\ and\ \bibinfo {author} {\bibfnamefont {J.}~\bibnamefont
  {Cardy}},\ }\href {\doibase 10.1088/1742-5468/2004/06/p06002} {\bibfield
  {journal} {\bibinfo  {journal} {J. Stat. Mech. Theory Exp.}\ }\textbf
  {\bibinfo {volume} {2004}},\ \bibinfo {pages} {P06002} (\bibinfo {year}
  {2004})}\BibitemShut {NoStop}%
\bibitem [{\citenamefont {Calabrese}\ and\ \citenamefont
  {Cardy}(2005)}]{calabrese2005evolution}%
  \BibitemOpen
  \bibfield  {author} {\bibinfo {author} {\bibfnamefont {P.}~\bibnamefont
  {Calabrese}}\ and\ \bibinfo {author} {\bibfnamefont {J.}~\bibnamefont
  {Cardy}},\ }\href {\doibase 10.1088/1742-5468/2005/04/p04010} {\bibfield
  {journal} {\bibinfo  {journal} {J. Stat. Mech. Theory Exp.}\ }\textbf
  {\bibinfo {volume} {2005}},\ \bibinfo {pages} {P04010} (\bibinfo {year}
  {2005})}\BibitemShut {NoStop}%
\bibitem [{\citenamefont {Eisert}\ and\ \citenamefont
  {Osborne}(2006)}]{eisert2006general}%
  \BibitemOpen
  \bibfield  {author} {\bibinfo {author} {\bibfnamefont {J.}~\bibnamefont
  {Eisert}}\ and\ \bibinfo {author} {\bibfnamefont {T.~J.}\ \bibnamefont
  {Osborne}},\ }\href {https://link.aps.org/doi/10.1103/PhysRevLett.97.150404}
  {\bibfield  {journal} {\bibinfo  {journal} {Phys. Rev. Lett.}\ }\textbf
  {\bibinfo {volume} {97}},\ \bibinfo {pages} {150404} (\bibinfo {year}
  {2006})}\BibitemShut {NoStop}%
\bibitem [{\citenamefont {Kim}\ and\ \citenamefont
  {Huse}(2013)}]{kim2013ballistic}%
  \BibitemOpen
  \bibfield  {author} {\bibinfo {author} {\bibfnamefont {H.}~\bibnamefont
  {Kim}}\ and\ \bibinfo {author} {\bibfnamefont {D.~A.}\ \bibnamefont {Huse}},\
  }\href {https://link.aps.org/doi/10.1103/PhysRevLett.111.127205} {\bibfield
  {journal} {\bibinfo  {journal} {Phys. Rev. Lett.}\ }\textbf {\bibinfo
  {volume} {111}},\ \bibinfo {pages} {127205} (\bibinfo {year}
  {2013})}\BibitemShut {NoStop}%
\bibitem [{\citenamefont {Calabrese}\ and\ \citenamefont
  {Cardy}(2016)}]{calabrese2016quantum}%
  \BibitemOpen
  \bibfield  {author} {\bibinfo {author} {\bibfnamefont {P.}~\bibnamefont
  {Calabrese}}\ and\ \bibinfo {author} {\bibfnamefont {J.}~\bibnamefont
  {Cardy}},\ }\href {\doibase 10.1088/1742-5468/2016/06/064003} {\bibfield
  {journal} {\bibinfo  {journal} {J. Stat. Mech. Theory Exp.}\ }\textbf
  {\bibinfo {volume} {2016}},\ \bibinfo {pages} {064003} (\bibinfo {year}
  {2016})}\BibitemShut {NoStop}%
\bibitem [{\citenamefont {Ho}\ and\ \citenamefont
  {Abanin}(2017)}]{ho2017entanglementdynamics}%
  \BibitemOpen
  \bibfield  {author} {\bibinfo {author} {\bibfnamefont {W.~W.}\ \bibnamefont
  {Ho}}\ and\ \bibinfo {author} {\bibfnamefont {D.~A.}\ \bibnamefont
  {Abanin}},\ }\href {https://link.aps.org/doi/10.1103/PhysRevB.95.094302}
  {\bibfield  {journal} {\bibinfo  {journal} {Phys. Rev. B}\ }\textbf {\bibinfo
  {volume} {95}},\ \bibinfo {pages} {094302} (\bibinfo {year}
  {2017})}\BibitemShut {NoStop}%
\bibitem [{\citenamefont {Nahum}\ \emph {et~al.}(2017)\citenamefont {Nahum},
  \citenamefont {Ruhman}, \citenamefont {Vijay},\ and\ \citenamefont
  {Haah}}]{nahum2017quantumentanglement}%
  \BibitemOpen
  \bibfield  {author} {\bibinfo {author} {\bibfnamefont {A.}~\bibnamefont
  {Nahum}}, \bibinfo {author} {\bibfnamefont {J.}~\bibnamefont {Ruhman}},
  \bibinfo {author} {\bibfnamefont {S.}~\bibnamefont {Vijay}}, \ and\ \bibinfo
  {author} {\bibfnamefont {J.}~\bibnamefont {Haah}},\ }\href
  {https://link.aps.org/doi/10.1103/PhysRevX.7.031016} {\bibfield  {journal}
  {\bibinfo  {journal} {Phys. Rev. X}\ }\textbf {\bibinfo {volume} {7}},\
  \bibinfo {pages} {031016} (\bibinfo {year} {2017})}\BibitemShut {NoStop}%
\bibitem [{\citenamefont {Mezei}\ and\ \citenamefont
  {Stanford}(2017)}]{mezei2017onentanglement}%
  \BibitemOpen
  \bibfield  {author} {\bibinfo {author} {\bibfnamefont {M.}~\bibnamefont
  {Mezei}}\ and\ \bibinfo {author} {\bibfnamefont {D.}~\bibnamefont
  {Stanford}},\ }\href {https://doi.org/10.1007/jhep05(2017)065} {\bibfield
  {journal} {\bibinfo  {journal} {J. High Energy Phys.}\ }\textbf {\bibinfo
  {volume} {2017}} (\bibinfo {year} {2017})}\BibitemShut {NoStop}%
\bibitem [{\citenamefont {Chiara}\ \emph {et~al.}(2006)\citenamefont {Chiara},
  \citenamefont {Montangero}, \citenamefont {Calabrese},\ and\ \citenamefont
  {Fazio}}]{dechiara2006entanglement}%
  \BibitemOpen
  \bibfield  {author} {\bibinfo {author} {\bibfnamefont {G.~D.}\ \bibnamefont
  {Chiara}}, \bibinfo {author} {\bibfnamefont {S.}~\bibnamefont {Montangero}},
  \bibinfo {author} {\bibfnamefont {P.}~\bibnamefont {Calabrese}}, \ and\
  \bibinfo {author} {\bibfnamefont {R.}~\bibnamefont {Fazio}},\ }\href
  {https://doi.org/10.1088/1742-5468/2006/03/p03001} {\bibfield  {journal}
  {\bibinfo  {journal} {J. Stat. Mech. Theory Exp.}\ }\textbf {\bibinfo
  {volume} {2006}},\ \bibinfo {pages} {P03001} (\bibinfo {year}
  {2006})}\BibitemShut {NoStop}%
\bibitem [{\citenamefont {\v{Z}nidari\v{c}}\ \emph {et~al.}(2008)\citenamefont
  {\v{Z}nidari\v{c}}, \citenamefont {Prosen},\ and\ \citenamefont
  {Prelov\v{s}ek}}]{znidaric2008manybody}%
  \BibitemOpen
  \bibfield  {author} {\bibinfo {author} {\bibfnamefont {M.}~\bibnamefont
  {\v{Z}nidari\v{c}}}, \bibinfo {author} {\bibfnamefont {T.}~\bibnamefont
  {Prosen}}, \ and\ \bibinfo {author} {\bibfnamefont {P.}~\bibnamefont
  {Prelov\v{s}ek}},\ }\href
  {https://link.aps.org/doi/10.1103/PhysRevB.77.064426} {\bibfield  {journal}
  {\bibinfo  {journal} {Phys. Rev. B}\ }\textbf {\bibinfo {volume} {77}},\
  \bibinfo {pages} {064426} (\bibinfo {year} {2008})}\BibitemShut {NoStop}%
\bibitem [{\citenamefont {Bardarson}\ \emph {et~al.}(2012)\citenamefont
  {Bardarson}, \citenamefont {Pollmann},\ and\ \citenamefont
  {Moore}}]{bardarson2012unbounded}%
  \BibitemOpen
  \bibfield  {author} {\bibinfo {author} {\bibfnamefont {J.~H.}\ \bibnamefont
  {Bardarson}}, \bibinfo {author} {\bibfnamefont {F.}~\bibnamefont {Pollmann}},
  \ and\ \bibinfo {author} {\bibfnamefont {J.~E.}\ \bibnamefont {Moore}},\
  }\href {https://link.aps.org/doi/10.1103/PhysRevLett.109.017202} {\bibfield
  {journal} {\bibinfo  {journal} {Phys. Rev. Lett.}\ }\textbf {\bibinfo
  {volume} {109}},\ \bibinfo {pages} {017202} (\bibinfo {year}
  {2012})}\BibitemShut {NoStop}%
\bibitem [{\citenamefont {Igl\'oi}\ \emph {et~al.}(2012)\citenamefont
  {Igl\'oi}, \citenamefont {Szatm\'ari},\ and\ \citenamefont
  {Lin}}]{igloi2012entanglement}%
  \BibitemOpen
  \bibfield  {author} {\bibinfo {author} {\bibfnamefont {F.}~\bibnamefont
  {Igl\'oi}}, \bibinfo {author} {\bibfnamefont {Z.}~\bibnamefont {Szatm\'ari}},
  \ and\ \bibinfo {author} {\bibfnamefont {Y.-C.}\ \bibnamefont {Lin}},\ }\href
  {https://link.aps.org/doi/10.1103/PhysRevB.85.094417} {\bibfield  {journal}
  {\bibinfo  {journal} {Phys. Rev. B}\ }\textbf {\bibinfo {volume} {85}},\
  \bibinfo {pages} {094417} (\bibinfo {year} {2012})}\BibitemShut {NoStop}%
\bibitem [{\citenamefont {Levine}\ \emph {et~al.}(2012)\citenamefont {Levine},
  \citenamefont {Bantegui},\ and\ \citenamefont {Burg}}]{levine2012full}%
  \BibitemOpen
  \bibfield  {author} {\bibinfo {author} {\bibfnamefont {G.~C.}\ \bibnamefont
  {Levine}}, \bibinfo {author} {\bibfnamefont {M.~J.}\ \bibnamefont
  {Bantegui}}, \ and\ \bibinfo {author} {\bibfnamefont {J.~A.}\ \bibnamefont
  {Burg}},\ }\href {https://link.aps.org/doi/10.1103/PhysRevB.86.174202}
  {\bibfield  {journal} {\bibinfo  {journal} {Phys. Rev. B}\ }\textbf {\bibinfo
  {volume} {86}},\ \bibinfo {pages} {174202} (\bibinfo {year}
  {2012})}\BibitemShut {NoStop}%
\bibitem [{\citenamefont {Zhao}\ \emph {et~al.}(2016)\citenamefont {Zhao},
  \citenamefont {Andraschko},\ and\ \citenamefont
  {Sirker}}]{zhao2016entanglement}%
  \BibitemOpen
  \bibfield  {author} {\bibinfo {author} {\bibfnamefont {Y.}~\bibnamefont
  {Zhao}}, \bibinfo {author} {\bibfnamefont {F.}~\bibnamefont {Andraschko}}, \
  and\ \bibinfo {author} {\bibfnamefont {J.}~\bibnamefont {Sirker}},\ }\href
  {https://link.aps.org/doi/10.1103/PhysRevB.93.205146} {\bibfield  {journal}
  {\bibinfo  {journal} {Phys. Rev. B}\ }\textbf {\bibinfo {volume} {93}},\
  \bibinfo {pages} {205146} (\bibinfo {year} {2016})}\BibitemShut {NoStop}%
\bibitem [{\citenamefont {Iemini}\ \emph {et~al.}(2016)\citenamefont {Iemini},
  \citenamefont {Russomanno}, \citenamefont {Rossini}, \citenamefont
  {Scardicchio},\ and\ \citenamefont {Fazio}}]{iemini2016signatures}%
  \BibitemOpen
  \bibfield  {author} {\bibinfo {author} {\bibfnamefont {F.}~\bibnamefont
  {Iemini}}, \bibinfo {author} {\bibfnamefont {A.}~\bibnamefont {Russomanno}},
  \bibinfo {author} {\bibfnamefont {D.}~\bibnamefont {Rossini}}, \bibinfo
  {author} {\bibfnamefont {A.}~\bibnamefont {Scardicchio}}, \ and\ \bibinfo
  {author} {\bibfnamefont {R.}~\bibnamefont {Fazio}},\ }\href
  {https://link.aps.org/doi/10.1103/PhysRevB.94.214206} {\bibfield  {journal}
  {\bibinfo  {journal} {Phys. Rev. B}\ }\textbf {\bibinfo {volume} {94}},\
  \bibinfo {pages} {214206} (\bibinfo {year} {2016})}\BibitemShut {NoStop}%
\bibitem [{\citenamefont {Huang}(2017)}]{huang2017entanglement}%
  \BibitemOpen
  \bibfield  {author} {\bibinfo {author} {\bibfnamefont {Y.}~\bibnamefont
  {Huang}},\ }\href
  {https://www.sciencedirect.com/science/article/pii/S0003491617300763}
  {\bibfield  {journal} {\bibinfo  {journal} {Ann. Phys.}\ }\textbf {\bibinfo
  {volume} {380}},\ \bibinfo {pages} {224} (\bibinfo {year}
  {2017})}\BibitemShut {NoStop}%
\bibitem [{\citenamefont {\v{Z}nidari\v{c}}(2018)}]{znidaric2018entanglement}%
  \BibitemOpen
  \bibfield  {author} {\bibinfo {author} {\bibfnamefont {M.}~\bibnamefont
  {\v{Z}nidari\v{c}}},\ }\href
  {https://link.aps.org/doi/10.1103/PhysRevB.97.214202} {\bibfield  {journal}
  {\bibinfo  {journal} {Phys. Rev. B}\ }\textbf {\bibinfo {volume} {97}},\
  \bibinfo {pages} {214202} (\bibinfo {year} {2018})}\BibitemShut {NoStop}%
\bibitem [{\citenamefont {Zhao}\ and\ \citenamefont
  {Sirker}(2019)}]{zhao2019logarithmic}%
  \BibitemOpen
  \bibfield  {author} {\bibinfo {author} {\bibfnamefont {Y.}~\bibnamefont
  {Zhao}}\ and\ \bibinfo {author} {\bibfnamefont {J.}~\bibnamefont {Sirker}},\
  }\href {https://link.aps.org/doi/10.1103/PhysRevB.100.014203} {\bibfield
  {journal} {\bibinfo  {journal} {Phys. Rev. B}\ }\textbf {\bibinfo {volume}
  {100}},\ \bibinfo {pages} {014203} (\bibinfo {year} {2019})}\BibitemShut
  {NoStop}%
\bibitem [{\citenamefont {Sierant}\ \emph {et~al.}()\citenamefont {Sierant},
  \citenamefont {Lewenstein}, \citenamefont {Scardicchio},\ and\ \citenamefont
  {Zakrzewski}}]{sierant2022stability}%
  \BibitemOpen
  \bibfield  {author} {\bibinfo {author} {\bibfnamefont {P.}~\bibnamefont
  {Sierant}}, \bibinfo {author} {\bibfnamefont {M.}~\bibnamefont {Lewenstein}},
  \bibinfo {author} {\bibfnamefont {A.}~\bibnamefont {Scardicchio}}, \ and\
  \bibinfo {author} {\bibfnamefont {J.}~\bibnamefont {Zakrzewski}},\ }\href
  {https://arxiv.org/abs/2203.15697} {}\Eprint
  {http://arxiv.org/abs/2203.15697} {2203.15697 [cond-mat.dis-nn]} \BibitemShut
  {NoStop}%
\bibitem [{\citenamefont {MacCormack}\ \emph {et~al.}(2021)\citenamefont
  {MacCormack}, \citenamefont {Tan}, \citenamefont {Kudler-Flam},\ and\
  \citenamefont {Ryu}}]{maccormack2021operator}%
  \BibitemOpen
  \bibfield  {author} {\bibinfo {author} {\bibfnamefont {I.}~\bibnamefont
  {MacCormack}}, \bibinfo {author} {\bibfnamefont {M.~T.}\ \bibnamefont {Tan}},
  \bibinfo {author} {\bibfnamefont {J.}~\bibnamefont {Kudler-Flam}}, \ and\
  \bibinfo {author} {\bibfnamefont {S.}~\bibnamefont {Ryu}},\ }\href
  {https://link.aps.org/doi/10.1103/PhysRevB.104.214202} {\bibfield  {journal}
  {\bibinfo  {journal} {Phys. Rev. B}\ }\textbf {\bibinfo {volume} {104}},\
  \bibinfo {pages} {214202} (\bibinfo {year} {2021})}\BibitemShut {NoStop}%
\bibitem [{\citenamefont {Serbyn}\ \emph {et~al.}(2013)\citenamefont {Serbyn},
  \citenamefont {Papi\ifmmode~\acute{c}\else \'{c}\fi{}},\ and\ \citenamefont
  {Abanin}}]{serbyn2013local}%
  \BibitemOpen
  \bibfield  {author} {\bibinfo {author} {\bibfnamefont {M.}~\bibnamefont
  {Serbyn}}, \bibinfo {author} {\bibfnamefont {Z.}~\bibnamefont
  {Papi\ifmmode~\acute{c}\else \'{c}\fi{}}}, \ and\ \bibinfo {author}
  {\bibfnamefont {D.~A.}\ \bibnamefont {Abanin}},\ }\href
  {https://link.aps.org/doi/10.1103/PhysRevLett.111.127201} {\bibfield
  {journal} {\bibinfo  {journal} {Phys. Rev. Lett.}\ }\textbf {\bibinfo
  {volume} {111}},\ \bibinfo {pages} {127201} (\bibinfo {year}
  {2013})}\BibitemShut {NoStop}%
\bibitem [{\citenamefont {Huse}\ \emph {et~al.}(2014)\citenamefont {Huse},
  \citenamefont {Nandkishore},\ and\ \citenamefont
  {Oganesyan}}]{huse2014phenomenology}%
  \BibitemOpen
  \bibfield  {author} {\bibinfo {author} {\bibfnamefont {D.~A.}\ \bibnamefont
  {Huse}}, \bibinfo {author} {\bibfnamefont {R.}~\bibnamefont {Nandkishore}}, \
  and\ \bibinfo {author} {\bibfnamefont {V.}~\bibnamefont {Oganesyan}},\ }\href
  {https://link.aps.org/doi/10.1103/PhysRevB.90.174202} {\bibfield  {journal}
  {\bibinfo  {journal} {Phys. Rev. B}\ }\textbf {\bibinfo {volume} {90}},\
  \bibinfo {pages} {174202} (\bibinfo {year} {2014})}\BibitemShut {NoStop}%
\bibitem [{\citenamefont {Ros}\ \emph {et~al.}(2015)\citenamefont {Ros},
  \citenamefont {Müller},\ and\ \citenamefont
  {Scardicchio}}]{ros2015integrals}%
  \BibitemOpen
  \bibfield  {author} {\bibinfo {author} {\bibfnamefont {V.}~\bibnamefont
  {Ros}}, \bibinfo {author} {\bibfnamefont {M.}~\bibnamefont {Müller}}, \ and\
  \bibinfo {author} {\bibfnamefont {A.}~\bibnamefont {Scardicchio}},\ }\href
  {https://www.sciencedirect.com/science/article/pii/S0550321314003836}
  {\bibfield  {journal} {\bibinfo  {journal} {Nucl. Phys. B.}\ }\textbf
  {\bibinfo {volume} {891}},\ \bibinfo {pages} {420} (\bibinfo {year}
  {2015})}\BibitemShut {NoStop}%
\bibitem [{\citenamefont
  {Imbrie}(2016{\natexlab{a}})}]{imbrie2016diagonalization}%
  \BibitemOpen
  \bibfield  {author} {\bibinfo {author} {\bibfnamefont {J.~Z.}\ \bibnamefont
  {Imbrie}},\ }\href {https://link.aps.org/doi/10.1103/PhysRevLett.117.027201}
  {\bibfield  {journal} {\bibinfo  {journal} {Phys. Rev. Lett.}\ }\textbf
  {\bibinfo {volume} {117}},\ \bibinfo {pages} {027201} (\bibinfo {year}
  {2016}{\natexlab{a}})}\BibitemShut {NoStop}%
\bibitem [{\citenamefont {Imbrie}(2016{\natexlab{b}})}]{imbrie2016onmanybody}%
  \BibitemOpen
  \bibfield  {author} {\bibinfo {author} {\bibfnamefont {J.~Z.}\ \bibnamefont
  {Imbrie}},\ }\href {https://doi.org/10.1007/s10955-016-1508-x} {\bibfield
  {journal} {\bibinfo  {journal} {J. Stat. Phys.}\ }\textbf {\bibinfo {volume}
  {163}},\ \bibinfo {pages} {998} (\bibinfo {year}
  {2016}{\natexlab{b}})}\BibitemShut {NoStop}%
\bibitem [{\citenamefont {Chanda}\ \emph {et~al.}(2020)\citenamefont {Chanda},
  \citenamefont {Sierant},\ and\ \citenamefont {Zakrzewski}}]{titas2020time}%
  \BibitemOpen
  \bibfield  {author} {\bibinfo {author} {\bibfnamefont {T.}~\bibnamefont
  {Chanda}}, \bibinfo {author} {\bibfnamefont {P.}~\bibnamefont {Sierant}}, \
  and\ \bibinfo {author} {\bibfnamefont {J.}~\bibnamefont {Zakrzewski}},\
  }\href {https://link.aps.org/doi/10.1103/PhysRevB.101.035148} {\bibfield
  {journal} {\bibinfo  {journal} {Phys. Rev. B}\ }\textbf {\bibinfo {volume}
  {101}},\ \bibinfo {pages} {035148} (\bibinfo {year} {2020})}\BibitemShut
  {NoStop}%
\bibitem [{\citenamefont {Sierant}\ and\ \citenamefont
  {Zakrzewski}()}]{sierant2021challenges}%
  \BibitemOpen
  \bibfield  {author} {\bibinfo {author} {\bibfnamefont {P.}~\bibnamefont
  {Sierant}}\ and\ \bibinfo {author} {\bibfnamefont {J.}~\bibnamefont
  {Zakrzewski}},\ }\href {https://arxiv.org/abs/2109.13608} {}\Eprint
  {http://arxiv.org/abs/2109.13608} {2109.13608 [cond-mat.dis-nn]} \BibitemShut
  {NoStop}%
\bibitem [{\citenamefont {Vosk}\ and\ \citenamefont
  {Altman}(2013)}]{vosk2013manybody}%
  \BibitemOpen
  \bibfield  {author} {\bibinfo {author} {\bibfnamefont {R.}~\bibnamefont
  {Vosk}}\ and\ \bibinfo {author} {\bibfnamefont {E.}~\bibnamefont {Altman}},\
  }\href {http://dx.doi.org/10.1103/PhysRevLett.110.067204} {\bibfield
  {journal} {\bibinfo  {journal} {Phys. Rev. Lett.}\ }\textbf {\bibinfo
  {volume} {110}},\ \bibinfo {pages} {067204} (\bibinfo {year}
  {2013})}\BibitemShut {NoStop}%
\bibitem [{\citenamefont {Vosk}\ and\ \citenamefont
  {Altman}(2014)}]{vosk2014dynamical}%
  \BibitemOpen
  \bibfield  {author} {\bibinfo {author} {\bibfnamefont {R.}~\bibnamefont
  {Vosk}}\ and\ \bibinfo {author} {\bibfnamefont {E.}~\bibnamefont {Altman}},\
  }\href {https://link.aps.org/doi/10.1103/PhysRevLett.112.217204} {\bibfield
  {journal} {\bibinfo  {journal} {Phys. Rev. Lett.}\ }\textbf {\bibinfo
  {volume} {112}},\ \bibinfo {pages} {217204} (\bibinfo {year}
  {2014})}\BibitemShut {NoStop}%
\bibitem [{\citenamefont {Herviou}\ \emph {et~al.}(2019)\citenamefont
  {Herviou}, \citenamefont {Bera},\ and\ \citenamefont
  {Bardarson}}]{herviou2019multiscale}%
  \BibitemOpen
  \bibfield  {author} {\bibinfo {author} {\bibfnamefont {L.}~\bibnamefont
  {Herviou}}, \bibinfo {author} {\bibfnamefont {S.}~\bibnamefont {Bera}}, \
  and\ \bibinfo {author} {\bibfnamefont {J.~H.}\ \bibnamefont {Bardarson}},\
  }\href {https://link.aps.org/doi/10.1103/PhysRevB.99.134205} {\bibfield
  {journal} {\bibinfo  {journal} {Phys. Rev. B}\ }\textbf {\bibinfo {volume}
  {99}},\ \bibinfo {pages} {134205} (\bibinfo {year} {2019})}\BibitemShut
  {NoStop}%
\bibitem [{\citenamefont {Vidal}\ and\ \citenamefont
  {Werner}(2002)}]{vidal2002computable}%
  \BibitemOpen
  \bibfield  {author} {\bibinfo {author} {\bibfnamefont {G.}~\bibnamefont
  {Vidal}}\ and\ \bibinfo {author} {\bibfnamefont {R.~F.}\ \bibnamefont
  {Werner}},\ }\href {https://link.aps.org/doi/10.1103/PhysRevA.65.032314}
  {\bibfield  {journal} {\bibinfo  {journal} {Phys. Rev. A}\ }\textbf {\bibinfo
  {volume} {65}},\ \bibinfo {pages} {032314} (\bibinfo {year}
  {2002})}\BibitemShut {NoStop}%
\bibitem [{\citenamefont {Shapourian}\ \emph {et~al.}(2017)\citenamefont
  {Shapourian}, \citenamefont {Shiozaki},\ and\ \citenamefont
  {Ryu}}]{shapourian2017partial}%
  \BibitemOpen
  \bibfield  {author} {\bibinfo {author} {\bibfnamefont {H.}~\bibnamefont
  {Shapourian}}, \bibinfo {author} {\bibfnamefont {K.}~\bibnamefont
  {Shiozaki}}, \ and\ \bibinfo {author} {\bibfnamefont {S.}~\bibnamefont
  {Ryu}},\ }\href {https://link.aps.org/doi/10.1103/PhysRevB.95.165101}
  {\bibfield  {journal} {\bibinfo  {journal} {Phys. Rev. B}\ }\textbf {\bibinfo
  {volume} {95}},\ \bibinfo {pages} {165101} (\bibinfo {year}
  {2017})}\BibitemShut {NoStop}%
\bibitem [{\citenamefont {Vasseur}\ \emph {et~al.}(2016)\citenamefont
  {Vasseur}, \citenamefont {Friedman}, \citenamefont {Parameswaran},\ and\
  \citenamefont {Potter}}]{vasseur2016particlehole}%
  \BibitemOpen
  \bibfield  {author} {\bibinfo {author} {\bibfnamefont {R.}~\bibnamefont
  {Vasseur}}, \bibinfo {author} {\bibfnamefont {A.~J.}\ \bibnamefont
  {Friedman}}, \bibinfo {author} {\bibfnamefont {S.~A.}\ \bibnamefont
  {Parameswaran}}, \ and\ \bibinfo {author} {\bibfnamefont {A.~C.}\
  \bibnamefont {Potter}},\ }\href
  {https://link.aps.org/doi/10.1103/PhysRevB.93.134207} {\bibfield  {journal}
  {\bibinfo  {journal} {Phys. Rev. B}\ }\textbf {\bibinfo {volume} {93}},\
  \bibinfo {pages} {134207} (\bibinfo {year} {2016})}\BibitemShut {NoStop}%
\bibitem [{\citenamefont {De~Tomasi}\ \emph {et~al.}()\citenamefont
  {De~Tomasi}, \citenamefont {Trapin}, \citenamefont {Heyl},\ and\
  \citenamefont {Bera}}]{detomasi2020anomalous}%
  \BibitemOpen
  \bibfield  {author} {\bibinfo {author} {\bibfnamefont {G.}~\bibnamefont
  {De~Tomasi}}, \bibinfo {author} {\bibfnamefont {D.}~\bibnamefont {Trapin}},
  \bibinfo {author} {\bibfnamefont {M.}~\bibnamefont {Heyl}}, \ and\ \bibinfo
  {author} {\bibfnamefont {S.}~\bibnamefont {Bera}},\ }\href
  {https://arxiv.org/abs/2001.04996} {}\Eprint
  {http://arxiv.org/abs/2001.04996} {2001.04996 [cond-mat.dis-nn]} \BibitemShut
  {NoStop}%
\bibitem [{\citenamefont {Ma}\ \emph {et~al.}(1979)\citenamefont {Ma},
  \citenamefont {Dasgupta},\ and\ \citenamefont {Hu}}]{ma1979random}%
  \BibitemOpen
  \bibfield  {author} {\bibinfo {author} {\bibfnamefont {S.-k.}\ \bibnamefont
  {Ma}}, \bibinfo {author} {\bibfnamefont {C.}~\bibnamefont {Dasgupta}}, \ and\
  \bibinfo {author} {\bibfnamefont {C.-k.}\ \bibnamefont {Hu}},\ }\href
  {https://link.aps.org/doi/10.1103/PhysRevLett.43.1434} {\bibfield  {journal}
  {\bibinfo  {journal} {Phys. Rev. Lett.}\ }\textbf {\bibinfo {volume} {43}},\
  \bibinfo {pages} {1434} (\bibinfo {year} {1979})}\BibitemShut {NoStop}%
\bibitem [{\citenamefont {Dasgupta}\ and\ \citenamefont
  {Ma}(1980)}]{dasgupta1980lowtemperature}%
  \BibitemOpen
  \bibfield  {author} {\bibinfo {author} {\bibfnamefont {C.}~\bibnamefont
  {Dasgupta}}\ and\ \bibinfo {author} {\bibfnamefont {S.-k.}\ \bibnamefont
  {Ma}},\ }\href {https://link.aps.org/doi/10.1103/PhysRevB.22.1305} {\bibfield
   {journal} {\bibinfo  {journal} {Phys. Rev. B}\ }\textbf {\bibinfo {volume}
  {22}},\ \bibinfo {pages} {1305} (\bibinfo {year} {1980})}\BibitemShut
  {NoStop}%
\bibitem [{\citenamefont {Doty}\ and\ \citenamefont
  {Fisher}(1992)}]{doty1992effects}%
  \BibitemOpen
  \bibfield  {author} {\bibinfo {author} {\bibfnamefont {C.~A.}\ \bibnamefont
  {Doty}}\ and\ \bibinfo {author} {\bibfnamefont {D.~S.}\ \bibnamefont
  {Fisher}},\ }\href {https://link.aps.org/doi/10.1103/PhysRevB.45.2167}
  {\bibfield  {journal} {\bibinfo  {journal} {Phys. Rev. B}\ }\textbf {\bibinfo
  {volume} {45}},\ \bibinfo {pages} {2167} (\bibinfo {year}
  {1992})}\BibitemShut {NoStop}%
\bibitem [{\citenamefont {Refael}\ and\ \citenamefont
  {Moore}(2004)}]{refael2004entanglement}%
  \BibitemOpen
  \bibfield  {author} {\bibinfo {author} {\bibfnamefont {G.}~\bibnamefont
  {Refael}}\ and\ \bibinfo {author} {\bibfnamefont {J.~E.}\ \bibnamefont
  {Moore}},\ }\href {http://dx.doi.org/10.1103/PhysRevLett.93.260602}
  {\bibfield  {journal} {\bibinfo  {journal} {Phys. Rev. Lett.}\ }\textbf
  {\bibinfo {volume} {93}},\ \bibinfo {pages} {260602} (\bibinfo {year}
  {2004})}\BibitemShut {NoStop}%
\bibitem [{\citenamefont {Laflorencie}(2005)}]{laflorencie2005scaling}%
  \BibitemOpen
  \bibfield  {author} {\bibinfo {author} {\bibfnamefont {N.}~\bibnamefont
  {Laflorencie}},\ }\href {https://link.aps.org/doi/10.1103/PhysRevB.72.140408}
  {\bibfield  {journal} {\bibinfo  {journal} {Phys. Rev. B}\ }\textbf {\bibinfo
  {volume} {72}},\ \bibinfo {pages} {140408} (\bibinfo {year}
  {2005})}\BibitemShut {NoStop}%
\bibitem [{\citenamefont {Refael}\ and\ \citenamefont
  {Moore}(2009)}]{refael2009criticality}%
  \BibitemOpen
  \bibfield  {author} {\bibinfo {author} {\bibfnamefont {G.}~\bibnamefont
  {Refael}}\ and\ \bibinfo {author} {\bibfnamefont {J.~E.}\ \bibnamefont
  {Moore}},\ }\href
  {https://iopscience.iop.org/article/10.1088/1751-8113/42/50/504010}
  {\bibfield  {journal} {\bibinfo  {journal} {J.~Phys. A}\ }\textbf {\bibinfo
  {volume} {42}},\ \bibinfo {pages} {504010} (\bibinfo {year}
  {2009})}\BibitemShut {NoStop}%
\bibitem [{\citenamefont {Ruggiero}\ \emph
  {et~al.}(2016{\natexlab{a}})\citenamefont {Ruggiero}, \citenamefont {Alba},\
  and\ \citenamefont {Calabrese}}]{ruggiero2016entanglement}%
  \BibitemOpen
  \bibfield  {author} {\bibinfo {author} {\bibfnamefont {P.}~\bibnamefont
  {Ruggiero}}, \bibinfo {author} {\bibfnamefont {V.}~\bibnamefont {Alba}}, \
  and\ \bibinfo {author} {\bibfnamefont {P.}~\bibnamefont {Calabrese}},\ }\href
  {http://dx.doi.org/10.1103/PhysRevB.94.035152} {\bibfield  {journal}
  {\bibinfo  {journal} {Phys. Rev. B}\ }\textbf {\bibinfo {volume} {94}},\
  \bibinfo {pages} {035152} (\bibinfo {year} {2016}{\natexlab{a}})}\BibitemShut
  {NoStop}%
\bibitem [{\citenamefont {Fagotti}\ \emph {et~al.}(2011)\citenamefont
  {Fagotti}, \citenamefont {Calabrese},\ and\ \citenamefont
  {Moore}}]{fagotti2011entanglement}%
  \BibitemOpen
  \bibfield  {author} {\bibinfo {author} {\bibfnamefont {M.}~\bibnamefont
  {Fagotti}}, \bibinfo {author} {\bibfnamefont {P.}~\bibnamefont {Calabrese}},
  \ and\ \bibinfo {author} {\bibfnamefont {J.~E.}\ \bibnamefont {Moore}},\
  }\href {https://link.aps.org/doi/10.1103/PhysRevB.83.045110} {\bibfield
  {journal} {\bibinfo  {journal} {Phys. Rev. B}\ }\textbf {\bibinfo {volume}
  {83}},\ \bibinfo {pages} {045110} (\bibinfo {year} {2011})}\BibitemShut
  {NoStop}%
\bibitem [{\citenamefont {Turkeshi}\ \emph
  {et~al.}(2020{\natexlab{a}})\citenamefont {Turkeshi}, \citenamefont
  {Ruggiero}, \citenamefont {Alba},\ and\ \citenamefont
  {Calabrese}}]{turkeshi2020entanglement}%
  \BibitemOpen
  \bibfield  {author} {\bibinfo {author} {\bibfnamefont {X.}~\bibnamefont
  {Turkeshi}}, \bibinfo {author} {\bibfnamefont {P.}~\bibnamefont {Ruggiero}},
  \bibinfo {author} {\bibfnamefont {V.}~\bibnamefont {Alba}}, \ and\ \bibinfo
  {author} {\bibfnamefont {P.}~\bibnamefont {Calabrese}},\ }\href
  {http://dx.doi.org/10.1103/PhysRevB.102.014455} {\bibfield  {journal}
  {\bibinfo  {journal} {Phys. Rev. B}\ }\textbf {\bibinfo {volume} {102}},\
  \bibinfo {pages} {014455} (\bibinfo {year} {2020}{\natexlab{a}})}\BibitemShut
  {NoStop}%
\bibitem [{\citenamefont {Turkeshi}\ \emph
  {et~al.}(2020{\natexlab{b}})\citenamefont {Turkeshi}, \citenamefont
  {Ruggiero},\ and\ \citenamefont {Calabrese}}]{turkeshi2020negativity}%
  \BibitemOpen
  \bibfield  {author} {\bibinfo {author} {\bibfnamefont {X.}~\bibnamefont
  {Turkeshi}}, \bibinfo {author} {\bibfnamefont {P.}~\bibnamefont {Ruggiero}},
  \ and\ \bibinfo {author} {\bibfnamefont {P.}~\bibnamefont {Calabrese}},\
  }\href {http://dx.doi.org/10.1103/PhysRevB.101.064207} {\bibfield  {journal}
  {\bibinfo  {journal} {Phys. Rev. B}\ }\textbf {\bibinfo {volume} {101}},\
  \bibinfo {pages} {064207} (\bibinfo {year} {2020}{\natexlab{b}})}\BibitemShut
  {NoStop}%
\bibitem [{\citenamefont {Kiefer-Emmanouilidis}\ \emph
  {et~al.}(2020{\natexlab{a}})\citenamefont {Kiefer-Emmanouilidis},
  \citenamefont {Unanyan}, \citenamefont {Sirker},\ and\ \citenamefont
  {Fleischhauer}}]{kiefer2020bounds}%
  \BibitemOpen
  \bibfield  {author} {\bibinfo {author} {\bibfnamefont {M.}~\bibnamefont
  {Kiefer-Emmanouilidis}}, \bibinfo {author} {\bibfnamefont {R.}~\bibnamefont
  {Unanyan}}, \bibinfo {author} {\bibfnamefont {J.}~\bibnamefont {Sirker}}, \
  and\ \bibinfo {author} {\bibfnamefont {M.}~\bibnamefont {Fleischhauer}},\
  }\href {\doibase 10.21468/SciPostPhys.8.6.083} {\bibfield  {journal}
  {\bibinfo  {journal} {SciPost Phys.}\ }\textbf {\bibinfo {volume} {8}},\
  \bibinfo {pages} {83} (\bibinfo {year} {2020}{\natexlab{a}})}\BibitemShut
  {NoStop}%
\bibitem [{\citenamefont {Igloi}\ and\ \citenamefont
  {Monthus}(2005)}]{igloi2005strong}%
  \BibitemOpen
  \bibfield  {author} {\bibinfo {author} {\bibfnamefont {F.}~\bibnamefont
  {Igloi}}\ and\ \bibinfo {author} {\bibfnamefont {C.}~\bibnamefont
  {Monthus}},\ }\href {\doibase 10.1016/j.physrep.2005.02.006} {\bibfield
  {journal} {\bibinfo  {journal} {Phys. Rep.}\ }\textbf {\bibinfo {volume}
  {412}},\ \bibinfo {pages} {277} (\bibinfo {year} {2005})}\BibitemShut
  {NoStop}%
\bibitem [{\citenamefont {Monthus}(2018{\natexlab{a}})}]{monthus2018strong}%
  \BibitemOpen
  \bibfield  {author} {\bibinfo {author} {\bibfnamefont {C.}~\bibnamefont
  {Monthus}},\ }\href {\doibase 10.1088/1751-8121/aac672} {\bibfield  {journal}
  {\bibinfo  {journal} {J. Phys. A}\ }\textbf {\bibinfo {volume} {51}},\
  \bibinfo {pages} {275302} (\bibinfo {year} {2018}{\natexlab{a}})}\BibitemShut
  {NoStop}%
\bibitem [{\citenamefont {Fisher}(1994)}]{fisher1994random}%
  \BibitemOpen
  \bibfield  {author} {\bibinfo {author} {\bibfnamefont {D.~S.}\ \bibnamefont
  {Fisher}},\ }\href {https://link.aps.org/doi/10.1103/PhysRevB.50.3799}
  {\bibfield  {journal} {\bibinfo  {journal} {Phys. Rev. B}\ }\textbf {\bibinfo
  {volume} {50}},\ \bibinfo {pages} {3799} (\bibinfo {year}
  {1994})}\BibitemShut {NoStop}%
\bibitem [{\citenamefont {Lieb}\ \emph {et~al.}(1961)\citenamefont {Lieb},
  \citenamefont {Schultz},\ and\ \citenamefont {Mattis}}]{lieb1961two}%
  \BibitemOpen
  \bibfield  {author} {\bibinfo {author} {\bibfnamefont {E.}~\bibnamefont
  {Lieb}}, \bibinfo {author} {\bibfnamefont {T.}~\bibnamefont {Schultz}}, \
  and\ \bibinfo {author} {\bibfnamefont {D.}~\bibnamefont {Mattis}},\ }\href
  {https://doi.org/10.1016/0003-4916(61)90115-4} {\bibfield  {journal}
  {\bibinfo  {journal} {Ann. Phys.}\ }\textbf {\bibinfo {volume} {16}},\
  \bibinfo {pages} {407} (\bibinfo {year} {1961})}\BibitemShut {NoStop}%
\bibitem [{\citenamefont {Dyson}(1953)}]{dyson1953the}%
  \BibitemOpen
  \bibfield  {author} {\bibinfo {author} {\bibfnamefont {F.~J.}\ \bibnamefont
  {Dyson}},\ }\href {https://link.aps.org/doi/10.1103/PhysRev.92.1331}
  {\bibfield  {journal} {\bibinfo  {journal} {Phys. Rev.}\ }\textbf {\bibinfo
  {volume} {92}},\ \bibinfo {pages} {1331} (\bibinfo {year}
  {1953})}\BibitemShut {NoStop}%
\bibitem [{\citenamefont {Sakurai}\ and\ \citenamefont
  {Napolitano}(2017)}]{sakurai}%
  \BibitemOpen
  \bibfield  {author} {\bibinfo {author} {\bibfnamefont {J.~J.}\ \bibnamefont
  {Sakurai}}\ and\ \bibinfo {author} {\bibfnamefont {J.}~\bibnamefont
  {Napolitano}},\ }\href {https://doi.org/10.1017/9781108499996} {\emph
  {\bibinfo {title} {Modern Quantum Mechanics}}}\ (\bibinfo  {publisher}
  {Cambridge University Press},\ \bibinfo {year} {2017})\BibitemShut {NoStop}%
\bibitem [{\citenamefont {Monthus}(2018{\natexlab{b}})}]{monthus2018floquet}%
  \BibitemOpen
  \bibfield  {author} {\bibinfo {author} {\bibfnamefont {C.}~\bibnamefont
  {Monthus}},\ }\href {https://doi.org/10.1088/1751-8121/aac672} {\bibfield
  {journal} {\bibinfo  {journal} {J. Phys. A}\ }\textbf {\bibinfo {volume}
  {51}},\ \bibinfo {pages} {275302} (\bibinfo {year}
  {2018}{\natexlab{b}})}\BibitemShut {NoStop}%
\bibitem [{\citenamefont {Bukov}\ \emph {et~al.}(2015)\citenamefont {Bukov},
  \citenamefont {D'Alessio},\ and\ \citenamefont
  {Polkovnikov}}]{bukov2015universal}%
  \BibitemOpen
  \bibfield  {author} {\bibinfo {author} {\bibfnamefont {M.}~\bibnamefont
  {Bukov}}, \bibinfo {author} {\bibfnamefont {L.}~\bibnamefont {D'Alessio}}, \
  and\ \bibinfo {author} {\bibfnamefont {A.}~\bibnamefont {Polkovnikov}},\
  }\href {https://doi.org/10.1080/00018732.2015.1055918} {\bibfield  {journal}
  {\bibinfo  {journal} {Adv. Phys.}\ }\textbf {\bibinfo {volume} {64}},\
  \bibinfo {pages} {139} (\bibinfo {year} {2015})}\BibitemShut {NoStop}%
\bibitem [{\citenamefont {Protopopov}\ \emph {et~al.}(2017)\citenamefont
  {Protopopov}, \citenamefont {Ho},\ and\ \citenamefont
  {Abanin}}]{protopopov2017effect}%
  \BibitemOpen
  \bibfield  {author} {\bibinfo {author} {\bibfnamefont {I.~V.}\ \bibnamefont
  {Protopopov}}, \bibinfo {author} {\bibfnamefont {W.~W.}\ \bibnamefont {Ho}},
  \ and\ \bibinfo {author} {\bibfnamefont {D.~A.}\ \bibnamefont {Abanin}},\
  }\href {https://link.aps.org/doi/10.1103/PhysRevB.96.041122} {\bibfield
  {journal} {\bibinfo  {journal} {Phys. Rev. B}\ }\textbf {\bibinfo {volume}
  {96}},\ \bibinfo {pages} {041122} (\bibinfo {year} {2017})}\BibitemShut
  {NoStop}%
\bibitem [{\citenamefont {Parameswaran}\ and\ \citenamefont
  {Vasseur}(2018)}]{Parameswaran_2018}%
  \BibitemOpen
  \bibfield  {author} {\bibinfo {author} {\bibfnamefont {S.~A.}\ \bibnamefont
  {Parameswaran}}\ and\ \bibinfo {author} {\bibfnamefont {R.}~\bibnamefont
  {Vasseur}},\ }\href {https://doi.org/10.1088/1361-6633/aac9ed} {\bibfield
  {journal} {\bibinfo  {journal} {Rep. Prog. Phys}\ }\textbf {\bibinfo {volume}
  {81}},\ \bibinfo {pages} {082501} (\bibinfo {year} {2018})}\BibitemShut
  {NoStop}%
\bibitem [{\citenamefont {Protopopov}\ \emph {et~al.}(2020)\citenamefont
  {Protopopov}, \citenamefont {Panda}, \citenamefont {Parolini}, \citenamefont
  {Scardicchio}, \citenamefont {Demler},\ and\ \citenamefont
  {Abanin}}]{protopopov2020nonabelian}%
  \BibitemOpen
  \bibfield  {author} {\bibinfo {author} {\bibfnamefont {I.~V.}\ \bibnamefont
  {Protopopov}}, \bibinfo {author} {\bibfnamefont {R.~K.}\ \bibnamefont
  {Panda}}, \bibinfo {author} {\bibfnamefont {T.}~\bibnamefont {Parolini}},
  \bibinfo {author} {\bibfnamefont {A.}~\bibnamefont {Scardicchio}}, \bibinfo
  {author} {\bibfnamefont {E.}~\bibnamefont {Demler}}, \ and\ \bibinfo {author}
  {\bibfnamefont {D.~A.}\ \bibnamefont {Abanin}},\ }\href
  {https://link.aps.org/doi/10.1103/PhysRevX.10.011025} {\bibfield  {journal}
  {\bibinfo  {journal} {Phys. Rev. X}\ }\textbf {\bibinfo {volume} {10}},\
  \bibinfo {pages} {011025} (\bibinfo {year} {2020})}\BibitemShut {NoStop}%
\bibitem [{\citenamefont {Ware}\ \emph {et~al.}(2021)\citenamefont {Ware},
  \citenamefont {Abanin},\ and\ \citenamefont {Vasseur}}]{vasseur3}%
  \BibitemOpen
  \bibfield  {author} {\bibinfo {author} {\bibfnamefont {B.}~\bibnamefont
  {Ware}}, \bibinfo {author} {\bibfnamefont {D.}~\bibnamefont {Abanin}}, \ and\
  \bibinfo {author} {\bibfnamefont {R.}~\bibnamefont {Vasseur}},\ }\href
  {https://link.aps.org/doi/10.1103/PhysRevB.103.094203} {\bibfield  {journal}
  {\bibinfo  {journal} {Phys. Rev. B}\ }\textbf {\bibinfo {volume} {103}},\
  \bibinfo {pages} {094203} (\bibinfo {year} {2021})}\BibitemShut {NoStop}%
\bibitem [{\citenamefont {Potter}\ and\ \citenamefont
  {Vasseur}(2016)}]{vasseur2}%
  \BibitemOpen
  \bibfield  {author} {\bibinfo {author} {\bibfnamefont {A.~C.}\ \bibnamefont
  {Potter}}\ and\ \bibinfo {author} {\bibfnamefont {R.}~\bibnamefont
  {Vasseur}},\ }\href {https://link.aps.org/doi/10.1103/PhysRevB.94.224206}
  {\bibfield  {journal} {\bibinfo  {journal} {Phys. Rev. B}\ }\textbf {\bibinfo
  {volume} {94}},\ \bibinfo {pages} {224206} (\bibinfo {year}
  {2016})}\BibitemShut {NoStop}%
\bibitem [{Note1()}]{Note1}%
  \BibitemOpen
  \bibinfo {note} {The most general case of random product state is a product
  of singlets (cf. Eq.~\protect \textup {\hbox {\mathsurround \z@ \protect
  \normalfont (\ignorespaces \ref {eq:singlets}\unskip \@@italiccorr )}}) and
  frozen degrees of freedom.}\BibitemShut {Stop}%
\bibitem [{\citenamefont {Nielsen}\ and\ \citenamefont
  {Chuang}(2009)}]{Nielsen2009}%
  \BibitemOpen
  \bibfield  {author} {\bibinfo {author} {\bibfnamefont {M.~A.}\ \bibnamefont
  {Nielsen}}\ and\ \bibinfo {author} {\bibfnamefont {I.~L.}\ \bibnamefont
  {Chuang}},\ }\href {https://doi.org/10.1017/cbo9780511976667} {\emph
  {\bibinfo {title} {Quantum Computation and Quantum Information}}}\ (\bibinfo
  {publisher} {Cambridge University Press},\ \bibinfo {year}
  {2009})\BibitemShut {NoStop}%
\bibitem [{\citenamefont {Berta}\ \emph {et~al.}(2015)\citenamefont {Berta},
  \citenamefont {Seshadreesan},\ and\ \citenamefont {Wilde}}]{berta2015renyi}%
  \BibitemOpen
  \bibfield  {author} {\bibinfo {author} {\bibfnamefont {M.}~\bibnamefont
  {Berta}}, \bibinfo {author} {\bibfnamefont {K.~P.}\ \bibnamefont
  {Seshadreesan}}, \ and\ \bibinfo {author} {\bibfnamefont {M.~M.}\
  \bibnamefont {Wilde}},\ }\href
  {https://link.aps.org/doi/10.1103/PhysRevA.91.022333} {\bibfield  {journal}
  {\bibinfo  {journal} {Phys. Rev. A}\ }\textbf {\bibinfo {volume} {91}},\
  \bibinfo {pages} {022333} (\bibinfo {year} {2015})}\BibitemShut {NoStop}%
\bibitem [{\citenamefont {Alba}\ and\ \citenamefont
  {Calabrese}(2017)}]{alba2017entanglement}%
  \BibitemOpen
  \bibfield  {author} {\bibinfo {author} {\bibfnamefont {V.}~\bibnamefont
  {Alba}}\ and\ \bibinfo {author} {\bibfnamefont {P.}~\bibnamefont
  {Calabrese}},\ }\href {http://dx.doi.org/10.1073/pnas.1703516114} {\bibfield
  {journal} {\bibinfo  {journal} {PNAS}\ }\textbf {\bibinfo {volume} {114}},\
  \bibinfo {pages} {7947} (\bibinfo {year} {2017})}\BibitemShut {NoStop}%
\bibitem [{\citenamefont {Alba}\ and\ \citenamefont
  {Calabrese}(2018)}]{alba2018entanglement}%
  \BibitemOpen
  \bibfield  {author} {\bibinfo {author} {\bibfnamefont {V.}~\bibnamefont
  {Alba}}\ and\ \bibinfo {author} {\bibfnamefont {P.}~\bibnamefont
  {Calabrese}},\ }\href {http://dx.doi.org/10.21468/SciPostPhys.4.3.017}
  {\bibfield  {journal} {\bibinfo  {journal} {SciPost Phys.}\ }\textbf
  {\bibinfo {volume} {4}},\ \bibinfo {pages} {017} (\bibinfo {year}
  {2018})}\BibitemShut {NoStop}%
\bibitem [{\citenamefont {Alba}\ and\ \citenamefont
  {Calabrese}(2019{\natexlab{a}})}]{alba2019quantum_information}%
  \BibitemOpen
  \bibfield  {author} {\bibinfo {author} {\bibfnamefont {V.}~\bibnamefont
  {Alba}}\ and\ \bibinfo {author} {\bibfnamefont {P.}~\bibnamefont
  {Calabrese}},\ }\href {http://dx.doi.org/10.1103/PhysRevB.100.115150}
  {\bibfield  {journal} {\bibinfo  {journal} {Phys. Rev. B}\ }\textbf {\bibinfo
  {volume} {100}},\ \bibinfo {pages} {115150} (\bibinfo {year}
  {2019}{\natexlab{a}})}\BibitemShut {NoStop}%
\bibitem [{\citenamefont {Alba}\ and\ \citenamefont
  {Calabrese}(2019{\natexlab{b}})}]{alba2019quantum}%
  \BibitemOpen
  \bibfield  {author} {\bibinfo {author} {\bibfnamefont {V.}~\bibnamefont
  {Alba}}\ and\ \bibinfo {author} {\bibfnamefont {P.}~\bibnamefont
  {Calabrese}},\ }\href {http://dx.doi.org/10.1209/0295-5075/126/60001}
  {\bibfield  {journal} {\bibinfo  {journal} {EPL}\ }\textbf {\bibinfo {volume}
  {126}},\ \bibinfo {pages} {60001} (\bibinfo {year}
  {2019}{\natexlab{b}})}\BibitemShut {NoStop}%
\bibitem [{\citenamefont {Kudler-Flam}\ \emph {et~al.}(2020)\citenamefont
  {Kudler-Flam}, \citenamefont {Shapourian},\ and\ \citenamefont
  {Ryu}}]{kudlerflam2020the}%
  \BibitemOpen
  \bibfield  {author} {\bibinfo {author} {\bibfnamefont {J.}~\bibnamefont
  {Kudler-Flam}}, \bibinfo {author} {\bibfnamefont {H.}~\bibnamefont
  {Shapourian}}, \ and\ \bibinfo {author} {\bibfnamefont {S.}~\bibnamefont
  {Ryu}},\ }\href {http://dx.doi.org/10.21468/SciPostPhys.8.4.063} {\bibfield
  {journal} {\bibinfo  {journal} {SciPost Phys.}\ }\textbf {\bibinfo {volume}
  {8}} (\bibinfo {year} {2020})}\BibitemShut {NoStop}%
\bibitem [{\citenamefont {Kudler-Flam}\ \emph {et~al.}(2021)\citenamefont
  {Kudler-Flam}, \citenamefont {Kusuki},\ and\ \citenamefont
  {Ryu}}]{kudlerflam2020thequasi}%
  \BibitemOpen
  \bibfield  {author} {\bibinfo {author} {\bibfnamefont {J.}~\bibnamefont
  {Kudler-Flam}}, \bibinfo {author} {\bibfnamefont {Y.}~\bibnamefont {Kusuki}},
  \ and\ \bibinfo {author} {\bibfnamefont {S.}~\bibnamefont {Ryu}},\ }\href
  {http://dx.doi.org/10.1007/JHEP03(2021)146} {\bibfield  {journal} {\bibinfo
  {journal} {J. High Energy Phys.}\ }\textbf {\bibinfo {volume} {2021}}
  (\bibinfo {year} {2021})}\BibitemShut {NoStop}%
\bibitem [{\citenamefont {Bertini}\ \emph {et~al.}()\citenamefont {Bertini},
  \citenamefont {Klobas},\ and\ \citenamefont {Lu}}]{bertini2022entanglement}%
  \BibitemOpen
  \bibfield  {author} {\bibinfo {author} {\bibfnamefont {B.}~\bibnamefont
  {Bertini}}, \bibinfo {author} {\bibfnamefont {K.}~\bibnamefont {Klobas}}, \
  and\ \bibinfo {author} {\bibfnamefont {T.-C.}\ \bibnamefont {Lu}},\ }\href
  {https://arxiv.org/abs/2203.17254} {}\Eprint
  {http://arxiv.org/abs/2203.17254} {2203.17254 [quant-ph]} \BibitemShut
  {NoStop}%
\bibitem [{\citenamefont {Shi}\ \emph {et~al.}()\citenamefont {Shi},
  \citenamefont {Dai},\ and\ \citenamefont {Lu}}]{shi2020entanglement}%
  \BibitemOpen
  \bibfield  {author} {\bibinfo {author} {\bibfnamefont {B.}~\bibnamefont
  {Shi}}, \bibinfo {author} {\bibfnamefont {X.}~\bibnamefont {Dai}}, \ and\
  \bibinfo {author} {\bibfnamefont {Y.-M.}\ \bibnamefont {Lu}},\ }\href
  {\doibase 10.48550/ARXIV.2012.00040} {}\Eprint
  {http://arxiv.org/abs/2012.00040} {2012.00040 [cond-mat.stat-mech]}
  \BibitemShut {NoStop}%
\bibitem [{\citenamefont {Sang}\ \emph {et~al.}(2021)\citenamefont {Sang},
  \citenamefont {Li}, \citenamefont {Zhou}, \citenamefont {Chen}, \citenamefont
  {Hsieh},\ and\ \citenamefont {Fisher}}]{sang2021entanglement}%
  \BibitemOpen
  \bibfield  {author} {\bibinfo {author} {\bibfnamefont {S.}~\bibnamefont
  {Sang}}, \bibinfo {author} {\bibfnamefont {Y.}~\bibnamefont {Li}}, \bibinfo
  {author} {\bibfnamefont {T.}~\bibnamefont {Zhou}}, \bibinfo {author}
  {\bibfnamefont {X.}~\bibnamefont {Chen}}, \bibinfo {author} {\bibfnamefont
  {T.~H.}\ \bibnamefont {Hsieh}}, \ and\ \bibinfo {author} {\bibfnamefont
  {M.~P.}\ \bibnamefont {Fisher}},\ }\href {\doibase
  10.1103/PRXQuantum.2.030313} {\bibfield  {journal} {\bibinfo  {journal} {PRX
  Quantum}\ }\textbf {\bibinfo {volume} {2}},\ \bibinfo {pages} {030313}
  (\bibinfo {year} {2021})}\BibitemShut {NoStop}%
\bibitem [{\citenamefont {Sharma}\ \emph {et~al.}(2022)\citenamefont {Sharma},
  \citenamefont {Turkeshi}, \citenamefont {Fazio},\ and\ \citenamefont
  {Dalmonte}}]{sharma2022}%
  \BibitemOpen
  \bibfield  {author} {\bibinfo {author} {\bibfnamefont {S.}~\bibnamefont
  {Sharma}}, \bibinfo {author} {\bibfnamefont {X.}~\bibnamefont {Turkeshi}},
  \bibinfo {author} {\bibfnamefont {R.}~\bibnamefont {Fazio}}, \ and\ \bibinfo
  {author} {\bibfnamefont {M.}~\bibnamefont {Dalmonte}},\ }\href {\doibase
  10.21468/SciPostPhysCore.5.2.023} {\bibfield  {journal} {\bibinfo  {journal}
  {SciPost Phys. Core}\ }\textbf {\bibinfo {volume} {5}},\ \bibinfo {pages}
  {23} (\bibinfo {year} {2022})}\BibitemShut {NoStop}%
\bibitem [{\citenamefont {Weinstein}\ \emph {et~al.}()\citenamefont
  {Weinstein}, \citenamefont {Bao},\ and\ \citenamefont
  {Altman}}]{weinstein2022}%
  \BibitemOpen
  \bibfield  {author} {\bibinfo {author} {\bibfnamefont {Z.}~\bibnamefont
  {Weinstein}}, \bibinfo {author} {\bibfnamefont {Y.}~\bibnamefont {Bao}}, \
  and\ \bibinfo {author} {\bibfnamefont {E.}~\bibnamefont {Altman}},\ }\href
  {https://arxiv.org/abs/2202.12905} {}\Eprint
  {http://arxiv.org/abs/2202.12905} {2202.12905 [quant-ph]} \BibitemShut
  {NoStop}%
\bibitem [{\citenamefont {Turkeshi}\ \emph {et~al.}()\citenamefont {Turkeshi},
  \citenamefont {Piroli},\ and\ \citenamefont
  {Schirò}}]{turkeshi2022enhanced}%
  \BibitemOpen
  \bibfield  {author} {\bibinfo {author} {\bibfnamefont {X.}~\bibnamefont
  {Turkeshi}}, \bibinfo {author} {\bibfnamefont {L.}~\bibnamefont {Piroli}}, \
  and\ \bibinfo {author} {\bibfnamefont {M.}~\bibnamefont {Schirò}},\ }\href
  {https://arxiv.org/abs/2205.07992} {}\Eprint
  {http://arxiv.org/abs/2205.07992} {2205.07992 [cond-mat.stat-mech]}
  \BibitemShut {NoStop}%
\bibitem [{\citenamefont {Alba}\ and\ \citenamefont
  {Carollo}()}]{alba2022logarithmic}%
  \BibitemOpen
  \bibfield  {author} {\bibinfo {author} {\bibfnamefont {V.}~\bibnamefont
  {Alba}}\ and\ \bibinfo {author} {\bibfnamefont {F.}~\bibnamefont {Carollo}},\
  }\href@noop {} {}\Eprint {http://arxiv.org/abs/2205.02139} {2205.02139
  [cond-mat.stat-mech]} \BibitemShut {NoStop}%
\bibitem [{\citenamefont {Plenio}(2005)}]{plenio2005logarithmic}%
  \BibitemOpen
  \bibfield  {author} {\bibinfo {author} {\bibfnamefont {M.~B.}\ \bibnamefont
  {Plenio}},\ }\href {https://link.aps.org/doi/10.1103/PhysRevLett.95.090503}
  {\bibfield  {journal} {\bibinfo  {journal} {Phys. Rev. Lett.}\ }\textbf
  {\bibinfo {volume} {95}},\ \bibinfo {pages} {090503} (\bibinfo {year}
  {2005})}\BibitemShut {NoStop}%
\bibitem [{Note2()}]{Note2}%
  \BibitemOpen
  \bibinfo {note} {Note, however, the fermionic negativity, as any entanglement
  measure, is independent of the choice of basis.}\BibitemShut {Stop}%
\bibitem [{\citenamefont {Shapourian}\ and\ \citenamefont
  {Ryu}(2019)}]{shapourian2019entanglement}%
  \BibitemOpen
  \bibfield  {author} {\bibinfo {author} {\bibfnamefont {H.}~\bibnamefont
  {Shapourian}}\ and\ \bibinfo {author} {\bibfnamefont {S.}~\bibnamefont
  {Ryu}},\ }\href {http://dx.doi.org/10.1103/PhysRevA.99.022310} {\bibfield
  {journal} {\bibinfo  {journal} {Phys. Rev. A}\ }\textbf {\bibinfo {volume}
  {99}},\ \bibinfo {pages} {022310} (\bibinfo {year} {2019})}\BibitemShut
  {NoStop}%
\bibitem [{\citenamefont {Shapourian}\ \emph {et~al.}(2019)\citenamefont
  {Shapourian}, \citenamefont {Ruggiero}, \citenamefont {Ryu},\ and\
  \citenamefont {Calabrese}}]{shapourian2019twisted}%
  \BibitemOpen
  \bibfield  {author} {\bibinfo {author} {\bibfnamefont {H.}~\bibnamefont
  {Shapourian}}, \bibinfo {author} {\bibfnamefont {P.}~\bibnamefont
  {Ruggiero}}, \bibinfo {author} {\bibfnamefont {S.}~\bibnamefont {Ryu}}, \
  and\ \bibinfo {author} {\bibfnamefont {P.}~\bibnamefont {Calabrese}},\ }\href
  {\doibase 10.21468/SciPostPhys.7.3.037} {\bibfield  {journal} {\bibinfo
  {journal} {SciPost Phys.}\ }\textbf {\bibinfo {volume} {7}},\ \bibinfo
  {pages} {37} (\bibinfo {year} {2019})}\BibitemShut {NoStop}%
\bibitem [{Note3()}]{Note3}%
  \BibitemOpen
  \bibinfo {note} {For comparison, we note that in Ref.~\cite
  {vosk2013manybody} the density of entanglement per pair $S_p$ was instead
  considered, with $S_p =s_p/\protect \qopname \relax o{log}2 \simeq
  0.557$).}\BibitemShut {Stop}%
\bibitem [{\citenamefont {Burrell}\ and\ \citenamefont
  {Osborne}(2007)}]{burrell2007bounds}%
  \BibitemOpen
  \bibfield  {author} {\bibinfo {author} {\bibfnamefont {C.~K.}\ \bibnamefont
  {Burrell}}\ and\ \bibinfo {author} {\bibfnamefont {T.~J.}\ \bibnamefont
  {Osborne}},\ }\href {https://link.aps.org/doi/10.1103/PhysRevLett.99.167201}
  {\bibfield  {journal} {\bibinfo  {journal} {Phys. Rev. Lett.}\ }\textbf
  {\bibinfo {volume} {99}},\ \bibinfo {pages} {167201} (\bibinfo {year}
  {2007})}\BibitemShut {NoStop}%
\bibitem [{\citenamefont {Coser}\ \emph {et~al.}(2014)\citenamefont {Coser},
  \citenamefont {Tonni},\ and\ \citenamefont
  {Calabrese}}]{coser2014entanglement}%
  \BibitemOpen
  \bibfield  {author} {\bibinfo {author} {\bibfnamefont {A.}~\bibnamefont
  {Coser}}, \bibinfo {author} {\bibfnamefont {E.}~\bibnamefont {Tonni}}, \ and\
  \bibinfo {author} {\bibfnamefont {P.}~\bibnamefont {Calabrese}},\ }\href
  {\doibase 10.1088/1742-5468/2014/12/p12017} {\bibfield  {journal} {\bibinfo
  {journal} {J. Stat. Mech. Theory Exp.}\ }\textbf {\bibinfo {volume} {2014}},\
  \bibinfo {pages} {P12017} (\bibinfo {year} {2014})}\BibitemShut {NoStop}%
\bibitem [{\citenamefont {Ruggiero}\ \emph
  {et~al.}(2016{\natexlab{b}})\citenamefont {Ruggiero}, \citenamefont {Alba},\
  and\ \citenamefont {Calabrese}}]{ruggiero2016negativity}%
  \BibitemOpen
  \bibfield  {author} {\bibinfo {author} {\bibfnamefont {P.}~\bibnamefont
  {Ruggiero}}, \bibinfo {author} {\bibfnamefont {V.}~\bibnamefont {Alba}}, \
  and\ \bibinfo {author} {\bibfnamefont {P.}~\bibnamefont {Calabrese}},\ }\href
  {http://dx.doi.org/10.1103/PhysRevB.94.195121} {\bibfield  {journal}
  {\bibinfo  {journal} {Phys. Rev. B}\ }\textbf {\bibinfo {volume} {94}},\
  \bibinfo {pages} {195121} (\bibinfo {year} {2016}{\natexlab{b}})}\BibitemShut
  {NoStop}%
\bibitem [{\citenamefont {Page}(1993)}]{page1993average}%
  \BibitemOpen
  \bibfield  {author} {\bibinfo {author} {\bibfnamefont {D.~N.}\ \bibnamefont
  {Page}},\ }\href {https://link.aps.org/doi/10.1103/PhysRevLett.71.1291}
  {\bibfield  {journal} {\bibinfo  {journal} {Phys. Rev. Lett.}\ }\textbf
  {\bibinfo {volume} {71}},\ \bibinfo {pages} {1291} (\bibinfo {year}
  {1993})}\BibitemShut {NoStop}%
\bibitem [{\citenamefont {Zyczkowski}\ and\ \citenamefont
  {Sommers}(2001)}]{zyczkowski2001induced}%
  \BibitemOpen
  \bibfield  {author} {\bibinfo {author} {\bibfnamefont {K.}~\bibnamefont
  {Zyczkowski}}\ and\ \bibinfo {author} {\bibfnamefont {H.-J.}\ \bibnamefont
  {Sommers}},\ }\href {https://doi.org/10.1088/0305-4470/34/35/335} {\bibfield
  {journal} {\bibinfo  {journal} {J. Phys. A}\ }\textbf {\bibinfo {volume}
  {34}},\ \bibinfo {pages} {7111} (\bibinfo {year} {2001})}\BibitemShut
  {NoStop}%
\bibitem [{\citenamefont {Mbeng}\ \emph {et~al.}(2017)\citenamefont {Mbeng},
  \citenamefont {Alba},\ and\ \citenamefont {Calabrese}}]{mbeng2017negativity}%
  \BibitemOpen
  \bibfield  {author} {\bibinfo {author} {\bibfnamefont {G.~B.}\ \bibnamefont
  {Mbeng}}, \bibinfo {author} {\bibfnamefont {V.}~\bibnamefont {Alba}}, \ and\
  \bibinfo {author} {\bibfnamefont {P.}~\bibnamefont {Calabrese}},\ }\href
  {https://iopscience.iop.org/article/10.1088/1751-8121/aa6734} {\bibfield
  {journal} {\bibinfo  {journal} {J. Phys. A}\ }\textbf {\bibinfo {volume}
  {50}},\ \bibinfo {pages} {194001} (\bibinfo {year} {2017})}\BibitemShut
  {NoStop}%
\bibitem [{\citenamefont {Blondeau-Fournier}\ \emph {et~al.}(2016)\citenamefont
  {Blondeau-Fournier}, \citenamefont {Castro-Alvaredo},\ and\ \citenamefont
  {Doyon}}]{blondeau2016universal}%
  \BibitemOpen
  \bibfield  {author} {\bibinfo {author} {\bibfnamefont {O.}~\bibnamefont
  {Blondeau-Fournier}}, \bibinfo {author} {\bibfnamefont {O.~A.}\ \bibnamefont
  {Castro-Alvaredo}}, \ and\ \bibinfo {author} {\bibfnamefont {B.}~\bibnamefont
  {Doyon}},\ }\href {https://doi.org/10.1088/1751-8113/49/12/125401} {\bibfield
   {journal} {\bibinfo  {journal} {J. Phys. A}\ }\textbf {\bibinfo {volume}
  {49}},\ \bibinfo {pages} {125401} (\bibinfo {year} {2016})}\BibitemShut
  {NoStop}%
\bibitem [{\citenamefont {Shapourian}\ \emph {et~al.}(2021)\citenamefont
  {Shapourian}, \citenamefont {Liu}, \citenamefont {Kudler-Flam},\ and\
  \citenamefont {Vishwanath}}]{shapourian2021entanglement}%
  \BibitemOpen
  \bibfield  {author} {\bibinfo {author} {\bibfnamefont {H.}~\bibnamefont
  {Shapourian}}, \bibinfo {author} {\bibfnamefont {S.}~\bibnamefont {Liu}},
  \bibinfo {author} {\bibfnamefont {J.}~\bibnamefont {Kudler-Flam}}, \ and\
  \bibinfo {author} {\bibfnamefont {A.}~\bibnamefont {Vishwanath}},\ }\href
  {https://link.aps.org/doi/10.1103/PRXQuantum.2.030347} {\bibfield  {journal}
  {\bibinfo  {journal} {PRX Quantum}\ }\textbf {\bibinfo {volume} {2}},\
  \bibinfo {pages} {030347} (\bibinfo {year} {2021})}\BibitemShut {NoStop}%
\bibitem [{\citenamefont {Peschel}\ and\ \citenamefont
  {Eisler}(2009)}]{Peschel_2009}%
  \BibitemOpen
  \bibfield  {author} {\bibinfo {author} {\bibfnamefont {I.}~\bibnamefont
  {Peschel}}\ and\ \bibinfo {author} {\bibfnamefont {V.}~\bibnamefont
  {Eisler}},\ }\href {https://doi.org/10.1088/1751-8113/42/50/504003}
  {\bibfield  {journal} {\bibinfo  {journal} {J. Phys. A}\ }\textbf {\bibinfo
  {volume} {42}},\ \bibinfo {pages} {504003} (\bibinfo {year}
  {2009})}\BibitemShut {NoStop}%
\bibitem [{Note4()}]{Note4}%
  \BibitemOpen
  \bibinfo {note} {The same does not hold true for the logarithmic negativity,
  since the partial transpose of a fermionic Gaussian state is \protect \textit
  {not} Gaussian.}\BibitemShut {Stop}%
\bibitem [{\citenamefont {Eisert}\ \emph {et~al.}(2018)\citenamefont {Eisert},
  \citenamefont {Eisler},\ and\ \citenamefont
  {Zimbor\'as}}]{eisert2018entanglement}%
  \BibitemOpen
  \bibfield  {author} {\bibinfo {author} {\bibfnamefont {J.}~\bibnamefont
  {Eisert}}, \bibinfo {author} {\bibfnamefont {V.}~\bibnamefont {Eisler}}, \
  and\ \bibinfo {author} {\bibfnamefont {Z.}~\bibnamefont {Zimbor\'as}},\
  }\href {https://link.aps.org/doi/10.1103/PhysRevB.97.165123} {\bibfield
  {journal} {\bibinfo  {journal} {Phys. Rev. B}\ }\textbf {\bibinfo {volume}
  {97}},\ \bibinfo {pages} {165123} (\bibinfo {year} {2018})}\BibitemShut
  {NoStop}%
\bibitem [{\citenamefont {Goldstein}\ and\ \citenamefont
  {Sela}(2018)}]{goldstein2018symmetryresolved}%
  \BibitemOpen
  \bibfield  {author} {\bibinfo {author} {\bibfnamefont {M.}~\bibnamefont
  {Goldstein}}\ and\ \bibinfo {author} {\bibfnamefont {E.}~\bibnamefont
  {Sela}},\ }\href {https://link.aps.org/doi/10.1103/PhysRevLett.120.200602}
  {\bibfield  {journal} {\bibinfo  {journal} {Phys. Rev. Lett.}\ }\textbf
  {\bibinfo {volume} {120}},\ \bibinfo {pages} {200602} (\bibinfo {year}
  {2018})}\BibitemShut {NoStop}%
\bibitem [{\citenamefont {Cornfeld}\ \emph {et~al.}(2018)\citenamefont
  {Cornfeld}, \citenamefont {Goldstein},\ and\ \citenamefont
  {Sela}}]{cornfeld2018imbalance}%
  \BibitemOpen
  \bibfield  {author} {\bibinfo {author} {\bibfnamefont {E.}~\bibnamefont
  {Cornfeld}}, \bibinfo {author} {\bibfnamefont {M.}~\bibnamefont {Goldstein}},
  \ and\ \bibinfo {author} {\bibfnamefont {E.}~\bibnamefont {Sela}},\ }\href
  {https://link.aps.org/doi/10.1103/PhysRevA.98.032302} {\bibfield  {journal}
  {\bibinfo  {journal} {Phys. Rev. A}\ }\textbf {\bibinfo {volume} {98}},\
  \bibinfo {pages} {032302} (\bibinfo {year} {2018})}\BibitemShut {NoStop}%
\bibitem [{\citenamefont {Murciano}\ \emph {et~al.}(2021)\citenamefont
  {Murciano}, \citenamefont {Bonsignori},\ and\ \citenamefont
  {Calabrese}}]{sara}%
  \BibitemOpen
  \bibfield  {author} {\bibinfo {author} {\bibfnamefont {S.}~\bibnamefont
  {Murciano}}, \bibinfo {author} {\bibfnamefont {R.}~\bibnamefont
  {Bonsignori}}, \ and\ \bibinfo {author} {\bibfnamefont {P.}~\bibnamefont
  {Calabrese}},\ }\href {\doibase 10.21468/SciPostPhys.10.5.111} {\bibfield
  {journal} {\bibinfo  {journal} {SciPost Phys.}\ }\textbf {\bibinfo {volume}
  {10}},\ \bibinfo {pages} {111} (\bibinfo {year} {2021})}\BibitemShut
  {NoStop}%
\bibitem [{\citenamefont {Parez}\ \emph {et~al.}()\citenamefont {Parez},
  \citenamefont {Bonsignori},\ and\ \citenamefont {Calabrese}}]{riccarda}%
  \BibitemOpen
  \bibfield  {author} {\bibinfo {author} {\bibfnamefont {G.}~\bibnamefont
  {Parez}}, \bibinfo {author} {\bibfnamefont {R.}~\bibnamefont {Bonsignori}}, \
  and\ \bibinfo {author} {\bibfnamefont {P.}~\bibnamefont {Calabrese}},\ }\href
  {https://arxiv.org/abs/2202.05309} {}\Eprint
  {http://arxiv.org/abs/2202.05309} {2202.05309 [cond-mat.stat-mech]}
  \BibitemShut {NoStop}%
\bibitem [{\citenamefont {Parez}\ \emph {et~al.}(2021)\citenamefont {Parez},
  \citenamefont {Bonsignori},\ and\ \citenamefont {Calabrese}}]{Parez_2021}%
  \BibitemOpen
  \bibfield  {author} {\bibinfo {author} {\bibfnamefont {G.}~\bibnamefont
  {Parez}}, \bibinfo {author} {\bibfnamefont {R.}~\bibnamefont {Bonsignori}}, \
  and\ \bibinfo {author} {\bibfnamefont {P.}~\bibnamefont {Calabrese}},\ }\href
  {https://doi.org/10.1088%2F1742-5468%2Fac21d7} {\bibfield  {journal}
  {\bibinfo  {journal} {J. Stat. Mech. Theory Exp.}\ }\textbf {\bibinfo
  {volume} {2021}},\ \bibinfo {pages} {093102} (\bibinfo {year}
  {2021})}\BibitemShut {NoStop}%
\bibitem [{\citenamefont {Kiefer-Emmanouilidis}\ \emph
  {et~al.}(2020{\natexlab{b}})\citenamefont {Kiefer-Emmanouilidis},
  \citenamefont {Unanyan}, \citenamefont {Fleischhauer},\ and\ \citenamefont
  {Sirker}}]{kiefer2020evidence}%
  \BibitemOpen
  \bibfield  {author} {\bibinfo {author} {\bibfnamefont {M.}~\bibnamefont
  {Kiefer-Emmanouilidis}}, \bibinfo {author} {\bibfnamefont {R.}~\bibnamefont
  {Unanyan}}, \bibinfo {author} {\bibfnamefont {M.}~\bibnamefont
  {Fleischhauer}}, \ and\ \bibinfo {author} {\bibfnamefont {J.}~\bibnamefont
  {Sirker}},\ }\href {https://link.aps.org/doi/10.1103/PhysRevLett.124.243601}
  {\bibfield  {journal} {\bibinfo  {journal} {Phys. Rev. Lett.}\ }\textbf
  {\bibinfo {volume} {124}},\ \bibinfo {pages} {243601} (\bibinfo {year}
  {2020}{\natexlab{b}})}\BibitemShut {NoStop}%
\bibitem [{\citenamefont {Kiefer-Emmanouilidis}\ \emph
  {et~al.}(2021{\natexlab{a}})\citenamefont {Kiefer-Emmanouilidis},
  \citenamefont {Unanyan}, \citenamefont {Fleischhauer},\ and\ \citenamefont
  {Sirker}}]{kiefer2021slow}%
  \BibitemOpen
  \bibfield  {author} {\bibinfo {author} {\bibfnamefont {M.}~\bibnamefont
  {Kiefer-Emmanouilidis}}, \bibinfo {author} {\bibfnamefont {R.}~\bibnamefont
  {Unanyan}}, \bibinfo {author} {\bibfnamefont {M.}~\bibnamefont
  {Fleischhauer}}, \ and\ \bibinfo {author} {\bibfnamefont {J.}~\bibnamefont
  {Sirker}},\ }\href {https://link.aps.org/doi/10.1103/PhysRevB.103.024203}
  {\bibfield  {journal} {\bibinfo  {journal} {Phys. Rev. B}\ }\textbf {\bibinfo
  {volume} {103}},\ \bibinfo {pages} {024203} (\bibinfo {year}
  {2021}{\natexlab{a}})}\BibitemShut {NoStop}%
\bibitem [{\citenamefont {Kiefer-Emmanouilidis}\ \emph
  {et~al.}(2021{\natexlab{b}})\citenamefont {Kiefer-Emmanouilidis},
  \citenamefont {Unanyan}, \citenamefont {Fleischhauer},\ and\ \citenamefont
  {Sirker}}]{kiefer2021unlimited}%
  \BibitemOpen
  \bibfield  {author} {\bibinfo {author} {\bibfnamefont {M.}~\bibnamefont
  {Kiefer-Emmanouilidis}}, \bibinfo {author} {\bibfnamefont {R.}~\bibnamefont
  {Unanyan}}, \bibinfo {author} {\bibfnamefont {M.}~\bibnamefont
  {Fleischhauer}}, \ and\ \bibinfo {author} {\bibfnamefont {J.}~\bibnamefont
  {Sirker}},\ }\href
  {https://www.sciencedirect.com/science/article/pii/S0003491621000877}
  {\bibfield  {journal} {\bibinfo  {journal} {Ann. Phys.}\ }\textbf {\bibinfo
  {volume} {435}},\ \bibinfo {pages} {168481} (\bibinfo {year}
  {2021}{\natexlab{b}})}\BibitemShut {NoStop}%
\bibitem [{\citenamefont {Kiefer-Emmanouilidis}\ \emph
  {et~al.}(2022)\citenamefont {Kiefer-Emmanouilidis}, \citenamefont {Unanyan},
  \citenamefont {Fleischhauer},\ and\ \citenamefont
  {Sirker}}]{kiefer2022particle}%
  \BibitemOpen
  \bibfield  {author} {\bibinfo {author} {\bibfnamefont {M.}~\bibnamefont
  {Kiefer-Emmanouilidis}}, \bibinfo {author} {\bibfnamefont {R.}~\bibnamefont
  {Unanyan}}, \bibinfo {author} {\bibfnamefont {M.}~\bibnamefont
  {Fleischhauer}}, \ and\ \bibinfo {author} {\bibfnamefont {J.}~\bibnamefont
  {Sirker}},\ }\href {\doibase 10.21468/SciPostPhys.12.1.034} {\bibfield
  {journal} {\bibinfo  {journal} {SciPost Phys.}\ }\textbf {\bibinfo {volume}
  {12}},\ \bibinfo {pages} {34} (\bibinfo {year} {2022})}\BibitemShut {NoStop}%
\bibitem [{\citenamefont {Luitz}\ and\ \citenamefont
  {Lev}(2020)}]{luitz2020absence}%
  \BibitemOpen
  \bibfield  {author} {\bibinfo {author} {\bibfnamefont {D.~J.}\ \bibnamefont
  {Luitz}}\ and\ \bibinfo {author} {\bibfnamefont {Y.~B.}\ \bibnamefont
  {Lev}},\ }\href {https://link.aps.org/doi/10.1103/PhysRevB.102.100202}
  {\bibfield  {journal} {\bibinfo  {journal} {Phys. Rev. B}\ }\textbf {\bibinfo
  {volume} {102}},\ \bibinfo {pages} {100202} (\bibinfo {year}
  {2020})}\BibitemShut {NoStop}%
\bibitem [{\citenamefont {Ghosh}\ and\ \citenamefont
  {\v{Z}nidari\v{c}}(2022)}]{znidaric2022resonance}%
  \BibitemOpen
  \bibfield  {author} {\bibinfo {author} {\bibfnamefont {R.}~\bibnamefont
  {Ghosh}}\ and\ \bibinfo {author} {\bibfnamefont {M.}~\bibnamefont
  {\v{Z}nidari\v{c}}},\ }\href
  {https://link.aps.org/doi/10.1103/PhysRevB.105.144203} {\bibfield  {journal}
  {\bibinfo  {journal} {Phys. Rev. B}\ }\textbf {\bibinfo {volume} {105}},\
  \bibinfo {pages} {144203} (\bibinfo {year} {2022})}\BibitemShut {NoStop}%
\bibitem [{\citenamefont {Lukin}\ \emph {et~al.}(2019)\citenamefont {Lukin},
  \citenamefont {Rispoli}, \citenamefont {Schittko}, \citenamefont {Tai},
  \citenamefont {Kaufman}, \citenamefont {Choi}, \citenamefont {Khemani},
  \citenamefont {L{\'{e} }onard},\ and\ \citenamefont {Greiner}}]{Lukin_2019}%
  \BibitemOpen
  \bibfield  {author} {\bibinfo {author} {\bibfnamefont {A.}~\bibnamefont
  {Lukin}}, \bibinfo {author} {\bibfnamefont {M.}~\bibnamefont {Rispoli}},
  \bibinfo {author} {\bibfnamefont {R.}~\bibnamefont {Schittko}}, \bibinfo
  {author} {\bibfnamefont {M.~E.}\ \bibnamefont {Tai}}, \bibinfo {author}
  {\bibfnamefont {A.~M.}\ \bibnamefont {Kaufman}}, \bibinfo {author}
  {\bibfnamefont {S.}~\bibnamefont {Choi}}, \bibinfo {author} {\bibfnamefont
  {V.}~\bibnamefont {Khemani}}, \bibinfo {author} {\bibfnamefont
  {J.}~\bibnamefont {L{\'{e} }onard}}, \ and\ \bibinfo {author} {\bibfnamefont
  {M.}~\bibnamefont {Greiner}},\ }\href
  {https://doi.org/10.1126%2Fscience.aau0818} {\bibfield  {journal} {\bibinfo
  {journal} {Science}\ }\textbf {\bibinfo {volume} {364}},\ \bibinfo {pages}
  {256} (\bibinfo {year} {2019})}\BibitemShut {NoStop}%
\bibitem [{\citenamefont {Neven}\ \emph {et~al.}(2021)\citenamefont {Neven},
  \citenamefont {Carrasco}, \citenamefont {Vitale}, \citenamefont {Kokail},
  \citenamefont {Elben}, \citenamefont {Dalmonte}, \citenamefont {Calabrese},
  \citenamefont {Zoller}, \citenamefont {Vermersch}, \citenamefont {Kueng},\
  and\ \citenamefont {Kraus}}]{neven2021symmetry}%
  \BibitemOpen
  \bibfield  {author} {\bibinfo {author} {\bibfnamefont {A.}~\bibnamefont
  {Neven}}, \bibinfo {author} {\bibfnamefont {J.}~\bibnamefont {Carrasco}},
  \bibinfo {author} {\bibfnamefont {V.}~\bibnamefont {Vitale}}, \bibinfo
  {author} {\bibfnamefont {C.}~\bibnamefont {Kokail}}, \bibinfo {author}
  {\bibfnamefont {A.}~\bibnamefont {Elben}}, \bibinfo {author} {\bibfnamefont
  {M.}~\bibnamefont {Dalmonte}}, \bibinfo {author} {\bibfnamefont
  {P.}~\bibnamefont {Calabrese}}, \bibinfo {author} {\bibfnamefont
  {P.}~\bibnamefont {Zoller}}, \bibinfo {author} {\bibfnamefont
  {B.}~\bibnamefont {Vermersch}}, \bibinfo {author} {\bibfnamefont
  {R.}~\bibnamefont {Kueng}}, \ and\ \bibinfo {author} {\bibfnamefont
  {B.}~\bibnamefont {Kraus}},\ }\href
  {https://doi.org/10.1038%2Fs41534-021-00487-y} {\bibfield  {journal}
  {\bibinfo  {journal} {npj Quantum Inf.}\ }\textbf {\bibinfo {volume} {7}}
  (\bibinfo {year} {2021})}\BibitemShut {NoStop}%
\bibitem [{\citenamefont {Vasseur}\ \emph {et~al.}(2015)\citenamefont
  {Vasseur}, \citenamefont {Potter},\ and\ \citenamefont
  {Parameswaran}}]{vasseur1}%
  \BibitemOpen
  \bibfield  {author} {\bibinfo {author} {\bibfnamefont {R.}~\bibnamefont
  {Vasseur}}, \bibinfo {author} {\bibfnamefont {A.~C.}\ \bibnamefont {Potter}},
  \ and\ \bibinfo {author} {\bibfnamefont {S.~A.}\ \bibnamefont
  {Parameswaran}},\ }\href
  {https://link.aps.org/doi/10.1103/PhysRevLett.114.217201} {\bibfield
  {journal} {\bibinfo  {journal} {Phys. Rev. Lett.}\ }\textbf {\bibinfo
  {volume} {114}},\ \bibinfo {pages} {217201} (\bibinfo {year}
  {2015})}\BibitemShut {NoStop}%
\bibitem [{\citenamefont {Pappalardi}\ \emph {et~al.}(2019)\citenamefont
  {Pappalardi}, \citenamefont {Calabrese},\ and\ \citenamefont
  {Parisi}}]{Pappalardi_2019}%
  \BibitemOpen
  \bibfield  {author} {\bibinfo {author} {\bibfnamefont {S.}~\bibnamefont
  {Pappalardi}}, \bibinfo {author} {\bibfnamefont {P.}~\bibnamefont
  {Calabrese}}, \ and\ \bibinfo {author} {\bibfnamefont {G.}~\bibnamefont
  {Parisi}},\ }\href {\doibase 10.1088/1742-5468/ab2903} {\bibfield  {journal}
  {\bibinfo  {journal} {Journal of Statistical Mechanics: Theory and
  Experiment}\ }\textbf {\bibinfo {volume} {2019}},\ \bibinfo {pages} {073102}
  (\bibinfo {year} {2019})}\BibitemShut {NoStop}%
\bibitem [{\citenamefont {Monthus}(2015)}]{Monthus_2015}%
  \BibitemOpen
  \bibfield  {author} {\bibinfo {author} {\bibfnamefont {C.}~\bibnamefont
  {Monthus}},\ }\href
  {https://doi.org/10.1088%2F1742-5468%2F2015%2F10%2Fp10024} {\bibfield
  {journal} {\bibinfo  {journal} {J. Stat. Mech. Theory Exp.}\ }\textbf
  {\bibinfo {volume} {2015}},\ \bibinfo {pages} {P10024} (\bibinfo {year}
  {2015})}\BibitemShut {NoStop}%
\bibitem [{\citenamefont {Schachenmayer}\ \emph {et~al.}(2013)\citenamefont
  {Schachenmayer}, \citenamefont {Lanyon}, \citenamefont {Roos},\ and\
  \citenamefont {Daley}}]{prxent}%
  \BibitemOpen
  \bibfield  {author} {\bibinfo {author} {\bibfnamefont {J.}~\bibnamefont
  {Schachenmayer}}, \bibinfo {author} {\bibfnamefont {B.~P.}\ \bibnamefont
  {Lanyon}}, \bibinfo {author} {\bibfnamefont {C.~F.}\ \bibnamefont {Roos}}, \
  and\ \bibinfo {author} {\bibfnamefont {A.~J.}\ \bibnamefont {Daley}},\ }\href
  {\doibase 10.1103/PhysRevX.3.031015} {\bibfield  {journal} {\bibinfo
  {journal} {Phys. Rev. X}\ }\textbf {\bibinfo {volume} {3}},\ \bibinfo {pages}
  {031015} (\bibinfo {year} {2013})}\BibitemShut {NoStop}%
\bibitem [{\citenamefont {Lerose}\ and\ \citenamefont
  {Pappalardi}(2020{\natexlab{a}})}]{pappalardi2}%
  \BibitemOpen
  \bibfield  {author} {\bibinfo {author} {\bibfnamefont {A.}~\bibnamefont
  {Lerose}}\ and\ \bibinfo {author} {\bibfnamefont {S.}~\bibnamefont
  {Pappalardi}},\ }\href {\doibase 10.1103/PhysRevResearch.2.012041} {\bibfield
   {journal} {\bibinfo  {journal} {Phys. Rev. Research}\ }\textbf {\bibinfo
  {volume} {2}},\ \bibinfo {pages} {012041} (\bibinfo {year}
  {2020}{\natexlab{a}})}\BibitemShut {NoStop}%
\bibitem [{\citenamefont {Lerose}\ and\ \citenamefont
  {Pappalardi}(2020{\natexlab{b}})}]{pappalardi3}%
  \BibitemOpen
  \bibfield  {author} {\bibinfo {author} {\bibfnamefont {A.}~\bibnamefont
  {Lerose}}\ and\ \bibinfo {author} {\bibfnamefont {S.}~\bibnamefont
  {Pappalardi}},\ }\href {\doibase 10.1103/PhysRevA.102.032404} {\bibfield
  {journal} {\bibinfo  {journal} {Phys. Rev. A}\ }\textbf {\bibinfo {volume}
  {102}},\ \bibinfo {pages} {032404} (\bibinfo {year}
  {2020}{\natexlab{b}})}\BibitemShut {NoStop}%
\bibitem [{\citenamefont {Pappalardi}\ \emph {et~al.}(2018)\citenamefont
  {Pappalardi}, \citenamefont {Russomanno}, \citenamefont {\ifmmode
  \check{Z}\else \v{Z}\fi{}unkovi\ifmmode~\check{c}\else \v{c}\fi{}},
  \citenamefont {Iemini}, \citenamefont {Silva},\ and\ \citenamefont
  {Fazio}}]{pappalardi4}%
  \BibitemOpen
  \bibfield  {author} {\bibinfo {author} {\bibfnamefont {S.}~\bibnamefont
  {Pappalardi}}, \bibinfo {author} {\bibfnamefont {A.}~\bibnamefont
  {Russomanno}}, \bibinfo {author} {\bibfnamefont {B.}~\bibnamefont {\ifmmode
  \check{Z}\else \v{Z}\fi{}unkovi\ifmmode~\check{c}\else \v{c}\fi{}}}, \bibinfo
  {author} {\bibfnamefont {F.}~\bibnamefont {Iemini}}, \bibinfo {author}
  {\bibfnamefont {A.}~\bibnamefont {Silva}}, \ and\ \bibinfo {author}
  {\bibfnamefont {R.}~\bibnamefont {Fazio}},\ }\href {\doibase
  10.1103/PhysRevB.98.134303} {\bibfield  {journal} {\bibinfo  {journal} {Phys.
  Rev. B}\ }\textbf {\bibinfo {volume} {98}},\ \bibinfo {pages} {134303}
  (\bibinfo {year} {2018})}\BibitemShut {NoStop}%
\bibitem [{\citenamefont {Carollo}\ and\ \citenamefont
  {Alba}(2022)}]{carollo2021emergent}%
  \BibitemOpen
  \bibfield  {author} {\bibinfo {author} {\bibfnamefont {F.}~\bibnamefont
  {Carollo}}\ and\ \bibinfo {author} {\bibfnamefont {V.}~\bibnamefont {Alba}},\
  }\href {\doibase 10.1103/PhysRevB.105.144305} {\bibfield  {journal} {\bibinfo
   {journal} {Phys. Rev. B}\ }\textbf {\bibinfo {volume} {105}},\ \bibinfo
  {pages} {144305} (\bibinfo {year} {2022})}\BibitemShut {NoStop}%
\bibitem [{\citenamefont {Wellnitz}\ \emph {et~al.}()\citenamefont {Wellnitz},
  \citenamefont {Preisser}, \citenamefont {Alba}, \citenamefont {Dubail},\ and\
  \citenamefont {Schachenmayer}}]{wellnitz2022rise}%
  \BibitemOpen
  \bibfield  {author} {\bibinfo {author} {\bibfnamefont {D.}~\bibnamefont
  {Wellnitz}}, \bibinfo {author} {\bibfnamefont {G.}~\bibnamefont {Preisser}},
  \bibinfo {author} {\bibfnamefont {V.}~\bibnamefont {Alba}}, \bibinfo {author}
  {\bibfnamefont {J.}~\bibnamefont {Dubail}}, \ and\ \bibinfo {author}
  {\bibfnamefont {J.}~\bibnamefont {Schachenmayer}},\ }\href
  {https://arxiv.org/abs/2201.05099} {\ }\Eprint
  {http://arxiv.org/abs/2201.05099} {2201.05099 [quant-ph]} \BibitemShut
  {NoStop}%
\bibitem [{\citenamefont {Monthus}(2017)}]{monthus2017dissipative}%
  \BibitemOpen
  \bibfield  {author} {\bibinfo {author} {\bibfnamefont {C.}~\bibnamefont
  {Monthus}},\ }\href {https://doi.org/10.1088/1742-5468/aa64f4} {\bibfield
  {journal} {\bibinfo  {journal} {J. Stat. Mech. Theory Exp.}\ }\textbf
  {\bibinfo {volume} {2017}},\ \bibinfo {pages} {043302} (\bibinfo {year}
  {2017})}\BibitemShut {NoStop}%
\bibitem [{\citenamefont {Zabalo}\ \emph {et~al.}()\citenamefont {Zabalo},
  \citenamefont {Wilson}, \citenamefont {Gullans}, \citenamefont {Vasseur},
  \citenamefont {Gopalakrishnan}, \citenamefont {Huse},\ and\ \citenamefont
  {Pixley}}]{zabalo2022disordered}%
  \BibitemOpen
  \bibfield  {author} {\bibinfo {author} {\bibfnamefont {A.}~\bibnamefont
  {Zabalo}}, \bibinfo {author} {\bibfnamefont {J.~H.}\ \bibnamefont {Wilson}},
  \bibinfo {author} {\bibfnamefont {M.~J.}\ \bibnamefont {Gullans}}, \bibinfo
  {author} {\bibfnamefont {R.}~\bibnamefont {Vasseur}}, \bibinfo {author}
  {\bibfnamefont {S.}~\bibnamefont {Gopalakrishnan}}, \bibinfo {author}
  {\bibfnamefont {D.~A.}\ \bibnamefont {Huse}}, \ and\ \bibinfo {author}
  {\bibfnamefont {J.~H.}\ \bibnamefont {Pixley}},\ }\href
  {https://arxiv.org/abs/2205.14002} {}\Eprint
  {http://arxiv.org/abs/2205.14002} {2205.14002 [quant-ph]} \BibitemShut
  {NoStop}%
\end{thebibliography}%

\end{document}